\documentclass{iopart}

\usepackage{iopams}
\usepackage{graphicx}

\usepackage{mymacros}

\begin{document}

\bibliographystyle{prsty}


\title[Continued-fraction methods for spin quantum-master equations]
{Solving spin quantum-master equations\\ 
with matrix continued-fraction methods:\\ 
application to superparamagnets}

\author{
J. L. Garc\'{\i}a-Palacios and D. Zueco
}

\address{
Dep.\ de F\'{\i}sica de la Materia Condensada e
Instituto de Ciencia de Materiales de~Arag\'on,
C.S.I.C.--Universidad de Zaragoza,
E-50009 Zaragoza, Spain
}


\date{\today}

\begin{abstract}
We implement continued-fraction techniques to solve exactly quantum
master equations for a spin with arbitrary $S$ coupled to a (bosonic)
thermal bath.
The full spin density matrix is obtained, so that along with
relaxation and thermoactivation, coherent dynamics is included
(precession, tunnel, etc.).
The method is applied to study isotropic spins and spins in a bistable
anisotropy potential (superparamagnets).
We present examples of static response, the dynamical susceptibility
including the contribution of the different relaxation modes, and of
spin resonance in transverse fields.
\end{abstract}



\section{
Introduction
}

In the field of {\em quantum dissipative systems} one studies a
subsystem consisting of a few relevant degrees of freedom coupled to
the surrounding medium (bath), which has a large number of
constituents (e.g., photons, phonons, electrons, nuclei)
\cite{weiss,guibascal98,datpur2004}.
The (sub)system is not necessarily microscopic, but it can be a
mesoscopic system (a Josephson junction, a magnetic molecular cluster,
etc.)  described by a few {\em collective\/} variables (phase across
the junction, net spin) which under certain conditions can display
quantum behaviour.
Various fundamental problems can be addressed, like dissipation and
quantum mechanics, decoherence, quantum Brownian motion, or the
quantum-to-classical transition.
The interaction with the bath, apart from producing dissipation,
fluctuations, and decoherence, enables the system to interchange
energy, momentum, and correlations with its environment and eventually
relax to thermal equilibrium.
For these reasons the study of open quantum systems is of
interest in many areas of physics and chemistry.

{\em Classical\/} open systems are handled as stochastic systems by
means of Langevin and Fokker--Planck equations
\cite{risken}.
This approach provides both a theoretical frame and computational
tools, e.g., Langevin molecular-dynamics simulations.
For few-variable systems, a powerful technique to solve Fokker--Planck
equations is Risken's {\em continued-fraction method\/} \cite{risken}
(a relative of Grad's moment approach for solving kinetic equations
\cite{balescu2}).
The non-equilibrium distribution $\W$ is expanded in a basis of
functions and the coupled equations for the expansion coefficients
$\ec_{\ir}$ derived.
An appropriate basis choice can render finite the coupling range
(e.g., with the equation for $\ec_{\ir}$ involving $\ec_{\ir-1}$,
$\ec_{\ir}$, and $\ec_{\ir+1}$).
Then these recurrences can be solved by iterating a simple algorithm,
which has a structure akin to a continued fraction.
This method provides numerically exact results in problems of Brownian
motion in external potentials, where closed form solutions are scarce
(such classical problems are structurally similar to solve a
Schr\"odinger-type equation).

It results more delicate to deal with {\em quantum\/} dissipative
systems \cite{weiss}.
First, phenomenological or non-standard quantisation poses problems
with basic quantum-mechanical principles \cite{calleg83}.
Thus, one has to model the environment in a simple way (set of
oscillators, 2-state systems, etc.), quantise it together with the
system, and eventually trace over the bath variables.
However, the resulting reduced descriptions are difficult to manage
except in simple cases---free particle (or in a uniform field),
harmonic oscillator, and few-state systems (e.g., $S=1/2$ spins).
In general, the exact {\em path-integral\/} expressions for the
(reduced) density operator $\dm(t)$ are not easy to handle
\cite{feynman-pathints,pathints-ingold}; besides, the
propagating function is highly oscillatory, rendering numerical
methods unstable at long times \cite{sto2003}.
{\em Quantum Langevin equations} (Heisenberg equations for $x$ and $p$
including operator fluctuations) are of limited use beyond linear
systems \cite{forlewoco88,bhattacharjee}.
Finally, under certain conditions (typically weak system-bath
coupling), the density matrix obeys a {\em quantum master equation}
\cite{kargra97,ank2003qme}.
But again these equations can only be solved in a few problems.

Due to their performance in classical systems (both translational
\cite{junris85,feretal93pa} and rotational \cite{cofkalwal94}),
continued-fraction techniques were adapted to several problems of
quantum Brownian motion in non-trivial potentials.
This was done exploiting pseudo-probability representations of $\dm$
\cite{lee95}.
Shibata and co-workers \cite{shiuch93} applied continued fractions to
solve master equations for isotropic spins ($\Hs=-\Bz\Sz$); Vogel and
Risken tackled similarly quantum non-linear optical problems
\cite{vogris88}.
In phase-space problems, using the Wigner representation of $\dm$, the
continued-fraction method for the Klein--Kramers equation was adapted
\cite{garzue2004} to quantum master equations of Caldeira--Leggett
type \cite{calleg83,anahal95} (explicit recurrences were presented for
polynomial and periodic potentials).
As these approaches do not rely on the Hamiltonian eigenstructure they
are applicable to demanding problems with (partially) continuous
spectrum.

\begin{figure}[!t]
\centerline{
\includegraphics[width=7.2cm]{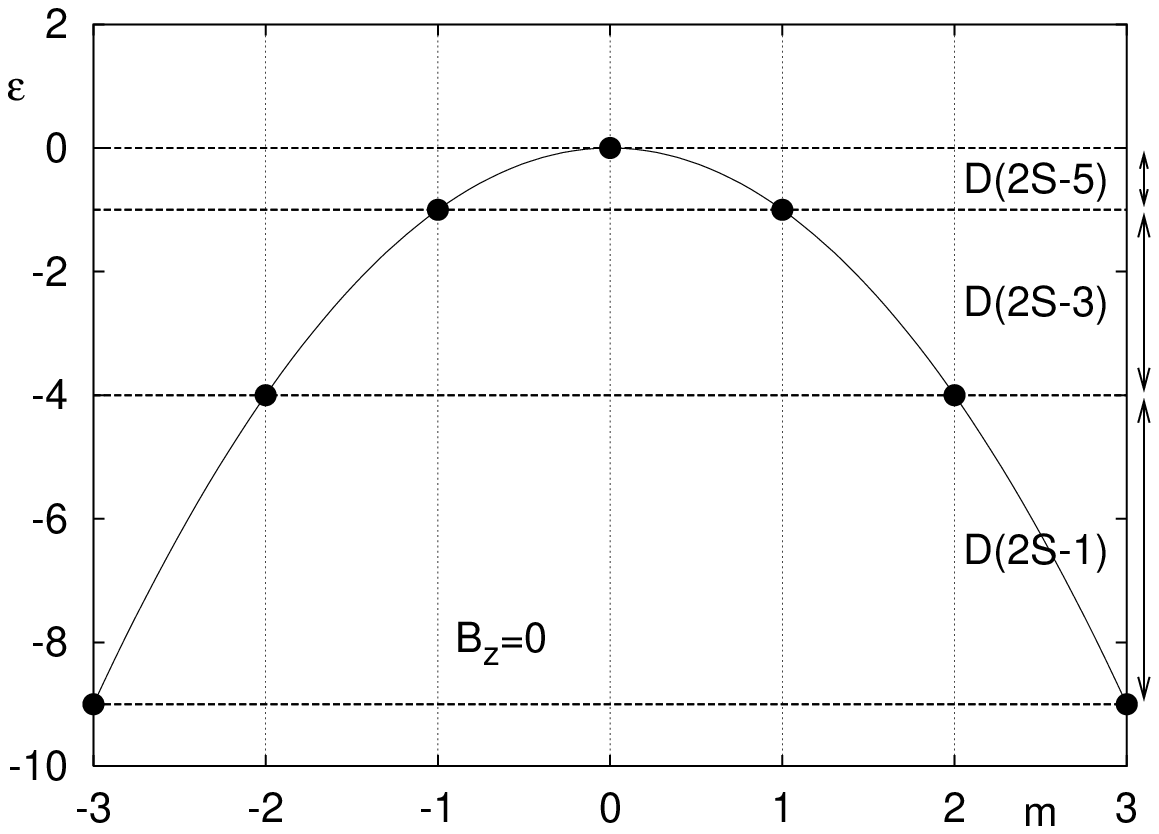}
\hspace*{-3.ex}
\includegraphics[width=7.2cm]{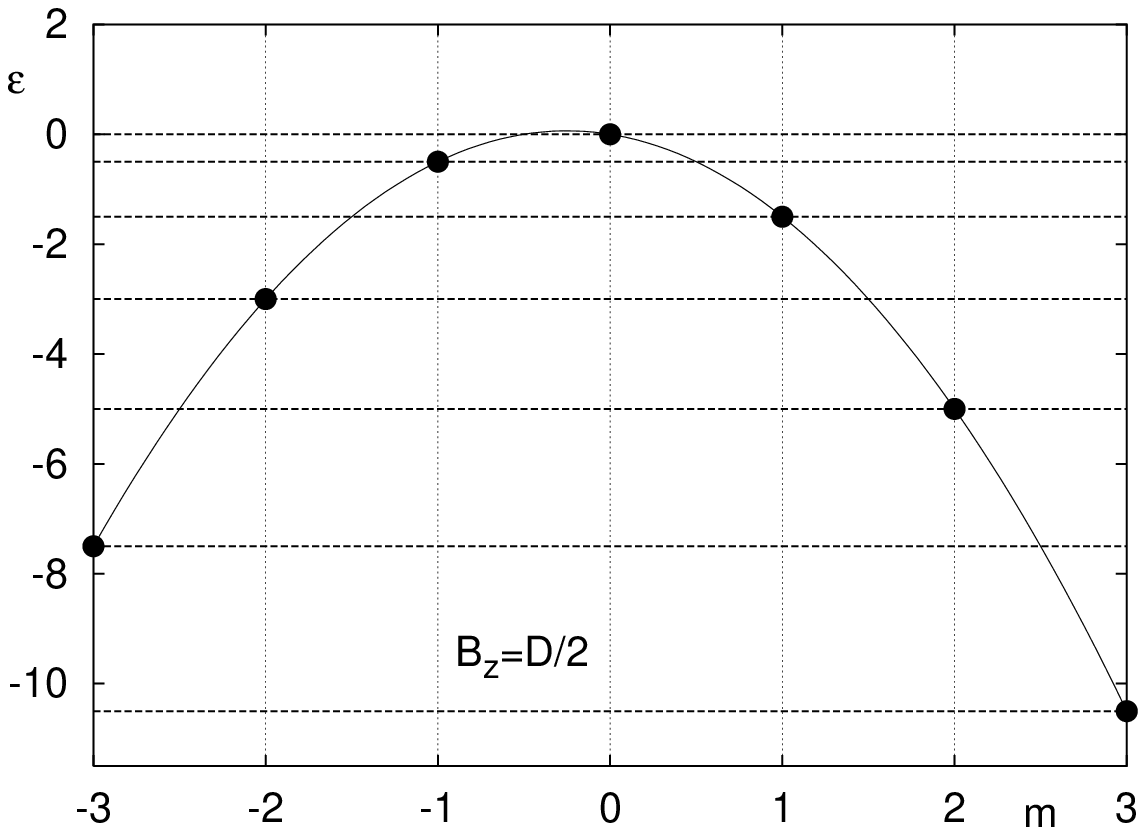}
}
\vspace*{-1.ex}
\centerline{
\includegraphics[width=7.2cm]{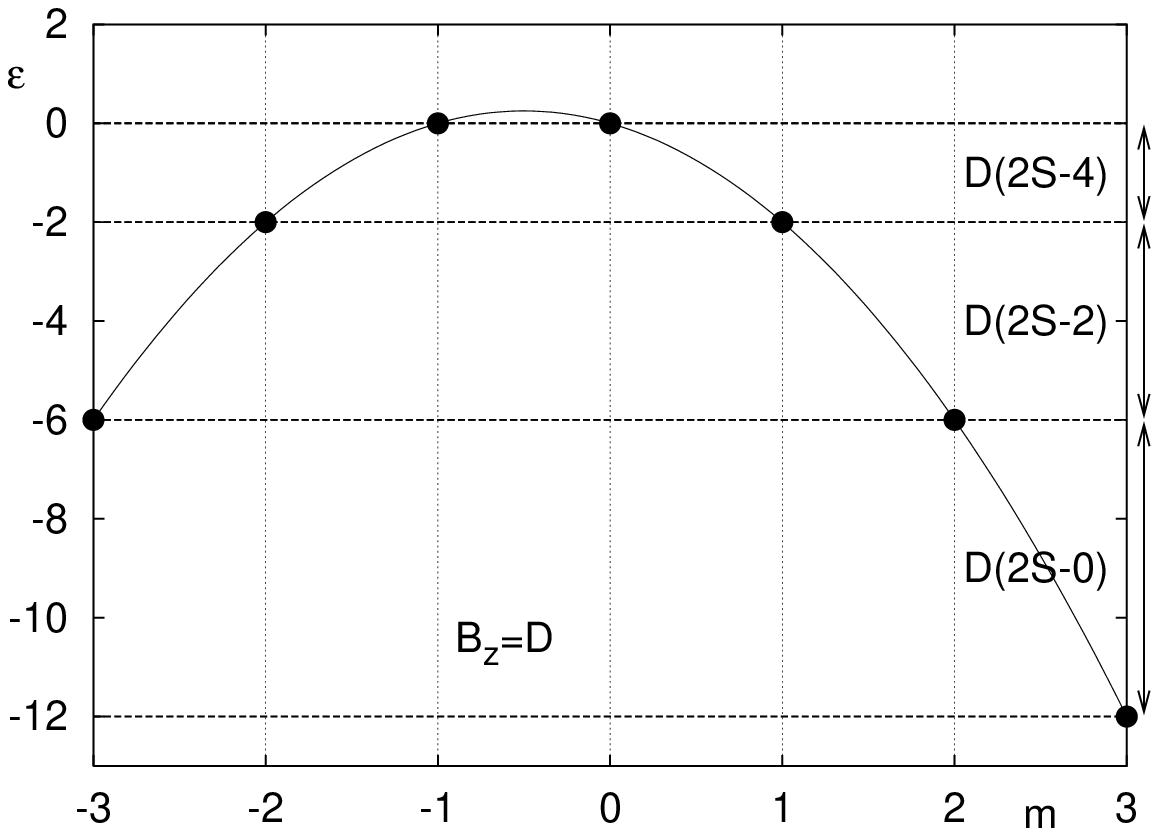}
\hspace*{-3.ex}
\includegraphics[width=7.2cm]{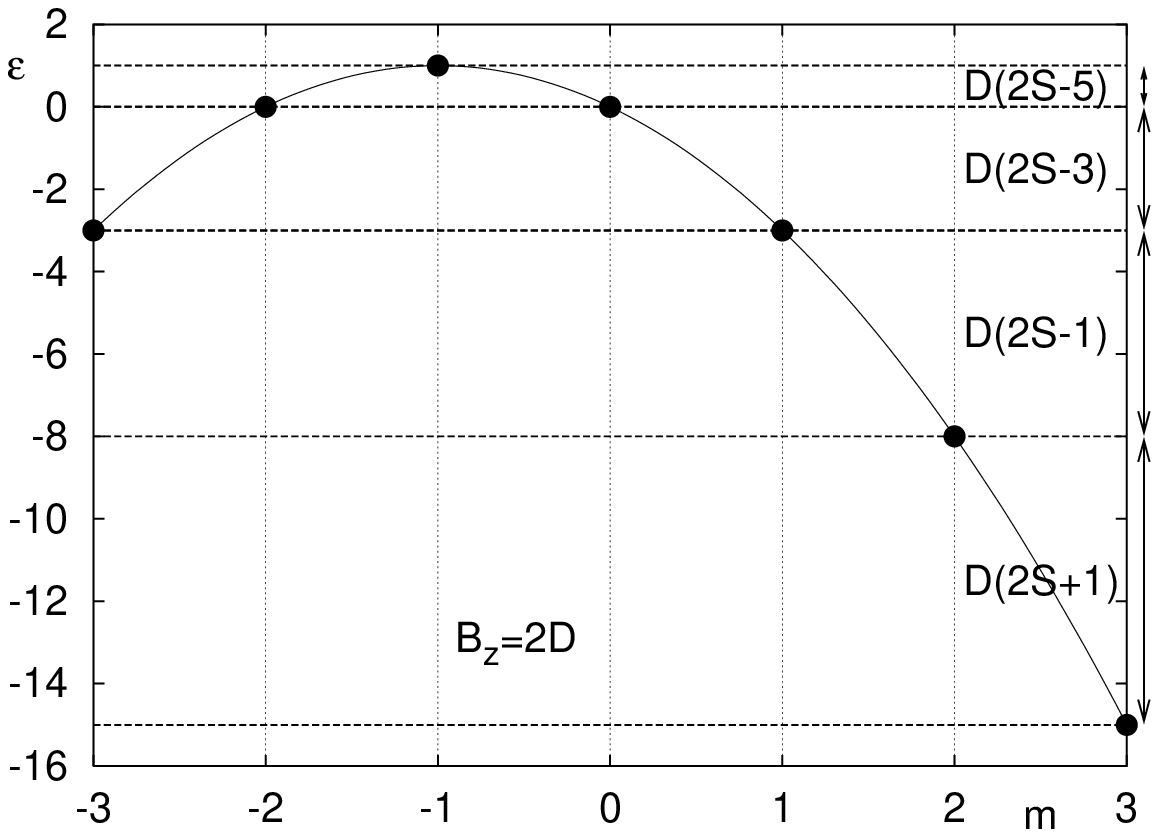}
}
\vspace*{-1.ex}
\caption{
Bistable energy levels of an anisotropic spin,
$\el_{\m}=-\K\m^{2}-\Bz\m$, with $S=3$ and $\K=1$ at zero field (top
left).
An external field lifts the degeneracy $\el_{-\m}=\el_{\m}$ (top right).
Degeneracies are restored at multiples of $\K$ (lower panels).
}
\label{fig:levels}
\end{figure}

Here we consider the following quantum-dissipative system: a spin in
the magnetic anisotropy potential ($\Hs=-\K\,\Sz^{2}-\B\cdot\vS$)
coupled to a boson (or bosonizable) thermal bath.
For $\K=0$ we recover the familiar isotropic spin with its equispaced
Zeeman spectrum; in a sense, the rotational equivalent of the harmonic
oscillator.
The anisotropy term $-\K\,\Sz^{2}$ makes the problem tougher, say, the
spin analogue of Brownian motion in double-well or periodic
potentials.
The environmental disturbances may indeed provoke a Brownian-type
``reversal'' of the spin, overcoming the potential barriers
(Fig.~\ref{fig:levels}).
In spite of the analogies, however, dissipative spin dynamics presents
essential differences with translational problems, due to the
underlying angular-momentum commutation relations
$[\Si\,,\Sj]=\iu\,\epsilon_{ijk}\Sk$
\cite{linmohses83,jay91,gar99}.

Our Hamiltonian describes paramagnets and superparamagnets---small
solids or clusters with a sizable net spin ($S\sim10^{1}$--$10^{4}$).
For large $S$ the physics is approximately classical (as in magnetic
nanoparticles) and described by a {\em rotational\/} Fokker--Planck
equation after Brown \cite{bro63} and Kubo--Hashitsume
\cite{kubhas70}.
As the spin decreases, quantum effects come to the fore, as in
magnetic molecular clusters where $S\sim5$--$25$ \cite{blupra2004}.
The discreteness of the energy levels sensibly affects the
thermoactivation processes, while the spin reversal may also occur by
tunnelling when the field brings into resonance states at both sides
of the barrier (Fig.~\ref{fig:levels}).
An appealing dynamical description is given by master equations of
{\em Pauli\/} type, for the diagonal elements of $\dm$ (``balance'' or
``gain--loss'' equations) \cite{bhattacharjee,pake}.
These provide insight, while more refined treatments are less intuitive
and difficult to apply.
Nevertheless, to take into account {\em coherent\/} dynamics, like
tunnel oscillations or the spin precession, one also needs
off-diagonal density-matrix elements.
But, as usual, solving the master equations for the full density
matrix is not easy and several (often drastic) simplifications are
required.

In this work, following the spirit of
Refs.~\cite{shiuch93,vogris88,garzue2004}, we will solve master
equations for non-interacting spins in contact with a dissipative
bath by means of continued-fraction techniques.
Exact methods available are limited to small spins ($S\lesssim5$--$10$
\cite{wur98}).
Our aim is to tackle arbitrary $S$, from the extreme quantum cases,
$S=1/2$ and~$1$, to values as close as possible to the classical
domain.
This approach differs from those giving some continued-fraction
expression for a certain quantity (typically relying on
linear-response theory; see the review \cite{balleetog2003}), in that
here the {\em full solution\/} of the density-matrix equation is
obtained by {\em matrix\/} continued-fraction methods.
In this theoretical-computational frame we can study spins in a wide
range of $S$ ($\lesssim100$--$200$) including their full dynamics:
relaxation and thermoactivation, precession and coherence, as well as
their possible interplay.

The manuscript is organised as follows.
We discuss the isolated spin and present the basic formalism in next
section.
In Sec.~\ref{sec:dissipative} we introduce dissipative equations for a
spin coupled to a bosonic bath, following the compact approach of
Garanin {\em et al.} \cite{gar91llb,garchu97} with Heisenberg
equations of motion for the Hubbard operators $\Xnm=\nket\mbra$.
Master equations in the Markovian regime are discussed in
Sec.~\ref{DME:markov} and fully specified for several spin problems in
Sec.~\ref{sec:DME:SisoSani}.
In Sec.~\ref{sec:pert} we derive the chain of equations resulting
from the perturbative treatment of probing fields (applicable to the
non-linear response).
In Sec.~\ref{sec:CF-DME} we manipulate the index-coupling structure
of the density-matrix equations to obtain few-term recurrences
suitable for implementing the continued-fraction algorithm.
Numerous examples of its application to isotropic and anisotropic
spins (superparamagnets) are given in Secs.~\ref{sec:Siso}
and~\ref{sec:Sani}; we will check the results against exact formulae,
whenever possible, and test heuristic expressions.
We conclude with an assessment of our approach in
Sec.~\ref{sec:summary} and putting some auxiliary material and
discussing technical issues in the appendices.


\section{
Isolated spin (unitary dynamics) and Hubbard formalism 
}
\label{sec:unitary}



Our starting point is a spin Hamiltonian \cite{pake,white} including a
magnetic anisotropy term and the Zeeman coupling to the field
(units $\hbar=k_{\rm B}=g\mu_{\rm B}=1$)
%
\begin{equation}
\label{H:spin}
\Hs(\vS)
=
-\K\,\Sz^{2}
-
\B
\cdot
\vS
\;.
\end{equation}
This is the minimal model capturing the physics of (super)paramagnets.
The term $-\K\,\Sz^{2}$ has a bistable structure (for $\K>0$) with
minima at $\Sz=\pm S$ and barrier top at $\Sz=0$
(Fig.~\ref{fig:levels}).
Along with potential barriers (and degeneracies), an important
consequence of the anisotropy is an unequally spaced energy spectrum
(\ref{app:energetics}).%
\footnote{
Hamiltonian~(\ref{H:spin}) also describes a set of $N$ two-level
systems interacting uniformly (Lipkin--Meshkov--Glick model)
\cite{ulyzas92}.
The spectrum of $2^{N}$ eigenvalues splits in multiplets characterized
by a certain $S\leq N/2$ and described by a pseudo-spin Hamiltonian
$\Hs=-\K\,\Sz^{2}-\Bx\Sx$; the excitation energy of each two-level
element corresponds to the field and their coupling to the anisotropy
parameter.
This is a problem where the possibility of handling large values of
$S$ (large $N$) is important.
} 



Let us introduce the Hubbard (level-shift) operators
\cite[Ch.~1]{gamlev95}
%
\begin{equation}
\label{hubbard}
\Xnm
\equiv
\nket\mbra
\;.
\end{equation}
They form a complete system and any spin operator $\opA$ can be
expressed as
%
\begin{equation}
\label{A:hubbard:rep}
\opA
=
\tsum_{\n\m}
A_{\n\m}
\Xnm
\;,
\qquad
A_{\n\m}
=
\nbra\opA\mket
\;.
\end{equation}
The expansion coefficients are simply the matrix elements of $\opA$
in the basis defining the $\Xnm$.
(If not restricting ourselves to a multiplet with fixed $S$, we just
need to introduce the corresponding indices $|S\,\m\rangle$ and sum
over them.)
In particular, for the components of the spin operator one has
$\Si
=
\sum_{\n\m}
\nbra\Si\mket\Xnm$.
Now, if we use the {\em standard basis\/} of eigenstates of $\vS^{2}$
and $\Sz$, the required matrix elements are
$\nbra\Sz\mket
=
\m\,\delta_{\n\m}$
and
$\nbra\Scpm\mket
=
\lf_{\m}^{\pm}\,\delta_{\n,\m\pm1}$.
Here $\Scpm=\Sx\pm\iu\Sy$ are the ladder operators and
$\lf_{\m}^{\pm}=[S(S+1)-\m(\m\pm1)]^{1/2}$ the factors giving
$\Scpm\mket=\lf_{\m}^{\pm}|\m\pm1\rangle$. 
Then, the $\Si$ are represented by the single-sum forms
%
\begin{equation}
\label{S:hubbard}
\Sz
=
\sum_{\m}
\m\,
X_{\m}^{\m}
\;,
\qquad
\Scpm
=
\sum_{\m}
\lf_{\m}^{\pm}
X_{\m\pm1}^{\m}
\;.
\end{equation}
In general, $f(\Sz)=\sum_{\m}f(\m)\,X_{\m}^{\m}$ for any operator
function $f(\Sz)$; this gives the second-order ``moments'':
$\Sz^{2}
=
\sum_{\m}
\m^{2}\,X_{\m}^{\m}$
and
$\Sx^{2}+\Sy^{2}
=
\sum_{\m}
[S(S+1)-\m^{2}]
\,X_{\m}^{\m}$.

Concerning dynamics, the evolution in the {\em Heisenberg\/}
representation of an operator is governed by
$\iu(\drm\opA/\drm t)=[\opA,\Hs]$.
This plus Hamiltonian~(\ref{H:spin}) gives for $\Xnm$
%
\begin{equation}
\label{DME:closed}
\fl
\frac{\drm{}}{\drm t}\Xnm
=
\iu
\tf_{\n\m}
\Xnm
+
\tfrac{\iu}{2}
\Bcp
\big(\lf_{\m}^{+}\,X_{\n}^{\m+1}-\lf_{\n}^{-}\,X_{\n-1}^{\m}\big)
+
\tfrac{\iu}{2}
\Bcm
\big(\lf_{\m}^{-}\,X_{\n}^{\m-1}-\lf_{\n}^{+}\,X_{\n+1}^{\m}\big)
\quad
\end{equation}
(see \ref{app:hubbard}) where $\tf_{\n\m}$ is the frequency associated to
the $\m\to\n$ transition
%
\begin{equation}
\label{frequencies:levels}
\tf_{\n\m}
\equiv
\el_{\n}-\el_{\m}
\;,
\qquad
\el_{\m}
=
-\K\m^{2}-\Bz\m
\;,
\end{equation}
between levels of the diagonal part of the Hamiltonian.
In the absence of transverse fields the evolution is simply
$\Xnm(t)=\e^{\iu(t-t_{0})\tf_{\n\m}}\Xnm(t_{0})$.
In general, $\Bcpm=\Bx\pm\iu\By$ couples the dynamics of the diagonal
elements, $X_{\m}^{\m}$, with the adjacent sub-diagonals,
$X_{\m}^{\m\pm1}$ and $X_{\m\pm1}^{\m}$, and so on.

Finally, the density operator is expressed as
$\dm
=
\sum_{\n\m}
\dm_{\n\m}
\Xnm$.
The quantum-statistical average of $\Xnm$ then follows from the trace
formula
$\llangle\opA\rrangle
=
\Tr(\dm\,\opA)$
%
%
and reads
$\llangle\Xnm\rrangle=\dm_{\m\n}$
(\ref{app:hubbard}).
This is important as it enables working with the density-matrix
elements $\dm_{\n\m}=\nbra\dm\mket$ or the $\Xnm$ interchangeably,
since all the equations we are going to handle are linear in $X$ and
their averaging thus trivial.
%


\section{
Spin weakly coupled to a bosonic bath (dissipative dynamics)
}
\label{sec:dissipative}

We address now the dynamics of the spin taking into account the
coupling to its surroundings.
Let us consider a total Hamiltonian including that of the spin
$\Hs(\vS)$, a bath of bosons
$\Ham_{\rm b}=\sum_{\bi}\omega_{\bi}\,\ap_{\bi}\am_{\bi}$,
and their interaction
%
\begin{equation}
\label{H}
\Ham_{\rm tot}
=
\Hs(\vS)
+
\tsum_{\bi}
\Cint_{\bi}
F_{\bi}(\vS)
\left(\ap_{\bi}+\am_{-\bi}\right)
+
\Ham_{\rm b}
\;.
\end{equation}
Here $\Cint_{\bi}$ are coupling constants.
The spin-dependent part of the interaction $F_{\bi}(\vS)$ is typically
a low-degree polynomial of $\vS$ \cite{pake,dohful75}.
The coupling written is linear in the bath operators---only $1$-boson
processes are included, no Raman scattering involving two quanta (for
interactions non-linear in the bath variables see
Refs.~\cite{gar99,gar91llb,corweslin85}).
Note that no counterterms are included in $\Ham_{\rm tot}$ to
compensate for the coupling induced renormalization of the spin levels
\cite{calleg83}; we will address this point below.



The total spin-plus-bath system is unlikely to be in a pure state and
a density-matrix description is required.
For observables depending only on the spin, the required object is the
{\em reduced\/} density operator $\dm=\Tr_{\rm b}(\dm_{\rm tot})$,
where one traces the bath out.
For {\em weak\/} system-bath coupling a closed dynamical equation for
$\dm$ can be obtained by perturbation theory.
This is the case of many problems in quantum optics, chemical physics,
or magnetism \cite{weiss,datpur2004}.
The equation has the generic form
$\iu
(\drm\dm/\drm t)
=
[\Hs\,,\dm]
+
\iu\,R[\dm(\tp)]$,
where the {\em relaxation term\/} $R$ adds to the Von Neuman evolution
the effects of the bath.
%

In the Hubbard framework, the Heisenberg time evolution of
$\Xnm=\nket\mbra$ is governed by an analogous equation
\cite{gar91llb,garchu97}
%
\begin{equation}
\label{DME:nonmarkov}
\drm\Xnm/\drm t
=
-\iu
\big[
\Xnm
\,,
\Hs
\big]
+
R_{\n}^{\m}
\;.
\end{equation}
The commutator generates the isolated-spin unitary evolution
[Eq.~(\ref{DME:closed})] and $R_{\n}^{\m}$ accounts for the
dissipation.
When $F(\vS)$ does not depend on the boson index [this is transferable
to $\Cint_{\bi}$; see Eq.~(\ref{H})], the relaxation term can be written
as
%
\begin{equation}
\label{Rnonmarkov}
\fl
R_{\n}^{\m}
=
-
\int_{-\infty}^{t}
\!\drm{\tp}
\Big\{
\Ker(\tp-t)
\,
F(\tp)
\big[
F
\,,
\Xnm
\big]
-
\Ker(t-\tp)
\,
\big[
F
\,,
\Xnm
\big]
F(\tp)
\Big\}
\;.
\end{equation}
This form is equivalent to the standard dissipative terms for $\dm$
obtained by projection operators or cumulant expansions to second
order \cite[App.~1.A]{datpur2004}.
The {\em memory kernel\/} is the autocorrelation
$\Ker(\tp)
\equiv
\langle
B(t+\tp)B(t)
\rangle_{\rm b}$
of the bath operator
$B
=
\tsum_{\bi}
\Cint_{\bi}
(\ap_{\bi}+\am_{-\bi})$,
and reads
$\Ker(\tp)
=
\sum_{\bi}
|\Cint_{\bi}|^{2}
[n_{\bi}\,
\e^{+\iu\omega_{\bi}\tp}
+
(n_{\bi}+1)\,
\e^{-\iu\omega_{\bi}\tp}]$,
with $n_{\bi}=1/(\e^{\omega_{\bi}/\kT}-1)$ boson occupation numbers.
This is how the temperature enters in the formalism, as the bath is
assumed in equilibrium  at the initial time $\tp\to-\infty$.
The operators without argument in Eq.~(\ref{Rnonmarkov}) are evaluated
at $t$ whereas
$F(\tp)=\sum_{kl}F_{kl}\Xkl(\tp)$
introduces formally the previous history of the spin (cf.\ next
section).

It is convenient to introduce the (coupling weighted) {\em spectral
density\/} of bath modes
%
$J(\omega)
=
\tsum_{\bi}
|\Cint_{\bi}|^{2}
\,
\pi\delta(\omega-\omega_{\bi})$.
All quantities incorporating environmental effects can be expressed in
terms of $J(\omega)$.
For instance, the kernel $\Ker(\tp)$ is given by
%
\begin{equation}
\label{kernel:J}
\Ker(\tp)
=
\int_{0}^{\infty}\!\frac{\drm\omega}{\pi}
\,
J(\omega)
\big[
n_{\omega}\,\e^{+\iu\omega\tp}+(n_{\omega}+1)\,\e^{-\iu\omega\tp}
\big]
\;,
\end{equation}
with $n_{\omega}=1/(\e^{\omega/\kT}-1)$.
A common functional form for the spectral density is
$J(\omega)\propto\omega^{\io}$ (times a high-frequency cut-off at
$\wD$).
The bath is called {\em Ohmic\/} when $\io=1$; this is realised by
Kondo coupling to electron-hole pairs near the Fermi energy in solids
(an example of bosonizable excitations from the ground state of a
non-bosonic environment \cite{calleg83}).
For $\io>1$ the bath is called {\em super-Ohmic}; for instance,
interaction with photons or phonons in three dimensions gives
$J(\omega)\propto\omega^{3}$ \cite{weiss}.

We shall write $J(\omega)=\Wo\,\omega^{\io}$ with $\Wo$, determined by
the $|\Cint_{\bi}|^{2}$, an overall measure of the coupling strength
(classically, the damping parameter).
The characteristic ``width'' of the memory kernel, $\tau_{\rm b}$,
depends on the competition of $1/\kT$ and $1/\wD$, the bath bandwidth
(the Debye frequency for phonons).
The relaxation term~(\ref{Rnonmarkov}) was obtained treating the
coupling perturbatively to second order for small $\Wo\tau_{\rm b}$.


\section{
Markovian (time-local) density-matrix equations
}
\label{DME:markov}

Due to the integral term~(\ref{Rnonmarkov}) the master equation is
formally an integro-differential equation for $\Xnm$.
To second order in the coupling, however, the retarded dependences
$F(\tp)=\sum_{kl}F_{kl}\Xkl(\tp)$
can be replaced by their unitary evolution,
$F(\tp)=U(\tp-t)F(t)$.
Introducing these time dependencies in $R$ only operators
evaluated at $t$ do remain.

To illustrate, let us assume the Hamiltonian evolution simply given by
$\Xkl(\tp)=\e^{\iu(\tp-t)\tf_{kl}}\Xkl(t)$.
This can be plugged in $R_{\n}^{\m}$ and the resulting operator
combinations $\Xkl[F\,,\Xnm]$ expressed in the Hubbard basis and
simplified using $\Xnm\Xkl=\delta_{\m k}X_{\n}^{l}$.
This results in an equation of motion fully in terms of the $\Xnm(t)$,
and {\em linear\/} in them.
Carrying out these steps one actually gets the relaxation term
(\ref{app:Rmarkov:gen})
%
\begin{eqnarray}
\label{Rmarkov}
R_{\n}^{\m}
=
\sum_{\n'\m'}
\big[
&
{}-
\delta_{\m\m'}
\big(
\tsum_{\is}
\Wu_{\is\tm\n'}^{\ast}
F_{\n'\is}
F_{\is\n}
\big)
\nonumber\\[-2.ex]
&
{}+
\big(
\Wu_{\n\tm\n'}^{\ast}
+
\Wu_{\m\tm\m'}
\big)
\,
F_{\n'\n}
F_{\m\m'}
\\[-0.5ex]
& 
{}-
\delta_{\n\n'}
\big(
\tsum_{\is}
\Wu_{\is\tm\m'}
F_{\m\is}
F_{\is\m'}
\big)
\quad
\big]
\;
X_{\n'}^{\m'}
\nonumber
\;.
\end{eqnarray}
The coefficients include the matrix elements of the spin portion of
the coupling, $F_{\n\m}=\nbra F(\vS)\mket$, and the $\m\to\n$ (complex)
transition rates
%
\begin{equation}
\label{W:def}
\Wu_{\n\tm\m}
\equiv
\Wu(\tf_{\n\m})
\;,
\qquad
\Wu(\tf)
=
{\textstyle\int}_{0}^{\infty}\!\drm{\tp}\,
\e^{-\iu\tp\tf}
\,
\Ker(\tp)
\;,
\end{equation}
evaluated at the level differences $\tf_{\n\m}=\el_{\n}-\el_{\m}$.
The form of the rate function $\Wu(\tf)$ emerges directly from the
retarded dependences
$X(\tp)=\e^{-\iu(t-\tp)\tf}X(t)$,
which yield oscillating factors $\e^{-\iu\tp\tf}$ multiplying the
kernel $\Ker(\tp)$ in the integrand of Eq.~(\ref{Rnonmarkov}).
On the other hand, the quantum-statistical average of the above
$R_{\n}^{\m}$, using $\llangle\Xnm\rrangle=\dm_{\m\n}$, gives an
equation for the matrix elements $\nbra\dm\mket$ which coincides what
can be obtained from the standard, second order, relaxation terms for
$\dm$ \cite[App.~1.A]{datpur2004}.

Let us discuss when the conservative evolution of $\Xkl(\tp)$ can be
substituted by $\e^{\iu(\tp-t)\tf_{kl}}\Xkl(t)$ in our problem.
If one uses the basis of eigenstates of the full spin Hamiltonian
(including the transverse terms \cite{luibarfer98,pohsch2000}), such
evolution is {\em exact\/} (then $\Bcpm$ do not appear explicitly in
the equation of motion, but only via $\tf_{\n\m}$; \ref{app:hubbard}).
Further, if the transverse field is not too large, one can use the
angular-momentum basis; then
$\Xnm(\tp)\simeq\e^{\iu(\tp-t)\tf_{\n\m}}\Xnm(t)$ gives the dominant
Hamiltonian dependences, providing an {\em approximate\/} relaxation
term.%
\footnote{
%
Classically, such $R_{\n}^{\m}$ corresponds to use
$\boldsymbol{R}
=
-\lambda_{\rm LL}\,\vS\times(\vS\times\boldsymbol{z}B_{\eff}^{z})$
as relaxation term in the Landau-Lifshitz equation, fully keeping the
Hamiltonian precession
$\drm\vS/\drm t=(\vS\times\B_{\eff})+\boldsymbol{R}$.
(Here $\B_{\eff}=-\partial\Hs/\partial\vS$; the form
$\vS\times\partial\Hs/\partial\vS$ follows from
$\drm\Si/\drm t=\{\Si\,,\Hs\}$ via the Poisson brakets
$\{\Si\,,\Sj\}=\epsilon_{ijk}\Sk$ \cite{ulyzas92}).
The $z$-component of $B_{\eff}$ retained in $\boldsymbol{R}$
incorporates the anisotropy field $B_{\anis}\sim 2\K S$, the dominant
energy scale in superparamagnets.
} 
This way of getting a time-local relaxation, without resorting to
$T\to\infty$ approximations or semiclassical baths \cite{calleg83},
works when one explicitly knows the conservative evolution; apart from
simple spin problems, it also applies to the harmonic oscillator
\cite{hupazzha92,kohdithan97}.


\subsection{Relaxation term for couplings via $\Scpm$}
\label{DME:markov:lin}

In various important cases the coupling is realized through
$\Scpm=\Sx\pm\iu\Sy$.
For instance, $F\sim\Scpm$ appears in Kondo coupling to electron-hole
excitations and $F\sim\{\Sz,\,\Scpm\}$ in magnetoelastic interaction
with phonons \cite{wur98,garchu97}.
Thus we will consider the form
%
\begin{equation}
\label{F:lin}
F(\vS)
=
\eta_{+}
\big\{\Vz(\Sz),\,\Scm\big\}
+
\eta_{-}
\big\{\Vz(\Sz),\,\Scp\big\}
\;,
\end{equation}
where $\{A,\,B\}=A\,B+B\,A$ and $\eta_{\pm}=\eta_{x}\pm\iu\eta_{y}$
are scalars incorporating the symmetry of the interaction
(isotropic $\eta_{x}=\eta_{y}=1$; anisotropic $\eta_{x}=1$ and
$\eta_{y}=0$, etc.).
The matrix elements $F_{\n\m}=\nbra F\mket$ are then
(\ref{app:Rmarkov:lin})
%
\begin{equation}
\label{Fnm}
\fl
F_{\n\m}
=
\LF_{\m,\m-1}
\delta_{\n,\m-1}
+
\LF_{\m+1,\m}^{\ast}
\delta_{\n,\m+1}
\;,
\quad
\LF_{\m,\m'}
=
\eta_{+}
[\Vz(\m)+\Vz(\m')]
\lf_{\m,\m'}
\;,
\end{equation}
where $\LF_{\m,\m'}$ is an extended ladder factor with
$\lf_{\m,\m\pm1}=[S(S+1)-\m\,(\m\pm1)]^{1/2}=\lf_{\m}^{\pm}$.
Although the operator $\Vz(\Sz)$ commutes with $\Sz$, it modulates via
$[\Vz(\m)+\Vz(\m')]$ the matrix elements of $\Scpm$, the ultimate
responsible for transitions between the levels $\mket$.
The inclusion of a $\eta_{z}\Sz$ term in $F$ does not lead to
structurally new terms in the final equation and will not be
considered here (it produces ``dephasing'' but not dissipation
\cite[Ch.~10]{datpur2004}; in the language of magnetic resonance it
modifies $T_{2}$).

The particularization of the relaxation term~(\ref{Rmarkov}) to the
coupling~(\ref{F:lin}) is done in \ref{app:Rmarkov:lin}.
Invoking on it the {\em secular\/} approximation one is left with
%
\begin{eqnarray}
\label{Rmarkov:lin:invk}
R_{\n}^{\m}
&=&
\LF_{\n,\n-1}
\LF_{\m,\m-1}^{\ast}
\;
(\Wu_{\n\tm\n-1}^{\ast}+\Wu_{\m\tm\m-1})
\quad
X_{\n-1}^{\m-1}
\nonumber\\[-0.25ex]
&-&
\big(\;\;\;\;
|\LF_{\n,\n+1}|^{2}
\;
\Wu_{\n+1\tm\n}^{\ast}
+
|\LF_{\m,\m+1}|^{2}
\;
\Wu_{\m+1\tm\m}
\nonumber\\[-0.5ex]
& &
\;
+
|\LF_{\n,\n-1}|^{2}
\;
\Wu_{\n-1\tm\n}^{\ast}
+
|\LF_{\m,\m-1}|^{2}
\;
\Wu_{\m-1\tm\m}
\;\;
\big)
\quad
\Xnm
\\[-0.25ex]
&+&
\LF_{\n,\n+1}^{\ast}
\LF_{\m,\m+1}
\;
(\Wu_{\n\tm\n+1}^{\ast}+\Wu_{\m\tm\m+1})
\quad
X_{\n+1}^{\m+1}
\nonumber
\;.
\end{eqnarray}
This corresponds to the {\em rotating-wave\/} approximation familiar
in quantum optics, where counter-rotating, rapidly oscillating terms,
are averaged out (\ref{app:Rmarkov:sec}).
Such manipulations seem not to pose problems for very weak coupling,
while they simplify the treatment \cite{cat90,desher95}.
%
%
Besides, the illustration of the continued-fraction approach will be
cleaner disregarding non-secular terms in the master equation.

To conclude, it is argued that the imaginary parts of the relaxation
coefficients reflect a coupling-induced renormalisation of the levels,
not genuine relaxation.
In the bath-of-oscillators formalism this renormalisation is cancelled
out by including suitable ``counter-terms'' in the starting
system-plus-bath Hamiltonian \cite{calleg83}.
Here \cite{gar91llb,garchu97} one cancels them by omiting the
imaginary parts of $\Wu_{\n\tm\m}$, redefining
$\Wu(\tf)
\equiv
\mathrm{Re}[\int_{0}^{\infty}\!\drm{\tp}\,\e^{-\iu\tp\tf}\Ker(\tp)]$.
\footnote{
%
%
The formalism takes as previously assessed whether such a
renormalization is physically meaningful for a given coupling, and if
so (e.g., the Lamb shift), it is considered to be already included in
$\Hs(\vS)$.
This has the advantage of making of $\Hs$ the experimentally accesible
Hamiltonian, instead of the bare one which may be difficult to
determine.
} 
%
%


\subsection{
Elements and structure of the resulting density-matrix equation
}
\label{DME:structure}

The basic ingredients of the master equation will be the energy
differences $\tf_{\n\m}$, transverse fields $\Bcpm$, ladder factors,
coupling matrix elements $\nbra F\mket\sim\LF_{\m,\m'}$, and
transition rates $\Wu_{\n\tm\m}=\Wu(\tf_{\n\m})$.
All properties of the bath enter via the rate function, which can be
expressed in terms of the spectral density $J(\omega)$ (\ref{app:Ws})
%
\begin{equation}
\label{W:J}
\Wu(\tf>0)
=
J(\tf)\,n_{\tf}
\;,
\qquad
\Wu(\tf<0)
=J(|\tf|)(n_{|\tf|}+1)
\;.
\end{equation}
As $n_{\omega}=1/(\e^{\omega/\kT}-1)$ boson absorption and emission
rates are related by the {\em detailed balance\/} condition
$\Wu(\tf)
=
\e^{-\tf/\kT}\,\Wu(-\tf)$.
This ensures, under certain conditions, convergence to the Gibbs
distribution at long times \cite{cat90,desher95}.

The rates in the relaxation term~(\ref{Rmarkov:lin:invk}) involve
adjacent levels only, $\Wu_{\m\tm\m\pm1}=\Wu(\tf_{\m,\m\pm1})$, while
$R_{\n}^{\m}$ connects $\Xnm$ with $\Xkl$ differing in indices by at
most~$1$.
Thus in the sequel we will compactly write
%
\begin{equation}
\label{Rmarkov:generic}
R_{\n}^{\m}
=
\R_{\n,\n-1}^{\m,\m-1}\,
X_{\n-1}^{\m-1}
+
\R_{\n,\n}^{\m,\m}\,
\Xnm
+
\R_{\n,\n+1}^{\m,\m+1}\,
X_{\n+1}^{\m+1}
\;,
\end{equation}
which together with Eqs.~(\ref{DME:closed}) and~(\ref{DME:nonmarkov}),
gives the working equation
%
\begin{eqnarray}
\label{DME}
\fl
\dot\Xnm
&=
\iu\,
\tf_{\n\m}
\Xnm
+
\tfrac{\iu}{2}
\Bcp
\big(\lf_{\m}^{+}\,X_{\n}^{\m+1}-\lf_{\n}^{-}\,X_{\n-1}^{\m}\big)
+
\tfrac{\iu}{2}
\Bcm
\big(\lf_{\m}^{-}\,X_{\n}^{\m-1}-\lf_{\n}^{+}\,X_{\n+1}^{\m}\big)
\nonumber\\[-0.5ex]
\fl
&
{}+
\R_{\n,\n-1}^{\m,\m-1}\,
X_{\n-1}^{\m-1}
+
\R_{\n,\n}^{\m,\m}\,
\Xnm
+
\R_{\n,\n+1}^{\m,\m+1}\,
X_{\n+1}^{\m+1}
\;.
\end{eqnarray}
It is worth mentioning that the density-matrix equation was obtained
within a, though approximate, fully quantum treatment, not introducing
any phenomenological relaxation or assuming preconceived structures
for the equation.

Note finally that if the transverse field is set to zero $\Bcpm=0$,
the diagonal part of Eq.~(\ref{DME}) becomes a closed system of {\em
balance equations\/} for the level ``populations'' $N_{\m}\equiv
X_{\m}^{\m}$, as in the Pauli master-equation approach
%
\begin{equation}
\label{DME:mm}
\dot{N}_{\m}
=
\R_{\m}^{-}
\,
N_{\m-1}
+
\R_{\m}
\,
N_{\m}
+
\R_{\m}^{+}
\,
N_{\m+1}
\;.
\end{equation}
As $\tf_{\m\m}=0$, the Hamiltonian part does not show up and the
dynamics is purely relaxational
($\R_{\m}^{-}=\R_{\m,\m-1}^{\m,\m-1}$,
$\R_{\m}=\R_{\m,\m}^{\m,\m}$,
$\R_{\m}^{+}=\R_{\m,\m+1}^{\m,\m+1}$).
For example, in the case of an isotropic spin in a {\em static\/}
field we can always choose the $z$ axis so that $\Hs=-\Bz\Sz$.
However, if we want to study resonance phenomena, we need the full
equation~(\ref{DME}) to account for transverse probing fields.
Further, even if $\Bcpm\equiv0$ (or when using exact eigenstates of
the full Hamiltonian), the simple balance structure~(\ref{DME:mm}) is
broken by terms like $X_{\m-1}^{\m+1}$ or $X_{\m}^{\m-2}$ if not
resorting to the secular approximation (\ref{app:Rmarkov:lin}).
For these reasons we will focus on the full density-matrix
equation.


\section{
Density-matrix equations for specific spin-bath problems
}
\label{sec:DME:SisoSani}

Here we particularize equation~(\ref{DME}) to an isotropic spin with
truly linear coupling $F\sim\Scpm$, as that to electron-hole
excitations (an Ohmic bath), and to anisotropic spins with
$F\sim\{\Sz,\,\Scpm\}$, corresponding to interaction with phonons (a
bath from Ohmic to super-Ohmic depending on the lattice
dimensionality).
In the classical limit, the first coupling yields {\em field-type\/}
fluctuations in the spin Langevin equations \cite{bro63,kubhas70},
whereas the second produces {\em anisotropy-type\/} fluctuations
\cite{gar99,gar97,garishpan90e} (the spin analogue of force-type and
frequency-type fluctuations in mechanical systems \cite{linses81}).


\subsection{
Density-matrix equation for isotropic spins
}
\label{sec:DME:Siso}

Let us consider a spin $\Hs=-\B\cdot\vS$, with the linear
coupling $F=\half(\eta_{+}\Scm+\eta_{-}\Scp)$.
This corresponds to a constant $\Vz(\Sz)=1/4$ in Eq.~(\ref{F:lin}).
Then $\LF_{\m,\m'}=\half\eta_{+}\lf_{\m,\m'}$ in the matrix elements
$F_{\n\m}$ and hence
$\LF_{\n,\n'}\LF_{\m,\m'}^{\ast}
=
\tfrac{1}{4}|\eta|^{2}\lf_{\n,\n'}\lf_{\m,\m'}$,
with $|\eta|^{2}=\eta_{+}\eta_{-}$.
On the other hand, as the transition rates involve adjacent levels
$\Wu_{\m\tm\m\pm1}=\Wu(\tf_{\m,\m\pm1})$, and here
$\tf_{\m,\m\pm1}=\pm\Bz$, only two rates appear.
Introducing $\Wiso\equiv\Wu_{\m\tm\m-1}$ for the $\m-1\to\m$
transition (decay for $\Bz>0$), and
$\Wu_{\m-1\tm\m}=\Wiso\,\e^{-\Bz/\kT}$ by detailed balance, the
relaxation term~(\ref{Rmarkov:lin:invk}) reduces to
%
\begin{equation}
\label{Rmarkov:Siso}
\fl
R_{\n}^{\m}
=
\Wiso
\big\{
\lf_{\n-1}\lf_{\m-1}
X_{\n-1}^{\m-1}
-
\half
\big[
(\lf_{\n}^{2}+\lf_{\m}^{2})
+
\e^{-y}
(\lf_{\n-1}^{2}+\lf_{\m-1}^{2})
\big]
\Xnm
+
\e^{-y}\lf_{\n}\lf_{\m}
X_{\n+1}^{\m+1}
\big\}
\quad
\end{equation}
Here we have introduced $y=\Bz/\kT$, the single factor
$\lf_{\m}=[S(S+1)-\m(\m+1)]^{1/2}$ (then $\lf_{\m}^{+}=\lf_{\m}$ and
$\lf_{\m}^{-}=\lf_{\m-1}$), and assumed isotropic coupling
$\eta_{x}=\eta_{y}=1$ ($|\eta|^{2}$ simply rescales $\Wiso$).
Plugging this relaxation term and the Zeeman transition frequency
$\tf_{\n\m}=-\Bz(\n-\m)$ into the master equation~(\ref{DME}), one
finally gets
%
\begin{eqnarray}
\label{DME:Siso}
\fl
&
\dot{X}_{\n}^{\m}
=
-
\iu
\Bz(\n\!-\!\m)
\Xnm
+
\tfrac{\iu}{2}
\Bcp
\big(\lf_{\m}X_{\n}^{\m+1}\!-\!\lf_{\n-1}X_{\n-1}^{\m}\big)
+
\tfrac{\iu}{2}
\Bcm
\big(\lf_{\m-1}X_{\n}^{\m-1}\!-\!\lf_{\n}X_{\n+1}^{\m}\big)
\nonumber\\
\fl
&
{}
+\Wiso
\Big\{
\lf_{\n-1}\lf_{\m-1}
X_{\n-1}^{\m-1}
-
\half
\big[
(\lf_{\n}^{2}+\lf_{\m}^{2})
+
\e^{-y}
(\lf_{\n-1}^{2}+\lf_{\m-1}^{2})
\big]
\Xnm
+
\e^{-y}\lf_{\n}\lf_{\m}
X_{\n+1}^{\m+1}
\Big\}
\qquad
\end{eqnarray}
This equation corresponds to the master equation of Garanin
\cite{gar91llb} (he also included $2$-boson Raman processes).
For $\Bcpm=0$ the diagonal elements obey the balance
equations~(\ref{DME:mm}) with
$\R_{\m}^{-}=\Wiso\lf_{\m-1}^{2}$,
$\R_{\m}=-\Wiso(\lf_{\m}^{2}+\e^{-y}\lf_{\m-1}^{2})$,
and
$\R_{\m}^{+}=\Wiso\e^{-y}\lf_{\m}^{2}$.

In an Ohmic bath $J(\omega)=\Wo\,\omega$ the rate function is
$\Wu(\tf)=\Wo\,\tf/(\e^{\tf/\kT}-1)$ (this form is valid for both
signs of $\tf$; see \ref{app:Ws}).
Then $\Wiso=\Wo\,\kT\,y/(1-\e^{-y})$ which in the limit $y\to0$ goes
over the classical rotational-diffusion constant $\Wiso\to\Wo\,\kT$
(independent of $\Bz$); thus $\Wo$ from $J(\omega)$ plays the role of
the Landau-Lifshitz damping parameter.
Actually, taking the limit $S\to\infty$ in the balance equations
\cite[App.~A]{zuegar2006} one gets the classical Fokker-Planck
equation (in a longitudinal field), and the correspondence is
established as $2\Wo S\to\lambda_{\rm LL}$.


\subsection{
Density-matrix equation for superparamagnets
}
\label{sec:DME:Sani}

Next we consider anisotropic spins, $\Hs=-\K\,\Sz^{2}-\B\cdot\vS$,
with a coupling linear in $\Scpm$ but with $\Sz$-dependent
``coefficients'', as it occurs in spin-lattice interactions.
There
$F
=
\half
\{\Sz,\,\eta_{+}\Scm+\eta_{-}\Scp\}$,
corresponding to $\Vz(\Sz)=\Sz/2$ in Eq.~(\ref{F:lin}).
For $|\eta|^{2}=2$ this gives
$\LF_{\n,\n'}\LF_{\m,\m'}^{\ast}
=
\half\lfb_{\n,\n'}\lfb_{\m,\m'}$,
with the modulated factors $\lfb_{\m,\m'}=(\m+\m')\lf_{\m,\m'}$.
Then the density-matrix equation~(\ref{DME}) goes over
%
\begin{eqnarray}
\label{DME:Sani}
\fl
\dot{X}_{\n}^{\m}
&=
\iu
\tf_{\n\m}
\Xnm
+
\tfrac{\iu}{2}
\Bcp
\big(
\lf_{\m}
X_{\n}^{\m+1}
-
\lf_{\n-1}
X_{\n-1}^{\m}
\big)
+
\tfrac{\iu}{2}
\Bcm
\big(
\lf_{\m-1}
X_{\n}^{\m-1}
-
\lf_{\n}
X_{\n+1}^{\m}
\big)
\nonumber\\
\fl
&+
\half
\lfb_{\n-1}\lfb_{\m-1}
(\Wu_{\n\tm\n-1}+\Wu_{\m\tm\m-1})
X_{\n-1}^{\m-1}
+
\half
\lfb_{\n}\lfb_{\m}
(\Wu_{\n\tm\n+1}+\Wu_{\m\tm\m+1})
X_{\n+1}^{\m+1}
\nonumber\\
\fl
&-
\half
\big(
\lfb_{\n}^{2}
\Wu_{\n+1\tm\n}
+
\lfb_{\m}^{2}
\Wu_{\m+1\tm\m}
+
\lfb_{\n-1}^{2}
\Wu_{\n-1\tm\n}
+
\lfb_{\m-1}^{2}
\Wu_{\m-1\tm\m}
\big)
\Xnm
\;,
\end{eqnarray}
where $\tf_{\n\m}=-[\K(\n+\m)+\Bz](\n-\m)$ and we have introduced the
corresponding 1-index notation $\lfb_{\m}=(2\m+1)\lf_{\m}$
(\ref{app:ladder}).
This equation was derived in Ref.~\cite{garchu97} for the study of the
archetypal magnetic molecular cluster Mn$_{12}$.
The replacement $\lf_{\m}\to\lfb_{\m}$ (not affecting the Hamiltonian
part) accounts for the $\Sz$-dependent coupling and results in an
extra $\m$ dependence of the relaxation term with respect to the case
$F\sim\Scpm$.
It can be seen as a level-dependent ``damping''
$\Wo_{{\rm eff}}(\m)\sim\Wo\,(2\m+1)^{2}$,
decreasing as the anisotropy barrier $\m\sim0$ is approached
(Fig.~\ref{fig:lf}); a spin analogue of position-dependent damping in
translational Brownian motion.

The rates $\Wu_{\n\tm\m}$ involve adjacent levels but no
simplification arises here due to the non-equispaced spectrum of
anisotropic spins $\tf_{\m,\m\pm1}=\pm[\K(2\m\pm1)+\Bz]$.
To get the rate function $\Wu(\tf)$ one can compute the distribution
of bath excitations
$J(\omega)
=
\tsum_{\bi}
|\Cint_{\bi}|^{2}
\,
\pi\delta(\omega-\omega_{\bi})$
with a Debye phonon model $\omega_{\boldsymbol{k}s}=v_{s}\,|k|$ and
replace $\sum_{\bi}\to\sum_{s}\int\!\drm^{d}\boldsymbol{k}$,
integrating over wave-vectors and summing over branches.
For magneto-elastic coupling one has
$|\Cint_{\bi}|^{2}\sim\omega_{\bi}$ so that
$\drm^{d}\boldsymbol{k}\times|\Cint_{\boldsymbol{k}s}|^{2}
\sim
|k|^{d-1}\times|k|$
gives spectral densities $J\propto\omega^{\io}$ evolving from Ohmic
$\io=1$, for phonons in one dimension, to super-Ohmic
$J\propto\omega^{3}$ for $d=3$.
The corresponding relaxation functions are
%
\begin{equation}
\label{W:DB}
\Wu(\tf)
=
\Wo\,\tf^{\io}/(\e^{\tf/\kT}-1)
\;,
\quad
\forall
\tf
\;.
\end{equation}
This unified form, instead of Eq.~(\ref{W:J}), is discussed in
\ref{app:Ws}.
Note that phonon velocities $v_{s}$, coupling constants, etc., are
subsumed in $\Wo$ from $J(\omega)=\Wo\,\omega^{\io}$.


\section{
Response to probing fields:
perturbative density-matrix equations
}
\label{sec:pert}

With the master equations one can describe the non-equilibrium
evolution from one stationary state to another.
A system can be made to ``relax'' either by subjecting it to a
``force'' (a magnetic, electric, stress field, etc.) or by removing it
after having kept it for a long time.
Then the question is how the infusion or withdrawal of energy is
shared by the system's degrees of freedom.
Alternatively one can apply a force oscillating with frequency $\w$;
this provides a time scale $1/\w$ whose competition with the intrinsic
scales of the system permits to analyse its different dynamical modes
\cite{dattagupta}.

To reflect intrinsic properties the probe should be suitably small,
not altering the nature of the studied system.
This has the advantage of allowing the use of perturbation theory
in the treatment.
In this section we will replace $\B$ by $\B+\Del\B$ in the spin
density-matrix equation, and treat it perturbatively in $\Del\B$,
getting a chain of coupled equations.
Each level will be tackled sequentially with the continued-fraction
treatment of Sec.~\ref{sec:CF-DME}, giving the spin response to the
perturbation (susceptibilities).

To alleviate the notation, we first write the density-matrix equation
(\ref{DME}) in the following compact form (including all $\n$ and
$\m$)
%
\begin{equation}
\label{DME:compact}
\dot{\X}
=
\iu
\tf(\B)
\X
+
\R(\B)
\X
\;.
\end{equation}
Here $\iu\tf\,\X$ stands for {\em all\/} the Hamiltonian part (we put
the ``$\iu$'' to remind us of this) and $\R\,\X$ for the relaxation
term.
The field enters via $\iu\tf$ linearly [Eq.~(\ref{DME:closed})] and
non-linearly through $\tf_{\m,\m\pm1}=\pm[\K(2\m\pm1)+\Bz]$ in
the rates $\Wu(\tf)\sim\tf^{\io}\,n_{\tf}$.
Let us now augment $\B$ by a probing field,
$\B
\to
\B
+
\dB(t)\,\boldsymbol{u}$,
with $\boldsymbol{u}=(\gx,\gy,\gz)$ a unit vector along $\Del\B$ and
$\dB(t)=\dB\cos(\w t)$ its magnitude.
To obtain {\em linear\/} susceptibilities, we expand field-dependent
parameters $f=f(\Bx,\By,\Bx)$ to first order:
$f\simeq f_{0}+f_{1}\,\dB(t)$,
with $f_{\iw}=(\drm^{\iw}f/\drm \dB^{\iw})/\iw!$ derivatives with
respect to the {\em amplitude\/}
$\drm /\drm \dB=\sum_{i}\gp_{i}\partial_{B_{i}}$.
For the coefficients of the master equation this gives
$\iu\tf
=
\iu\tf_{0}
+
\iu\tf_{1}\dB(t)$
and
$\R
\simeq
\R_{0}
+
\R_{1}\dB(t)$
(the former is exact as $\tf$ is linear in $\B$).

Although modulated quantities like $\iu\tf$ and $\R$ have the
parametric time dependence $f(t)=f[B+\dB(t)]$ (\ref{app:Rmarkov:sec}),
our dynamical variable $\X(t)$ does not need to evolve as some
function of $B+\dB(t)$.
Thus, we seek for a solution of the form [no $(t)$ in $\dB$]
%
\begin{equation}
\label{expansions:X}
\X(t)
\simeq
\X_{0}(t)\,
+\,
\X_{1}(t)\,\dB
\;.
\end{equation}
We compute now $\iu\tf\times\X$ and $\R\times\X$ to first order,
replace them in the dynamical equation (\ref{DME:compact}), and equate
terms with the same power of $\dB$, getting ($\X_{-1}\equiv0$)
%
\begin{equation}
\label{DME:chain:wt}
\dot{\X}_{\iw}
=
\big(
\iu\tf_{0}+\R_{0}
\big)
\X_{\iw}
+
\big(
\iu\tf_{1}+\R_{1}
\big)
\cos(\w t)
\,
\X_{\iw-1}
\;.
\end{equation}
The perturbative structure is clear: original equation with
unperturbed coefficients (first term) plus their derivatives
$(\,\cdot\,)_{1}$ times the previous order result.
Thus the lower level $\X_{\iw-1}$ acts as a forcing (source) term when
solving the equation for $\X_{\iw}$.

To get the long-time {\em stationary response}, when all transients
have died out, we introduce Fourier expansions (subindex for order in
$\dB$, superindex for the harmonic)
%
\begin{equation}
\label{X:fourier}
\X_{\iw}(t)
=
\tfrac{1}{2^{\iw}}
\tsum_{\iwp}
\X_{\iw}^{(\iwp)}\e^{\iwp\iu\w t}
\;,
\end{equation}
and go order by order.
The harmonics $\e^{\iwp\iu\w t}$ generated at each $\iw$ will coincide
with those of the forcing $\cos(\w t)\,\X_{\iw-1}(t)$, because
Eq.~(\ref{DME:chain:wt}) is linear in $\X_{\iw}$ {\em and\/}
$\X_{\iw}$ itself is not multiplied by oscillating terms (additive
sources).
The  {\em zeroth\/} order equation
$\dot{\X}_{0}
=
(\iu\tf_{0}+\R_{0})
\X_{0}$
has no sources.
Then, only $\iwp=0$ is left in the Fourier series,
$\X_{0}=\X_{0}^{(0)}$, and the static response obeys
$0
=
(\iu\tf_{0}+\R_{0})
\X_{0}^{(0)}$.
At {\em first\/} order the forcing is
$\cos(\w t)\,\X_{0}
=
\tfrac{1}{2}(\e^{+\iu\w t}+\e^{-\iu\w t})
\,
\X_{0}^{(0)}$
and only the harmonics $\iwp=\pm1$ get excited
($\X_{1}^{(0)}\equiv0$).
Thus we plug
$\X_{1}
=
\tfrac{1}{2}
(\X_{1}^{(1)}\e^{+\iu\w t}
+
\X_{1}^{(-1)}\e^{-\iu\w t})$
into Eq.~(\ref{DME:chain:wt}) and equate Fourier coefficients at both
sides, getting the remaining equations 
%
\begin{equation}
\label{DME:chain:0:1}
\left\{
\begin{array}{ccccc}
0
&=&
\big(
\iu\tf_{0}+\R_{0}
\big)
\X_{0}^{(0)}
&\!+\!&
0\qquad
\\
\pm\iu\w\,
\X_{1}^{(\pm1)}
&=&
\big(
\iu\tf_{0}+\R_{0}
\big)
\X_{1}^{(\pm1)}
&\!+\!&
\big(
\iu\tf_{1}+\R_{1}
\big)
\X_{0}^{(0)}
\end{array}
\right.
\end{equation}
These equations are to be solved sequentially by the
continued-fraction method, with the previous order acting as a forcing
on the next.

Some final remarks.
Technically, we use an {\em equation-of-motion\/} approach to obtain
the response, not relying on Kubo-type linear-response-theory
expressions \cite{dattagupta}.
This allows proceeding systematically to higher orders to get {\em
non-linear suceptibilities} (harmonics of the excitation generated by
non-linearities).
Then, one finds terms of the form
$\tsum_{l\geq2}^{\iw}\,\R_{l}\cos^{l}\!(\w t)\X_{\iw-l}$ in the
equation for $\X_{\iw}$, due to the non-linearity of the relaxation
term [which includes $\Wu(\tf)=\Wo\,\tf^{\io}/(\e^{\tf/\kT}-1)$].
In quantum Brownian motion as described by the Caldeira--Leggett
equation \cite{calleg83,anahal95}, due to the high-$T$ approximations
plus Ohmic bath (corresponding to $\Wu\simeq\Wo\,\kT$), the relaxation
term does not depend on the system potential:
$R
=
-\iu\gamma
\big[
\half
(x-y)
(\partial_{x}-\partial_{y})
+
\kT
\,
(x-y)^2
\big]
\dm(x,y)$
or $R=\gamma\,\partial_{p}(p+\kT\,\partial_{p})\W(x,p)$ in the Wigner
representation.
In particular, $R$ does not depend on the forcing, and such
higher-derivative terms vanish ($\R_{l\geq2}\equiv0$).
They are also absent in classical spins and dipoles
\cite{garsve2000,raiste97}, where the relaxation term {\em does\/}
depend on the field,
$\boldsymbol{R}=-\lambda_{\rm LL}\,\vS\times(\vS\times\B_{\eff})$,
but linearly.
Anyway, as we have seen, the $\R_{l\geq2}$ terms do not affect the
calculation of the linear responses.


\section{
Continued-fraction methods for spin density-matrix equations
}
\label{sec:CF-DME}

To solve the master equations we will cast them into the form of
$3$-term recurrence relations suitable to apply the continued-fraction
method \cite{risken}.
This is related with schemes of solution by tri-diagonalization, like
the Lanczos algorithm or the recursion method \cite{hay80}.
In Brownian motion problems it shares elements with the expansion into
complete sets (Grad's) approach for solving kinetic equations
\cite{balescu2}.
The non-equilibrium distribution $\W$ is expanded in a basis of
functions (Hermite or Laguerre polynomials, plane waves, spherical
harmonics, etc.) and the partial-differential kinetic equation is
transformed into a {\em set of coupled equations\/} for the expansion
coefficients $\ec_{\ir}$.
%
%
Approximate solutions can be obtained by truncating the hierarchy of
equations at various levels.
But as a matter of fact, to obtain manageable equations the
truncation needs to be performed at a low level.
In the continued-fraction variant, instead of truncating directly, one
seeks for bases in which the range of index coupling is short
(ideally, the equation for $\ec_{\ir}$ involves $\ec_{\ir-1}$,
$\ec_{\ir}$, and $\ec_{\ir+1}$).
%
%
Then these recurrences between the $\ec_{\ir}$ can be solved exactly
by iterating a simple algorithm, which has the structure of a
continued-fraction.%
\footnote{
%
%
For a nice brief survey of the relation between ordinary series
expansions, orthogonal polynomials, recursions, and continued
fractions, see Ref.~\cite{benmil94}.
} 

This technique was exploited to solve classical Fokker--Planck
equations for few-variable systems in external potentials
\cite{junris85,feretal93pa,cofkalwal94}.
Compared with direct simulations, continued-fraction methods have
several {\em shortcomings}: (i) the basis choice is quite
problem specific, (ii) the stability and convergence of the algorithm
may fail in some ranges of parameters, and (iii) it does not return
trajectories (which always provide helpful insight).
When the method works, however, its {\em advantages\/} are valuable:
(i) no statistical errors, (ii) special aptness to get stationary
solutions (long times), (iii) high efficiency, allowing to explore
parameter ranges out of reach of simulation, and (iv) the obtaining of
the distribution $\W$ (also insightful).

This, together with the lack of quantum Langevin simulations,
motivated several adaptations of the continued-fraction approach to
quantum problems \cite{shiuch93,vogris88,garzue2004}.
The master equation was transformed into Fokker--Planck-like form
using pseudo-probability representations $\W$ of the density matrix,
such $\W$ expanded in appropriate bases, and the recurrences for the
coefficients derived from the kinetic equation and solved by continued
fractions.
As this approach is not based on the Hamiltonian eigenstructure, it is
invaluable for systems including continuous parts in the spectrum
(e.g., Morse and periodic potentials).
Notwithstanding this, for systems with discrete levels only, the
density-matrix equation for $\dm_{\n\m}$ already has an
index-recurrence structure and such transformation-expansion protocol
(often cumbersome) may be bypassed.
This is our purpose in this section (cf.\
Refs.~\cite[Sec.~V]{allaribam77,naretal75} for $S=1/2$).


\subsection{Index-coupling structure and vector $3$-term recurrences}
\label{ICS-3RR}

Let us begin writing the master equation~(\ref{DME}) compactly as
$\dot{X}_{\n}^{\m}
=
\sum_{\n'\m'}
\Q_{\n,\n'}^{\m,\m'}\,
X_{\n'}^{\m'}$,
with $\n'=\n-1,~\n,~\n+1$ and $\m'=\m-1,~\m,~\m+1$, and with
coefficients
\begin{equation}
\label{Qs:gral}
\fl
\begin{array}{lcllcllcl}
\Q_{\n,\n-1}^{\m,\m-1}
&=&
\R_{\n,\n-1}^{\m,\m-1}
&
\Q_{\n,\n-1}^{\m,\m}
&=&
-\tfrac{\iu}{2}
\Bcp
\lf_{\n}^{-}
&
\Q_{\n,\n-1}^{\m,\m+1}
&\equiv&
0
\\
\Q_{\n,\n}^{\m,\m-1}
&=&
\tfrac{\iu}{2}
\Bcm
\lf_{\m}^{-}
&
\Q_{\n,\n}^{\m,\m}
&=&
\iu\,
\tf_{\n\m}
+
\R_{\n,\n}^{\m,\m}
&
\Q_{\n,\n}^{\m,\m+1}
&=&
\tfrac{\iu}{2}
\Bcp
\lf_{\m}^{+}
\\
\Q_{\n,\n+1}^{\m,\m-1}
&\equiv&
0
&
\Q_{\n,\n+1}^{\m,\m}
&=&
-\tfrac{\iu}{2}
\Bcm
\lf_{\n}^{+}
&
\Q_{\n,\n+1}^{\m,\m+1}
&=&
\R_{\n,\n+1}^{\m,\m+1}
\end{array}
\end{equation}
The matrix associated to this linear system has dimensions
$(2S+1)^{2}\!\times\!(2S+1)^{2}$.
For small spins it can be solved directly.
However, already at $S=3$ the size is $49\!\times\!49$, becoming
$441\!\times\!441$ for $S=10$ (Mn$_{12}$ or Fe$_{8}$),
%
%
and rising to $1156\!\times\!1156$ for $S=33/2$ (Fe$_{19}$) and to
$2704\!\times\!2704$ for $S=51/2$ (Mn$_{25}$).
Thus, if one is tempted to study mesoscopic spins in this way, let
alone pursue the classical limit, soon faces large matrices.

The problem gets simpler if $\Bcpm\equiv0$, as the system splits into
uncoupled recurrences inside each sub-diagonal
$\dot X_{\m}^{\m+k}
=
F[
X_{\m-1}^{(\m-1)+k},
X_{\m}^{\m+k},
X_{\m+1}^{(\m+1)+k}]$.
These can be solved separately by {\em scalar\/} continued fractions,
as in the approach of Shibata~\cite{shiuch93}.
Nevertheless, in problems involving coherent dynamics the diagonals
are typically coupled and such a strategy is not applicable.
They remain coupled even when $\Bcpm=0$ (or using the exact
eigenstates of the full Hamiltonian) if not resorting to the
rotating-wave approximation [Eq.~(\ref{Rmarkov:lin})].

This calls for more generic methods.
Our aim is to retain the spirit of Shibata's approach by converting
the 2-index recurrence
$\dot\Xnm
=
\sum
\Q_{\n,\n'}^{\m,\m'}\,
X_{\n'}^{\m'}$,
into a 1-index {\em vector\/} recurrence.
To this end, we first rewrite the equations highlighting the
index-coupling structure
\begin{equation}
\label{DME:expl}
\fl
\begin{array}{llllllllll}
\dot{X}_{\n}^{\m}
&=&
\Q_{\n,\n-1}^{\m,\m-1}
&
X_{\n-1}^{\m-1}
&+&
\Q_{\n,\n}^{\m,\m-1}
&
X_{\n}^{\m-1}
&+&
0
&
X_{\n+1}^{\m-1}
\\
&+&
\Q_{\n,\n-1}^{\m,\m}
&
X_{\n-1}^{\m}
&+&
\Q_{\n,\n}^{\m,\m}
&
\Xnm
&+&
\Q_{\n,\n+1}^{\m,\m}
&
X_{\n+1}^{\m}
\\
&+&
0
&
X_{\n-1}^{\m+1}
&+&
\Q_{\n,\n}^{\m,\m+1}
&
X_{\n}^{\m+1}
&+&
\Q_{\n,\n+1}^{\m,\m+1}
&
X_{\n+1}^{\m+1}
\end{array}
\;.
\end{equation}
Forgeting for a moment about the upper indices $\m'$, we see that the
equation for $X_{\n}^{(\cdot)}$ only involves $X_{\n-1}^{(\cdot)}$
(first column), $X_{\n}^{(\cdot)}$ itself (second), and
$X_{\n+1}^{(\cdot)}$ (last).
Thus, if we build up ``vectors'' $\mc_{\n}$ including all
$X_{\n}^{\m}$ for a given $\n$ we will have a $3$-term recurrence
between them, with some matrix coefficients $\mQ_{\n,\n'}$.
That is,
$\dot{\mc}_{\n}
=
\mQ_{\n,\n-1}\mc_{\n-1}+\mQ_{\n,\n}\mc_{\n}+\mQ_{\n,\n+1}\mc_{\n+1}$.
This could be tackled by {\em matrix\/} continued fractions, replacing
the original $(2S+1)^{2}\times(2S+1)^{2}$ problem by one with $2S+1$
steps but with matrices $(2S+1)\times(2S+1)$ (\ref{app:RR-CF}).
This approach actually converts the $\sim S^{4}$ problem into an
overall $\sim S^{3}$ one and reduces significantly the storage
demands.
Compared with previous exact techniques, this will allow us to increase
the possible values of $S$ dramatically (up to $\sim100$--$200$).



Explicitly, the required $(2S+1)$-tuples $\mc_{\n}$ and
$(2S+1)\times(2S+1)$ matrices $\mQ_{\n,\n'}$ comprise (with averages
in $(\mc_{\n})_{\m}=\langle\Xnm\rangle$ one deals directly with
$\dm_{\n\m}$)
%
\begin{eqnarray}
\label{vectors:expl}
\fl
\hspace*{1.em}
\mc_{\n}
&=&
\left(
\begin{array}{l}
X_{\n}^{-S}
\\[-0.5ex]
\;\;\vdots
\\
X_{\n}^{S}
\end{array}
\right)
\quad
\begin{array}{c}
\mbox{\Large $\uparrow$}
\\[0.ex]
{\scriptstyle 2S+1}
\\[0.ex]
\mbox{\Large $\downarrow$}
\end{array}
\;,
\qquad
\big(\mc_{\n}\big)_{\m}
=
X_{\n}^{\m}
\\[1.ex]
\nonumber
\fl
\mQ_{\n,\n'}
&=&
\left(
\begin{array}{ccccccc}
\Q_{\n,\n'}^{-S\quad,-S}
\!\!&\!\!
\Q_{\n,\n'}^{-S\quad,-S+1}
\!\!&\!\!
0
\!\!&\!\!
{}
\!\!&\!\!
{}
\!\!&\!\!
{}
\!\!&\!\!
{}
\\[0.5ex]
\Q_{\n,\n'}^{-S+1,-S}
\!\!&\!\!
\Q_{\n,\n'}^{-S+1,-S+1}
\!\!&\!\!
\Q_{\n,\n'}^{-S+1,-S+2}
\!\!&\!\!
0
\!\!&\!\!
{}
\!\!&\!\!
\mbox{\Large $0$}
\!\!&\!\!
{}
\\[0.5ex]
0
\!\!&\!\!
\Q_{\n,\n'}^{-S+2,-S+1}
\!\!&\!\!
\Q_{\n,\n'}^{-S+2,-S+2}
\!\!&\!\!
\Q_{\n,\n'}^{-S+2,-S+3}
\!\!&\!\!
0
\!\!&\!\!
{}
\!\!&\!\!
{}
\\[0.5ex]
{}
\!\!&\!\!
\hspace{-2.5em}
\ddots
\!\!&\!\!
\hspace{-2.5em}
\ddots
\!\!&\!\!
\hspace{-2.5em}
\ddots
\!\!&\!\!
\hspace{-2.5em}
\ddots
\!\!&\!\!
\hspace{-2.5em}
\ddots
\!\!&\!\!
{}
\\[0.5ex]
{}
\!\!&\!\!
{}
\!\!&\!\!
0
\!\!&\!\!
\Q_{\n,\n'}^{\m,\m-1}
\!\!&\!\!
\Q_{\n,\n'}^{\m,\m}
\!\!&\!\!
\Q_{\n,\n'}^{\m,\m+1}
\!\!&\!\!
0
\\[0.5ex]
{}
\!\!&\!\!
\mbox{\Large $0$}
\!\!&\!\!
{}
\!\!&\!\!
\hspace{-2.5em}
\ddots
\!\!&\!\!
\hspace{-2.5em}
\ddots
\!\!&\!\!
\hspace{-2.5em}
\ddots
\!\!&\!\!
\hspace{-2.5em}
\ddots
\\[0.5ex]
{}
\!\!&\!\!
{}
\!\!&\!\!
{}
\!\!&\!\!
{}
\!\!&\!\!
0
\!\!&\!\!
\Q_{\n,\n'}^{S,S-1}
\!\!&\!\!
\Q_{\n,\n'}^{S,S}
\end{array}
\right)
\end{eqnarray}
Then, introducing the custom notation
$\mQ_{\n}^{-}
\equiv
\mQ_{\n,\n-1}$,
$\mQ_{\n}
\equiv
\mQ_{\n,\n}$,
and
$\mQ_{\n}^{+}
\equiv
\mQ_{\n,\n+1}$,
the density-matrix equation~(\ref{DME:expl}) is transformed into the
$3$-term canonical form:
%
\begin{equation}
\label{dcdt:matrix}
\dot{\mc}_{\n}
=
\mQ_{\n}^{-}
\mc_{\n-1}
+
\mQ_{\n}
\mc_{\n}
+
\mQ_{\n}^{+}
\mc_{\n+1}
\;,
\qquad
\n=-S,\cdots,S
\;.
\end{equation}
It is worth emphasizing the simple attainment of the sought recursion
form in comparison with the Fokker--Planck-like approach with
pseudo-distributions (as a consequence of the discrete spectrum).
On the other hand, if the original equation had included couplings
with $X_{\n\pm2}^{\m\pm2}$ we would have arrived at $5$-term vector
recursions (e.g., for biaxial spins, or not invoking secular
approximations).
Nevertheless, recurrences involving more than $3$ terms can be
``folded'' into $3$-term form by introducing appropriate block vectors
and matrices (\ref{app:Rmarkov:sec}).
Alternatively, if the elements breaking the $3$-term structure are
suitably small, they can be treated iteratively (avoiding to enlarge
dimensions), in a way similar to the forcing terms below.


\subsection{The forcing terms (iterative calculations)}
\label{sec:forcing}

The equations
$\iw\iu\w\,\X_{\iw}
\sim
(\iu\tf_{0}+\R_{0})\X_{\iw}
+
(\iu\tf_{1}+\R_{1})\X_{\iw-1}$
of Sec.~\ref{sec:pert} can be manipulated analogously, as they inherit
the structure of the original master equation. 
%
%
The main novelty are the forcing terms
$(\iu\tf_{1}+\R_{1})\X_{\iw-1}$, which will be handled here.

Recall that bold letters in
Eqs.~(\ref{X:fourier})--(\ref{DME:chain:0:1}) standed for all $\Xnm$
elements, to which we will have to add indices for perturbative order
and harmonic.
Then to avoid too baroque expressions we introduce the simplified
notation 
$\boldsymbol{\Xy}
\equiv
\X_{0}^{(0)}$
and
$\boldsymbol{\Xz}
\equiv
\X_{1}^{(1)}$.
The $0$th order equation~(\ref{X:fourier}) can then be written as
[cf.\ Eq.~(\ref{DME})]
%
\begin{eqnarray}
\label{DME:chain:0:ex}
\fl
0
=&
\iu\,
\tf_{\n\m}^{0}
\Xy_{\n}^{\m}
+
\tfrac{\iu}{2}
\Bcp^{0}
\big(
\lf_{\m}^{+}
\Xy_{\n}^{\m+1}
\!-\!
\lf_{\n}^{-}
\Xy_{\n-1}^{\m}
\big)
+
\tfrac{\iu}{2}
\Bcm^{0}
\big(
\lf_{\m}^{-}
\Xy_{\n}^{\m-1}
\!-\!
\lf_{\n}^{+}
\Xy_{\n+1}^{\m}
\big)
\nonumber\\
\fl
&+
(\R_{0})_{\n,\n-1}^{\m,\m-1}\,
\Xy_{\n-1}^{\m-1}
+
(\R_{0})_{\n,\n}^{\m,\m}\,
\Xy_{\n}^{\m}
+
(\R_{0})_{\n,\n+1}^{\m,\m+1}\,
\Xy_{\n+1}^{\m+1}
\;,
\end{eqnarray}
where the index $0$ stands for absence of probing field.
The first line corresponds to $\iu\tf_{0}\X_{0}^{(0)}$ and the second
to $\R_{0}\X_{0}^{(0)}$.
Similarly, the first order Eq.~(\ref{DME:chain:0:1}) reads
%
\begin{eqnarray}
\label{DME:chain:1:ex}
\fl
\iu\w\,
\Xz_{\n}^{\m}
=&
\iu\,
\tf_{\n\m}^{0}
\Xz_{\n}^{\m}
+
\tfrac{\iu}{2}
\Bcp^{0}
\big(
\lf_{\m}^{+}
\Xz_{\n}^{\m+1}
\!-\!
\lf_{\n}^{-}
\Xz_{\n-1}^{\m}
\big)
+
\tfrac{\iu}{2}
\Bcm^{0}
\big(
\lf_{\m}^{-}
\Xz_{\n}^{\m-1}
\!-\!
\lf_{\n}^{+}
\Xz_{\n+1}^{\m}
\big)
\nonumber\\
\fl
&+
\big[
(\R_{0})_{\n,\n-1}^{\m,\m-1}\,
\Xz_{\n-1}^{\m-1}
+
(\R_{0})_{\n,\n}^{\m,\m}\,
\Xz_{\n}^{\m}
+
(\R_{0})_{\n,\n+1}^{\m,\m+1}\,
\Xz_{\n+1}^{\m+1}
\big]
+
\eF_{\n}^{\m}(\boldsymbol{\Xy})
\;.
\end{eqnarray}
Again the custom terms stand for $\iu\tf_{0}\X_{1}^{(1)}$ and
$\R_{0}\X_{1}^{(1)}$ while the sources $\eF_{\n}^{\m}$ account for
$(\iu\tf_{1}+\R_{1})\X_{0}^{(0)}$.
This is determined by the previous order result
$\boldsymbol{\Xy}=\X_{0}^{(0)}$ and the field derivatives of $\iu\tf$
and $\R$.
Using 
$\Del\B\sim \dB\,(\gx,\gy,\gz)$
and
$\tf_{\n\m}=-[\K(\n+\m)+\Bz](\n-\m)$
we obtain 
%
\begin{eqnarray}
\label{DME:chain:forcing:y}
\fl
\eF_{\n}^{\m}
=&
-\iu\,
(\n-\m)
\gz
\Xy_{\n}^{\m}
+
\tfrac{\iu}{2}
\,u_{+}
\left(
\lf_{\m}^{+}
\Xy_{\n}^{\m+1}
\!-\!
\lf_{\n}^{-}
\Xy_{\n-1}^{\m}
\right)
+
\tfrac{\iu}{2}
\,u_{-}
\left(
\lf_{\m}^{-}
\Xy_{\n}^{\m-1}
\!-\!
\lf_{\n}^{+}
\Xy_{\n+1}^{\m}
\right)
\nonumber\\
\fl
&+
\gz
\big[
(\R_{1})_{\n,\n-1}^{\m,\m-1}\,
\Xy_{\n-1}^{\m-1}
+
(\R_{1})_{\n,\n}^{\m,\m}\,
\Xy_{\n}^{\m}
+
(\R_{1})_{\n,\n+1}^{\m,\m+1}\,
\Xy_{\n+1}^{\m+1}
\big]
\;,
\end{eqnarray}
where $u_{\pm}=\gx\pm\iu\gy$ and
$(\R_{1})_{\n,\n'}^{\m,\m'}\equiv\drm(\R_{\n,\n'}^{\m,\m'})/\drm\Bz$,
because $\R$ only depends on $\Bz$ (recall that our relaxation term is
approximate in $\Bcpm$; Sec.~\ref{DME:markov}).

Now, to convert the $(2S+1)^{2}\times(2S+1)^{2}$
systems~(\ref{DME:chain:0:ex}) and~(\ref{DME:chain:1:ex}) into vector
recurrences, we proceed just as in Sec.~\ref{ICS-3RR} for the parts
$\iw\iu\w\,\X_{\iw}
\sim
(\iu\tf_{0}+\R_{0})\X_{\iw}$,
while we introduce appropriate forcing vectors $\mF_{\n}$.
This gives
\begin{equation}
\label{RR:matrix}
\mQ_{\n}^{-}\mc_{\n-1}
+
\hat{\mQ}_{\n}\mc_{\n}
+
\mQ_{\n}^{+}\mc_{\n+1}
=
-\mF_{\n}
\;,
\qquad
\hat{\mQ}_{\n}
\equiv
\mQ_{\n}
-
\iw\iu\w\,
\mI
\;.
\end{equation}
Now
$(\mc_{\n})_{\m}=\Xy_{\n}^{\m}$ or $\Xz_{\n}^{\m}$
and
$(\mF_{\n})_{\m}
=
f_{\n}^{\m}$,
while
$(\mQ_{\n})_{\m\m'}
=
\Q_{\n,\n}^{\m,\m'}$
and
$(\mQ_{\n}^{\pm})_{\m\m'}
=
\Q_{\n,\n\pm1}^{\m,\m'}$
as above.
The modified central matrix $\hat{\mQ}_{\n}$ ($\mI$ is the identity)
incorporates the left-hand sides $\iw\iu\w\X_{\iw}$.
The source terms (absent for $\iw=0$) can also be written as
$\mF_{\n}
=
\drm_{\dB}\mQ_{\n}^{-}
\,
\mc_{\n-1}^{(\iw-1)}
+
\drm_{\dB}\mQ_{\n}
\,
\mc_{\n}^{(\iw-1)}
+
\drm_{\dB}\mQ_{\n}^{+}
\,
\mc_{\n+1}^{(\iw-1)}$,
with $\mc_{\n}^{(\iw-1)}$ the previous order result and the matrices
differentiated with respect to the probing field:
$(\,\cdot\,)_{1}\equiv\drm(\,\cdot\,)/\drm \dB$.

Equation~(\ref{RR:matrix}) has the canonical form permitting to apply
directly the continued-fraction algorithm of \ref{app:RR-CF}.
Besides, apart from perturbatively, the form~(\ref{RR:matrix}) also
follows from the original master equation
$\dot{\mc}_{\n}
=
\mQ_{\n}^{-}
\mc_{\n-1}
+
\mQ_{\n}
\mc_{\n}
+
\mQ_{\n}^{+}
\mc_{\n+1}$
through Laplace transformation (for $t$-independent $\mQ_{\n}$).
Then one just identifies $\hat{\mQ}_{\n}\equiv\mQ_{\n}-s\,\mI$ (i.e.,
$\iw\iu\w\to s$, the Laplace variable) and
$\mF_{\n}\equiv\mc_{\n}(t=0)\sim\dm(0)$.
This would allow tackling initial-value problems (not forgetting the
cautionary remarks of \ref{app:Rmarkov:sec}).


\subsection{
Spin response and susceptibilities
}

Once the recursions are solved we have {\em all\/} density-matrix
elements
$(\mc_{\n})_{\m}=\langle\Xnm\rangle=\dm_{\m\n}$
and {\em any\/} observable can be obtained from the trace formula
$\llangle\opA\rrangle
=
\tsum_{\n\m}\dm_{\n\m}A_{\m\n}$.
For instance
$\langle\Si\rangle=\sum_{\n\m}\nbra\Si\mket(\mc_{\n})_{\m}$,
which connects directly the spin response with the continued-fraction
results.
To get explicitly the response to $\Del B=\dB\cos(\w t)$ we insert the
expansion
$\Xnm(t)
\simeq
\Xy_{\n}^{\m}
+
\tfrac{\dB}{2}
(\Xz_{\n}^{\m}
\e^{+\,\iu\w t}
+
\tilde{\Xz}_{\n}^{\m}
\e^{-\,\iu\w t})$
into the above average [cf.~Eqs.~(\ref{expansions:X})
and~(\ref{X:fourier}), $X_{0}^{(0)}\to\Xy$ and $X_{1}^{(1)}\to\Xz$]
%
%
\begin{equation*}
\fl
\langle\Si\rangle
=
\tsum_{\n\m}
\nbra\Si\mket
\Xy_{\n}^{\m}
+
\tfrac{\dB}{2}
\big\{
\big[
\tsum_{\n\m}
\nbra\Si\mket
\Xz_{\n}^{\m} 
\big]
\,
\e^{+\,\iu\w t}
+
\big[
\tsum_{\n\m}
\nbra\Si\mket
\tilde{\Xz}_{\n}^{\m} 
\big]
\,
\e^{-\,\iu\w t}
\big]
\big\}
\;.
\end{equation*}
Comparison with
$\langle\Si\rangle
=
\langle\Si\rangle_{0}
+
\tfrac{\dB}{2}
(\chi\,\e^{+\iu\w t}
+
\chi^{\ast}\,\e^{-\iu\w t})$
gives the static response $\langle\Si\rangle_{0}$ and the dynamical
susceptibility $\chi(\w)$.
In terms of its real part $\chi'$ and imaginary part $\chi''$ the
time-dependent response is
$\langle\Si\rangle(t)
-
\langle\Si\rangle_{0}
=
\dB\,(\chi'\cos\w t+\chi''\sin\w t)$.
Proceeding to higher orders, as sketched in Sec.~\ref{sec:pert},
$X_{0}^{(0)}
\to 
X_{1}^{(1)}
\to 
X_{2}^{(2)}
\to 
X_{3}^{(3)}
\to
\cdots$,
the non-linear susceptibilites (higher harmonics) would follow
similarly:
$\chi^{(\iw)}(\w)
=
\tsum_{\n\m}
\nbra\Si\mket
[X_{\iw}^{(\iw)}]_{\n}^{\m}$.


\section{
Application to isotropic spins
}
\label{sec:Siso}

Now we proceed to apply the discussed methods to solve the
density-matrix equations for various systems.
We will start with Garanin's master equation~(\ref{DME:Siso}) for
isotropic spins, $\Hs=-\B\cdot\vS$, with a simple linear coupling to
the bath $F=\boldsymbol{\eta}\cdot\vS$.
The bath is assumed Ohmic, $J(\omega)=\Wo\,\omega$, with rate function
$\Wu(\tf)=\Wo\,\tf/(\e^{\tf/\kT}-1)$ (Sec.~\ref{sec:DME:Siso}).
There are several analytical results for the static and dynamical
response of isotropic spins, which will be used as benchmark for the
continued-fraction approach.


\subsection{
Matrix coefficients of the recurrences
}
\label{sec:Q:Siso}

Comparing the relaxation term~(\ref{Rmarkov:Siso}) with the generic
form~(\ref{Rmarkov:generic}) we identify the relaxation coefficients
$\R_{\n,\n'}^{\m,\m'}$ of the isotropic spin.
Plugging them in Eq.~(\ref{Qs:gral}) for the coefficients $\mQ_{\n}$
and the Zeeman level differences $\tf_{\n\m}=-(\n-\m)\Bz$, we have
\begin{eqnarray}
\label{QmQp:iso}
\fl
\mQ_{\n}^{-}
&
\left\{
\begin{array}{lcl}
\Q_{\n,\n-1}^{\m,\m-1}
&=&
\Wiso\;
\lf_{\n-1}\lf_{\m-1}
\\
\Q_{\n,\n-1}^{\m,\m}
&=&
-(\iu/2)
\Bcp
\lf_{\n-1}
\\
\Q_{\n,\n-1}^{\m,\m+1}
&\equiv&
0
\end{array}
\right.
\qquad
\mQ_{\n}^{+}
\left\{
\begin{array}{lclc}
\Q_{\n,\n+1}^{\m,\m-1}
&\equiv&
0
\\
\Q_{\n,\n+1}^{\m,\m}
&=&
-(\iu/2)
\Bcm
\lf_{\n}
\\
\Q_{\n,\n+1}^{\m,\m+1}
&=&
\Wiso\,\e^{-y}\;
\lf_{\n}\lf_{\m}
\end{array}
\right.
\\
\label{Q:iso}
\fl
\mQ_{\n}
&
\left\{
\begin{array}{lclc}
\Q_{\n,\n}^{\m,\m-1}
&=&
(\iu/2)
\Bcm
\lf_{\m-1}
\\
\Q_{\n,\n}^{\m,\m}
&=&
-\iu\,
(\n-\m)\Bz
-
(\Wiso/2)
\big[
(\lf_{\n}^{2}+\lf_{\m}^{2})
+
\e^{-y}
(\lf_{\n-1}^{2}+\lf_{\m-1}^{2})
\big]
\\
\Q_{\n,\n}^{\m,\m+1}
&=&
(\iu/2)
\Bcp
\lf_{\m}
\end{array}
\right.
\end{eqnarray}
Recall that $y=\Bz/\kT$, the decay rate is
$\Wiso=\Wo\,\kT\,y/(1-\e^{-y})$, while $\lf_{\m}=\lf_{\m}^{+}$ and
$\lf_{\m-1}=\lf_{\m}^{-}$ with
$\lf_{\m}^{\pm}=[S(S+1)-\m(\m\pm1)]^{1/2}$.
From these coefficients one also gets the derivatives
$\drm_{\dB}\mQ_{\n}$ for the treatment of probing fields,
$\B\to\B+\dB(t)\,\boldsymbol{u}$, completing the specification of the
3-term recurrences~(\ref{dcdt:matrix}) and~(\ref{RR:matrix}).


\subsection{
Thermal-equilibrium response
}
\label{sec:Siso:static}


\paragraph{
Analytical results.
}

The statics of isotropic spins $\Hs=-\Bz\Sz$ can be studied in full
analytically, giving us the opportunity to test the continued-fraction
solution of the master equation.
The {\em magnetisation\/} $\Mz\equiv\langle\Sz\rangle$ is given by the
Brillouin function $\brill_{S}$:
%
\begin{equation}
\label{Mz:brillouin}
\fl
\Mz
=
S\,
\brill_{S}(\xi)
\;,
\qquad
\brill_{S}(\xi)
\equiv
(1\!+\!\tfrac{1}{2S})\,
\cth\big[(1\!+\!\tfrac{1}{2S})\xi\big]
-
\tfrac{1}{2S}\,
\cth\big(\tfrac{1}{2S}\,\xi\big)
\;,
\end{equation}
with the scaled field variable $\xi=S\Bz/\kT$ ($=S\,y$).
For $S\Bz\gg\kT$, $\brill_{S}\to1$ and saturation $\Mz\simeq S$ is
reached.
The {\em longitudinal susceptibility\/}
$\chi_{\|}\equiv\partial\langle\Sz\rangle/\partial\Bz$
follows using $(\cth x)'=1-\cth^{2}x$ and
$\partial_{\Bz}=\partial_{y}/\kT$:
%
\begin{equation}
\label{chi:brillouin}
\fl
\chi_{\|}
=
S(S+1)/\kT
-
\big\{
(S+\half)^{2}
\cth^{2}
\big[(1\!+\!\tfrac{1}{2S})\xi\big]
-
\tfrac{1}{4}\,
\cth^{2}
\big(\tfrac{1}{2S}\,\xi\big)
\big\}/\kT
\;.
\end{equation}
Expanding the hyperbolic cotangent as $\cth x \simeq1/x+x/3$, the
second term goes over $-2S(S+1)/3\kT$ as $\Bz\to0$, completing the
Curie law $\chi_{0}=S(S+1)/3\kT$ for the initial (zero-bias)
susceptibility of isotropic spins.
On the other hand, the response to a probing field perperdicular to
$\Bz$ is given by the {\em transverse susceptibility}:
%
\begin{equation}
\label{chi:iso:trans}
\chi_{\perp}
=
\Mz/\Bz
\;,
\qquad
\Mz
=
\langle\Sz\rangle
\;.
\end{equation}
Here we used Van Vleck's formula \cite{carlin} to get
$\chi_{\perp}\equiv\chi_{xx}$ with the transverse-field dependent energy
levels
$\chi_{ii}=\Z^{-1}
\sum_{\m}
[\beta(\partial_{i}\el_{\m})^{2}
-\partial_{i}^{2}\el_{\m}]\,
\e^{-\bEm}$
where $\partial_{i}\el\equiv\partial\el/\partial B_{i}$.
For $S\Bz\ll\kT$, one has $\Mz\simeq\Bz\times S(S+1)/3\kT$, recovering
Curie's law from this side.
\begin{figure}[!tb]
\centerline{
\includegraphics[width=7.2cm]{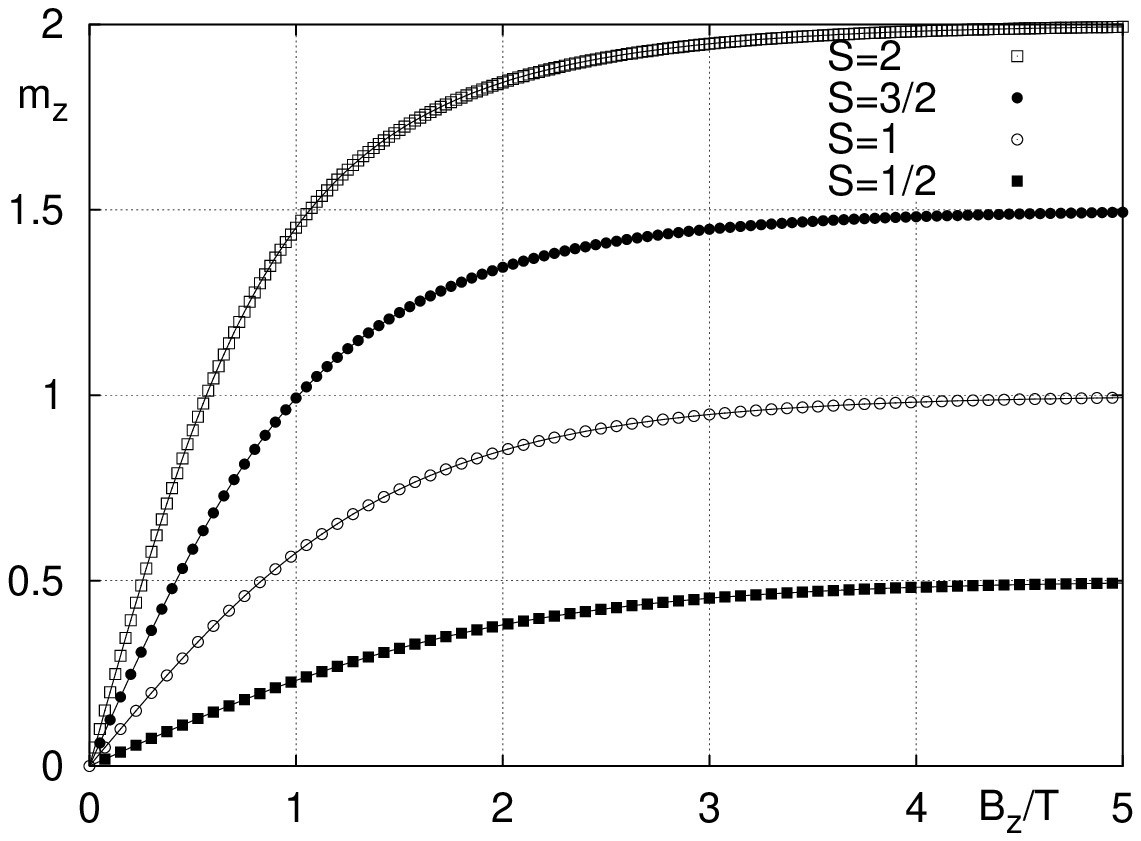}
\hspace*{-3.ex}
\includegraphics[width=7.2cm]{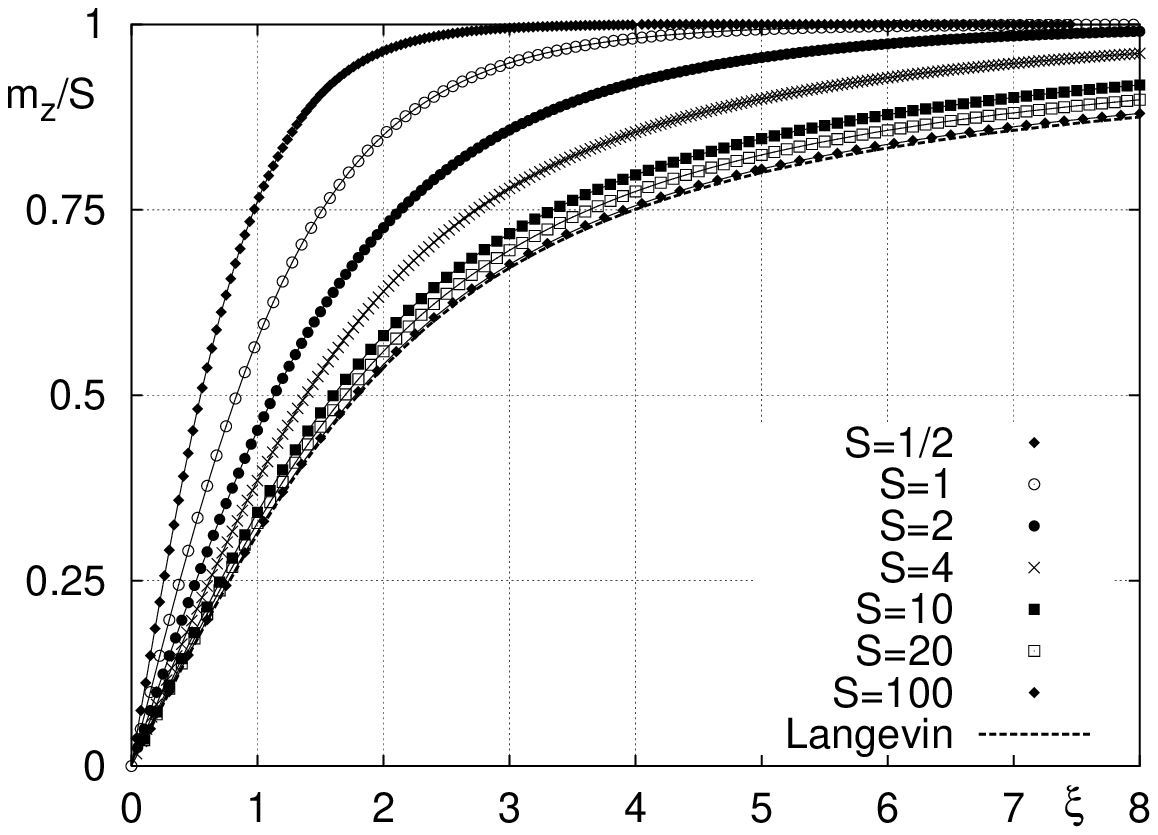}
}
\vspace*{-1.ex}
\caption{
Magnetisation $\Mz=\langle\Sz\rangle$ vs.\ field of isotropic spins.
Left panel: $\Mz$ vs.\ $\Bz/\kT$ for various small $S$.
Right panel: Reduced magnetisation $\Mz/S$ vs.\ $\xi=S\Bz/\kT$ for
increasingly large $S$.
The symbols are continued-fraction calculations and the lines
Brillouin functions~(\ref{Mz:brillouin}) [the dashed line is the
classical limit $L(\xi)=\cth(\xi)-1/\xi$].
}
\label{fig:Mz:iso}
\centerline{
\includegraphics[width=7.2cm]{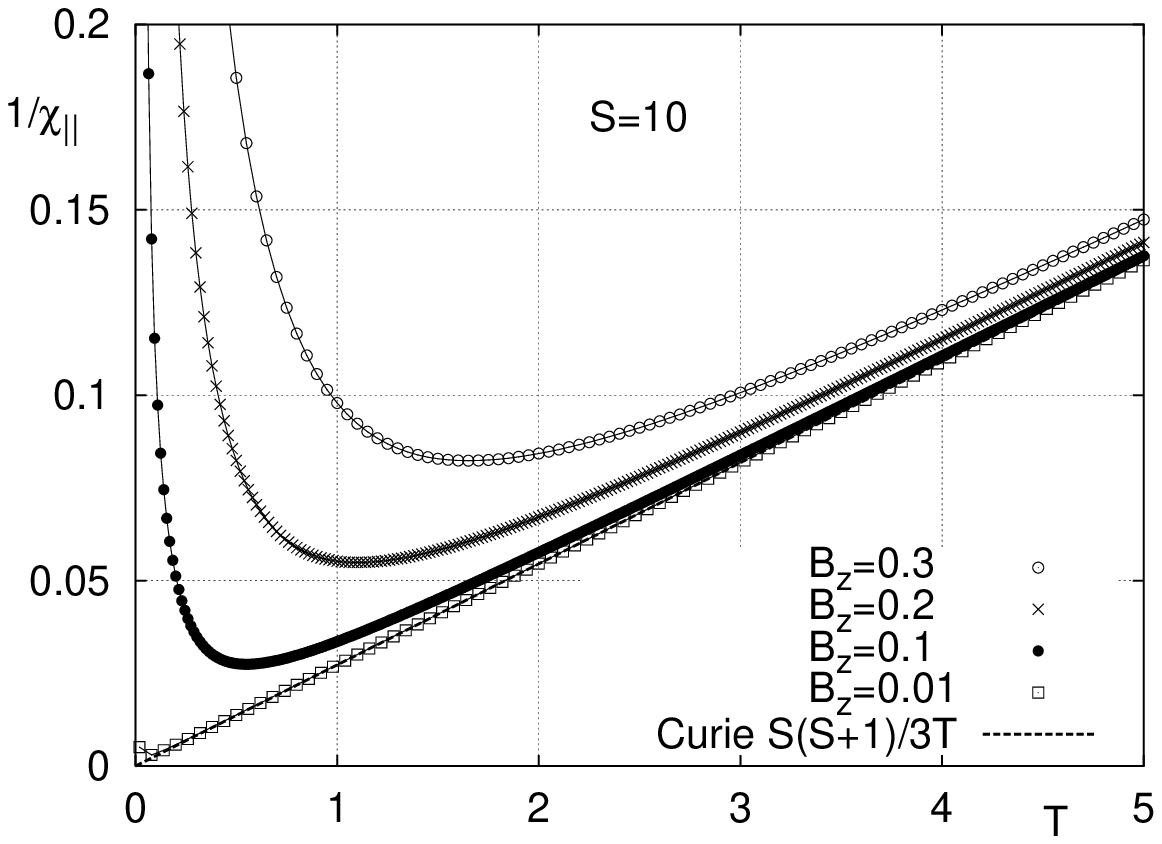}
\hspace*{-3.ex}
\includegraphics[width=7.2cm]{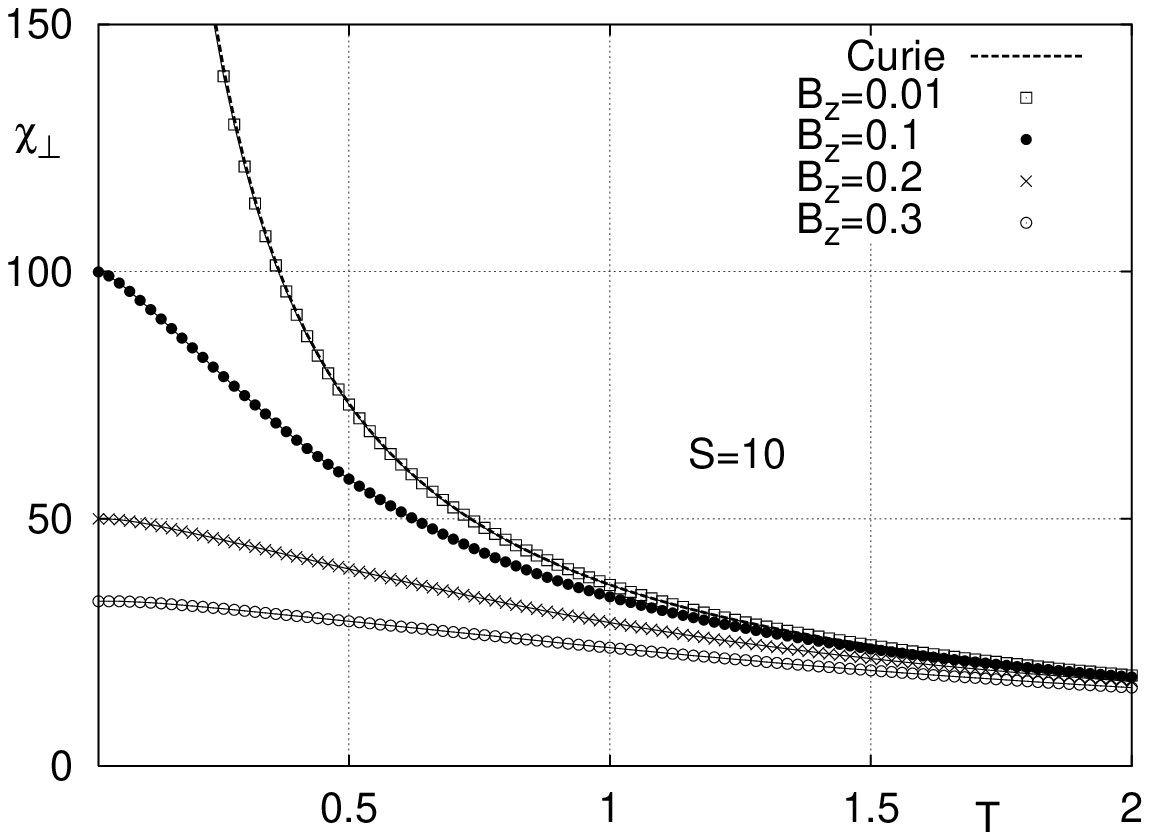}
}
\vspace*{-1.ex}
\caption{
Temperature dependence of the susceptibilities of an isotropic $S=10$
spin in various external fields.
Left: reciprocal longitudinal response $1/\chi_{\|}$; the lines are
Curie--Brillouin susceptibilities~(\ref{chi:brillouin}).
Right: transverse response; lines $\chi_{\perp}=\Mz/\Bz$
[Eq.~(\ref{chi:iso:trans})].
The symbols are continued-fraction results.
}
\label{fig:chi-eq-iso}
\end{figure}


\paragraph{
Numerical results.
}

Figure~\ref{fig:Mz:iso} shows the agreement of the continued-fraction
results with the analytical $\langle\Sz\rangle$.
The curves for small spins exhibit saturation to the corresponding $S$
at large fields.
Increasing $S$, up to $S=100$, we can follow the evolution towards the
{\em classical\/} Langevin magnetization
$\Mz/S\to\cth(\xi)-1/\xi$
[Eq.~(\ref{Mz:brillouin}) with the leading terms $1+1/2S\simeq1$ and
$\cth(\xi/2S)\simeq2S/\xi$].
At $S=20$ the result is already close to the classical asymptote.
However, this depends on the field range observed, for quantum
behaviour is found whenever the discreteness of the energy levels is
important, and this can be attained by increasing sufficiently $\Bz$.
\footnote{
For the statics we use a weak spin-bath coupling $\Wo=10^{-9}$ in the
density-matrix equation~(\ref{DME:Siso}).
We know that in the absence of transverse field its diagonals are
decoupled (after the rotating wave approximation).
Then detailed balance $\Wu(\tf)=\e^{-\tf/\kT}\,\Wu(-\tf)$ ensures that
the Gibbs distribution is its statiorary solution.
Thus the static continued-fraction results must be independent of
$\Wo$.
} 

The agreement of the magnetisations, which is numerically exact,
ensures agreement for the susceptibilities.
Still we have computed
$\chi_{\|}=\partial\Mz/\partial\Bz$
directly from the equilibrium fluctuations of the spin
$\chi_{\|}
=
(\langle\Sz^{2}\rangle
-
\langle\Sz\rangle^{2})/\kT$,
to check the proper continued-fraction obtainment of second-order
moments
$\langle\Si\Sj\rangle
=
\sum_{\n\m}\nbra\Si\Sj\mket(\mc_{\n})_{\m}$.
Figure~\ref{fig:chi-eq-iso} shows $1/\chi_{\|}$ vs.\ temperature for a
moderate spin.
In a small $\Bz$ there is a straight-line dependence in almost all the
range [Curie law $\chi_{0}^{-1}=3\kT/S(S+1)$].
Raising the field we observe deviations upwards (maximum in
$\chi_{\|}$) at sufficiently low $T$.
This occurs because $\chi_{\|}$ is the slope of the magnetisation
curve and, at high $\Bz/\kT$, saturation $\Mz\to S$ takes place and the
slope drops to zero.

Figure~\ref{fig:chi-eq-iso} also displays the transverse
susceptibility.
For a quantum spin $\chi_{\perp}$ is not easily expressed in terms of
averages in the absence of perturbation, so we resorted to
applying directly a small transverse field and computing
$\chi\sim\langle\Sx\rangle/\Bx$.
The agreement with Eq.~(\ref{chi:iso:trans}) is remarkable (recall
that the relaxation term is approximate in the transverse field,
Sec.~\ref{DME:markov}).
We see how $\chi_{\perp}$ is reduced as $T$ increases, approaching the
Curie regime.
At low $T$ the magnetization saturates, $\Mz\sim S$, and the curves
tend to the constant values $\chi_{\perp}(T=0)=S/\Bz$.


\subsection{
Dynamical response
}
\label{sec:Siso:dynamic}

Let us turn now to the dynamics of isotropic spins.
We will consider the response to probing fields $\dB\cos(\w t)$
parallel and perpendicular to the bias field $\Bz$.


\subsubsection{
Analytical results.
}
\label{Siso:formulae}

For $\Del\B\parallel\B$, on replacing $\Bz\to\Bz+\dB\cos(\w t)$ in the
balance equations~(\ref{DME:mm}) (with the coefficients of
Sec.~\ref{sec:DME:Siso}) and expanding to first order in $\dB$ one
obtains equations determining the {\em longitudinal\/} susceptibility.
They form a system of few coupled equations for small spins and can be
solved analytically \cite{gar91llb,zuegar2006,shi80I}.
For $S=1/2$ one finds the {\em Debye\/} form
[$(\,\cdot\,)'\equiv\drm(\,\cdot\,)/\drm y$; $y=\Bz/\kT$]
%
\begin{equation}
\label{chi:1/2}
\chi_{\|}(\w)
=
\frac{\Mz'}{\kT}
\frac{\rate}{\rate+\iu\w}
\;,
\qquad
\rate
=
\Wiso\,
(1+\e^{-y})
\;,
\end{equation}
where
$\Mz=\half\thrm(\half y)$
and $\Mz'/\kT$ is the equilibrium response.
The characteristic {\em relaxation time\/} is $\tau=1/\rate$.
As the decay rate is $\Wiso=\Wu(-y)$, with
$\Wu(y)=\Wo\,\kT\,y/(\e^{y}-1)$, one has $\rate=\Wu(-y)+\Wu(y)$ (see
\ref{app:Ws}).
Next, for $S=1$ the susceptibility can be written as \cite{gar91llb}
%
\begin{equation}
\label{chi:1}
\chi_{\|}(\w)
=
\frac{\Mz'}{\kT}
\frac
{\Lambda_{1}\Lambda_{2}+\iu\w\rateone}
{(\Lambda_{1}+\iu\w)(\Lambda_{2}+\iu\w)}
\qquad
\simeq
\;
\frac{\Mz'}{\kT}
\frac
{\rateone}
{\rateone+\iu\w}
\;.
\end{equation}
Here $\rateone=\rate\,(2\ch y+1)/(\ch y+2)$, with
$\rate=\Wu(-y)+\Wu(y)$ again, and $\Lambda_{1,2}$ are the non-zero
eigenvalues of the matrix associated to the system of balance
equations
$\Lambda_{1,2}
=
\rate\,
[2\ch(\half\,y)\mp1]/\ch(\half\,y)$
($\Lambda_{0}=0$ corresponds to the thermal equilibrium solution).
This formula can be expressed as the sum of two Debye terms (cf.\
Eq.~(\ref{chi:1:ani}) below).
But as $\Lambda_{1}$ is numerically close to $\rateone$, the
susceptibility is nearly single Debye with an effective relaxation
time $\tau\sim1/\rateone$.
\begin{figure}[!tb]
\centerline{
\includegraphics[width=7.2cm]{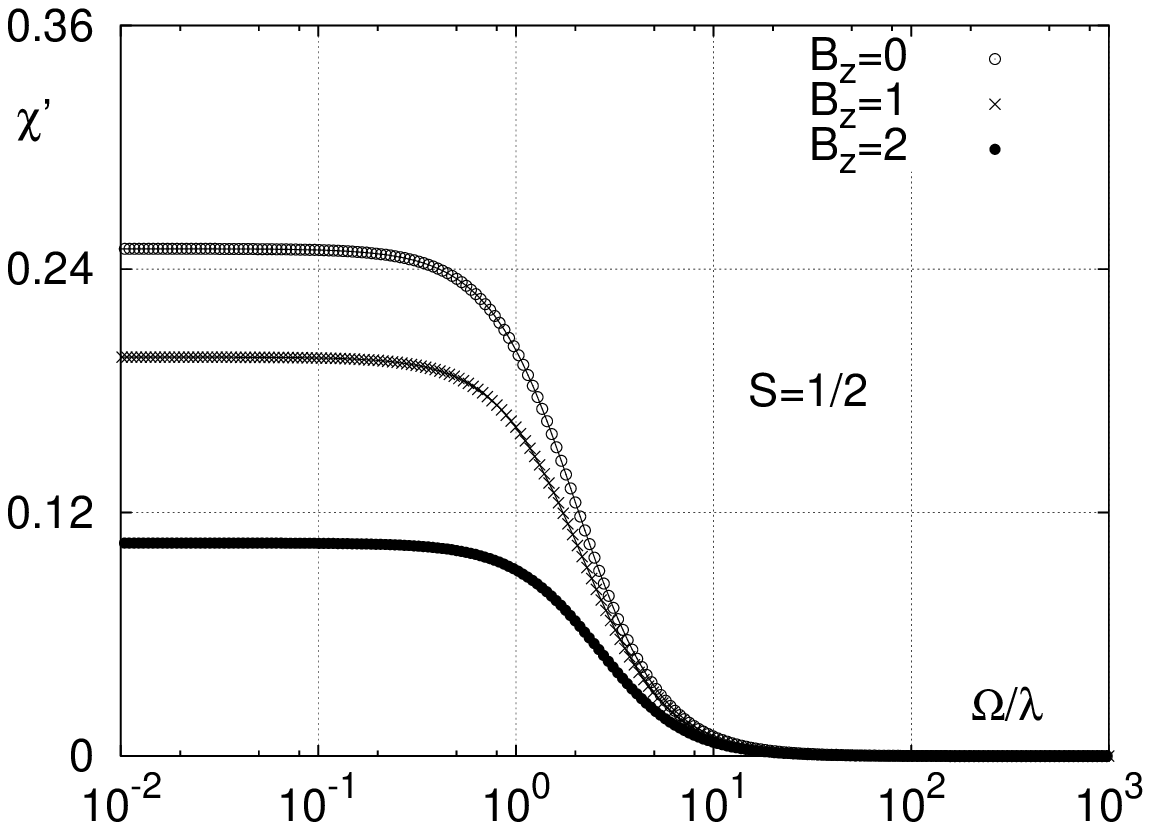}
\hspace*{-3.ex}
\includegraphics[width=7.2cm]{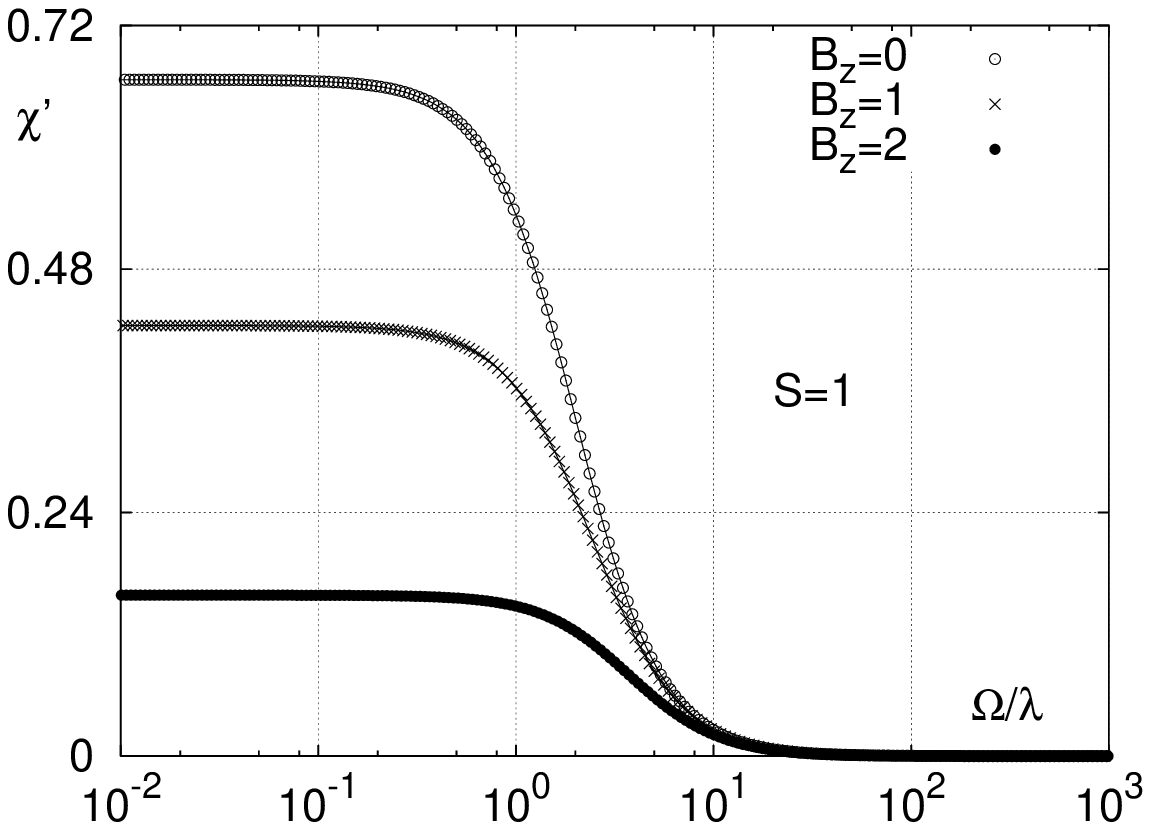}
}
\vspace*{-1.ex}
\centerline{
\includegraphics[width=7.2cm]{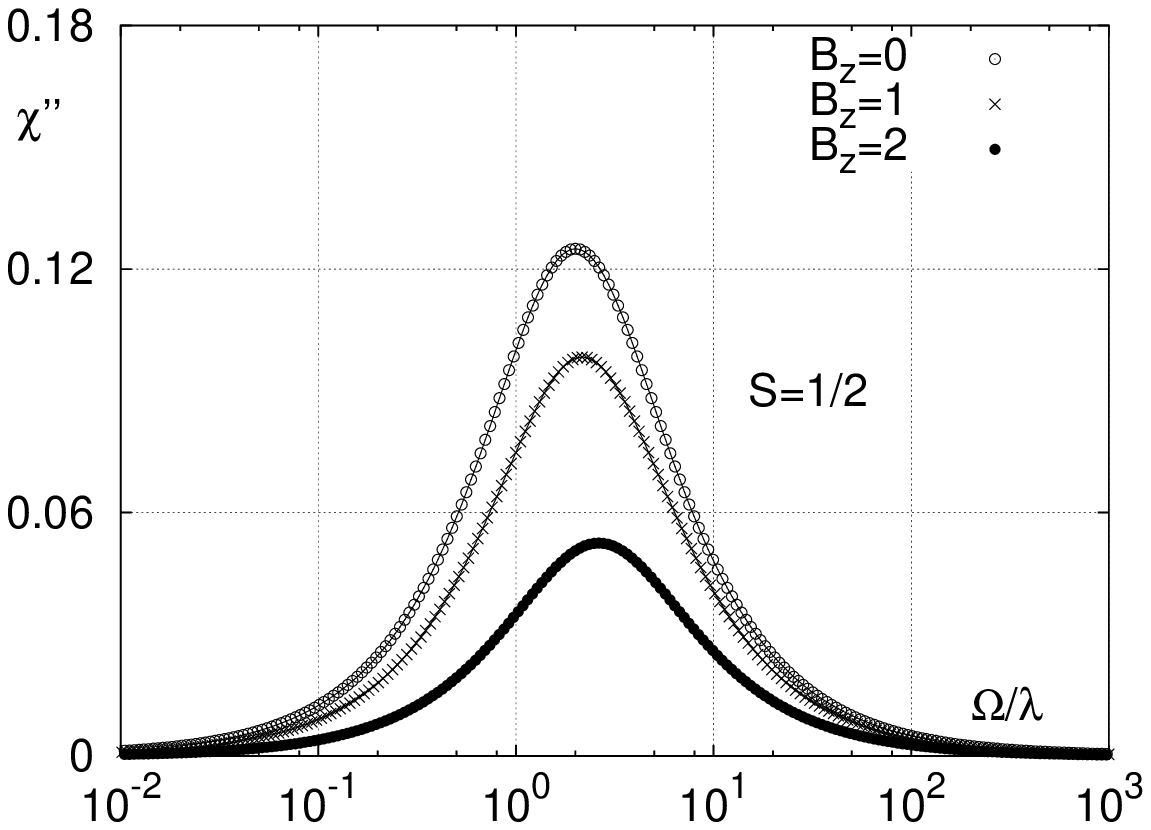}
\hspace*{-3.ex}
\includegraphics[width=7.2cm]{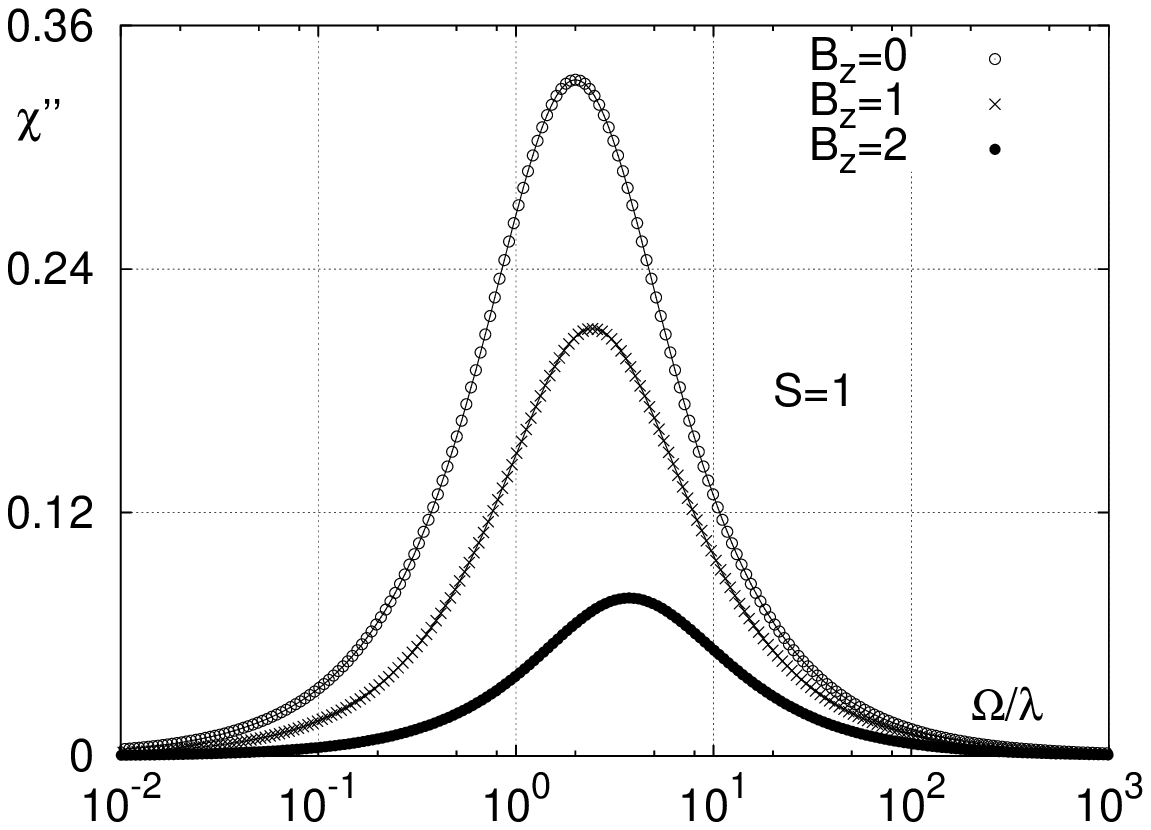}
}
\vspace*{-1.ex}
\centerline{
\includegraphics[width=7.2cm]{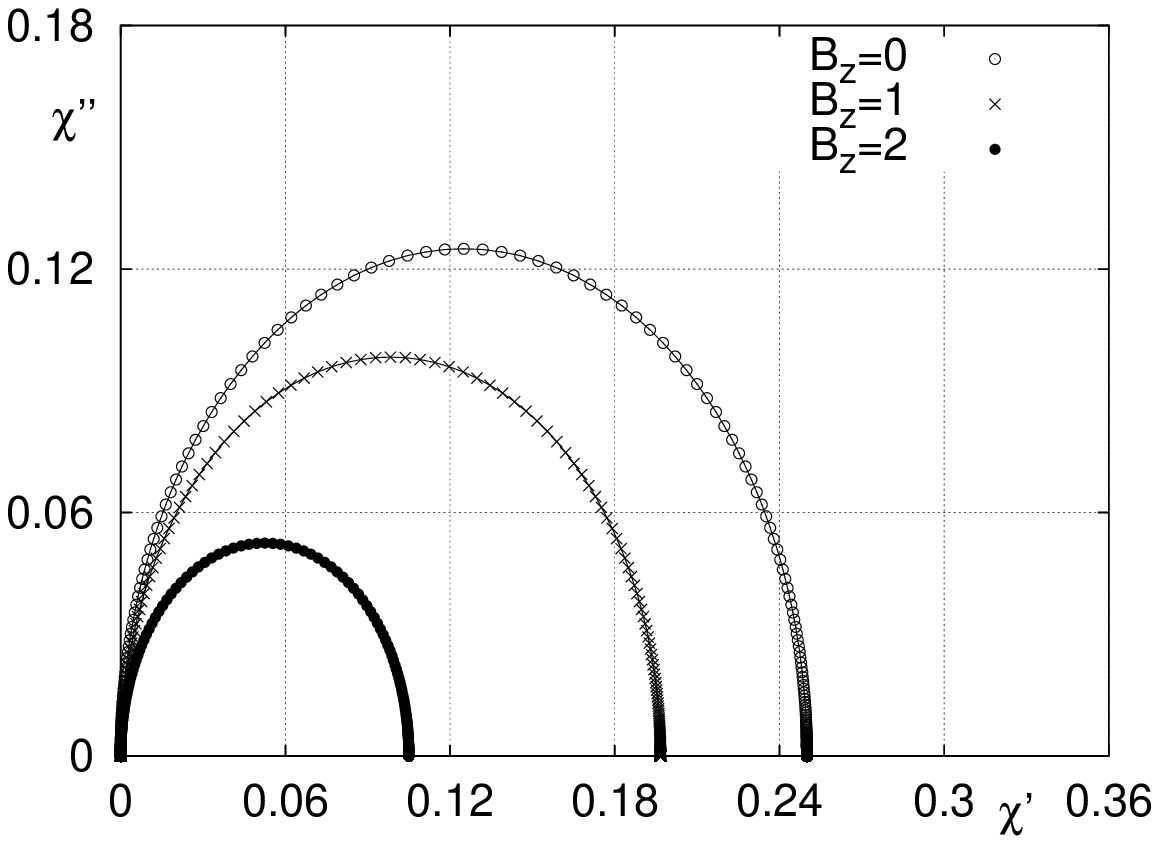}
\hspace*{-3.ex}
\includegraphics[width=7.2cm]{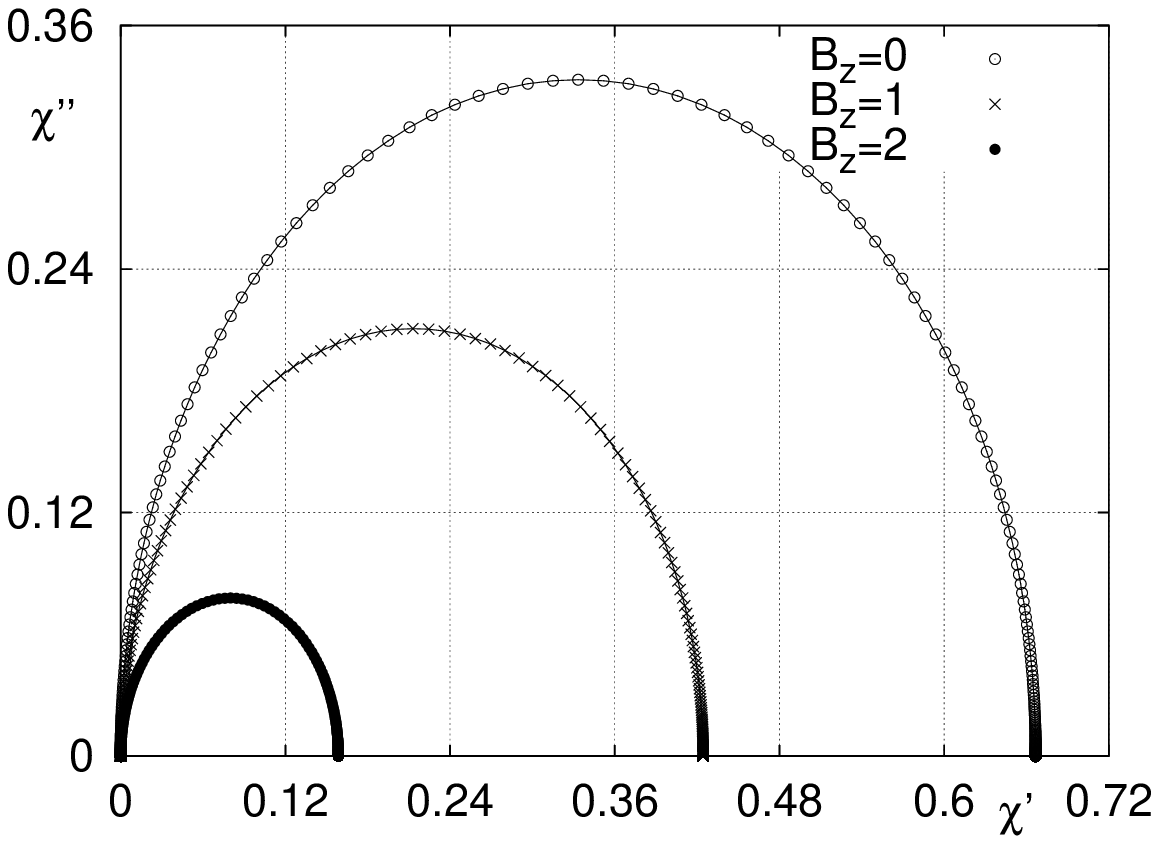}
}
\vspace*{-1.ex}
\caption{
Dynamical susceptibility $\chi_{\|}(\w)$ for $S=1/2$ (left panels) and
$S=1$ (right).
The temperature is $\kT=1$, $\Wo=10^{-9}$, and results in various
fields are shown.
The lines are the analytical expressions~(\ref{chi:1/2})
and~(\ref{chi:1}) and the symbols continued-fraction results.
The different panels are the real part $\chi'$ (top), imaginary part
$\chi''$ (middle) and Cole-Cole plot $\chi''$ vs.\ $\chi'$ (bottom
panels).
}
\label{fig:chi:iso:S0p5-S1:w}
\end{figure}

For arbitrary $S$ the longitudinal relaxation comprises $2S+1$ modes
\cite{shi80I} (the rank of the matrix associated to the balance
equations).
In general their amplitudes $a_{i}$ and rates
$0=\Lambda_{0}\leq\Lambda_{1}\leq\cdots\leq\Lambda_{2S}$
need to be obtained numerically \cite{zuegar2006}.
This makes it difficult to derive closed expressions for the
susceptibility,
$\chi(\w)
=
\chi
\sum_{i=1}^{2S}
a_{i}/(1+\iu\,\w/\Lambda_{i})$,
but still something can be said on the relaxation time.
An overall measure is provided by the {\em integral relaxation time},
$\tint\equiv\int_{0}^{\infty}\!\drm t\,\delta M(t)/\delta M(0)$,
where $\delta M(t)$ is the response to a small field change
$\delta\Bz$ at $t=0$ \cite{risken}.
Its advantage is that one can by-pass the computation of the time
constants and amplitudes \cite{gar97,garishpan90e,zuegar2006} by
getting $\tint$ directly from the low-frequency susceptibility
$\chi(\w)\simeq\chi(1-\iu\w\tint+\ldots)$
which can be obtained in closed form (see \ref{app:tauint} for
explicit expressions for $\tint$).

Finally, to have some analytical formula for the susceptibility at
arbitrary $S$, one introduces a {\em single-relaxation time\/}
approximation
\begin{equation}
\label{chi:SRT}
\chi_{\|}(\w)
=
\frac{\Mz'}{\kT}
\frac{1}{1+\iu\w\tint}
%
\;.
\end{equation}
This is possibly the most popular expression in the modelling of
relaxation experiments.
By construction this heuristic form is correct at low frequencies.
The accuracy for arbitrary $\w$ will have to be assessed in each
problem addressed.
\begin{figure}[!tb]
\centerline{
\includegraphics[width=7.2cm]{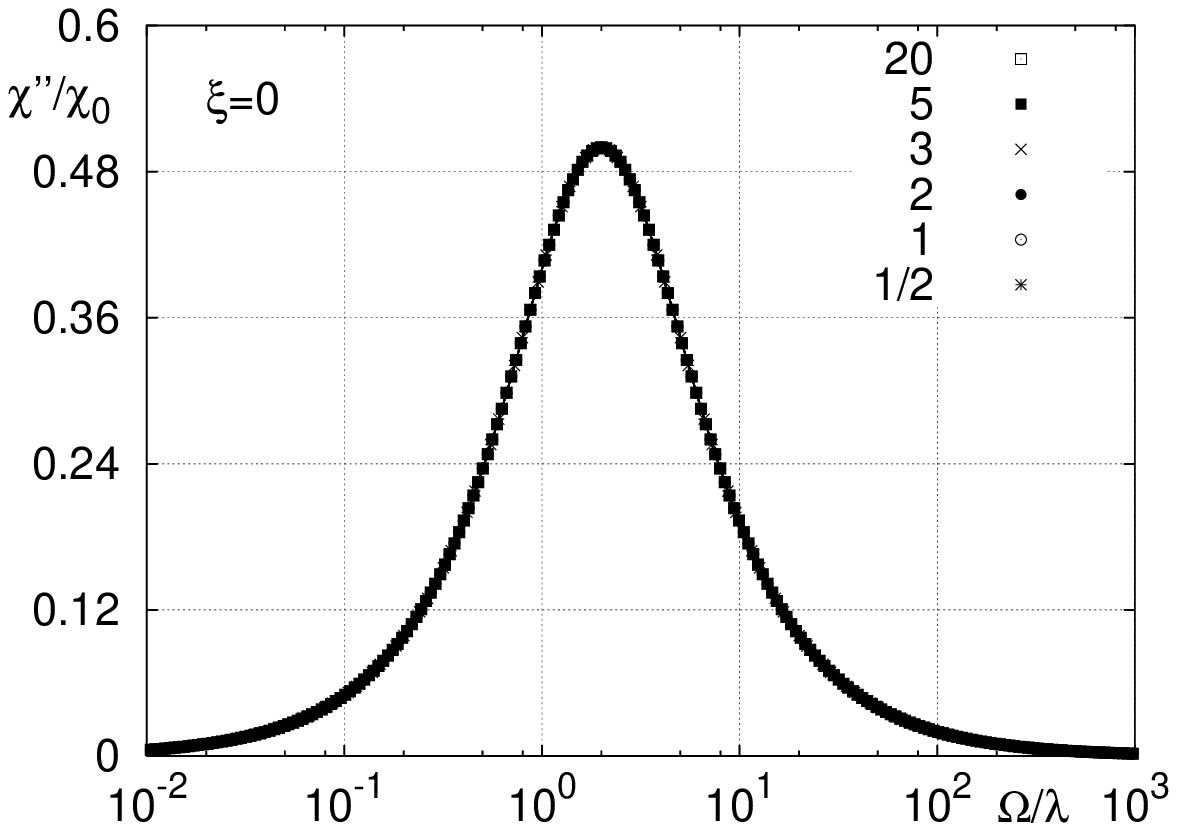}
\hspace*{-3.ex}
\includegraphics[width=7.2cm]{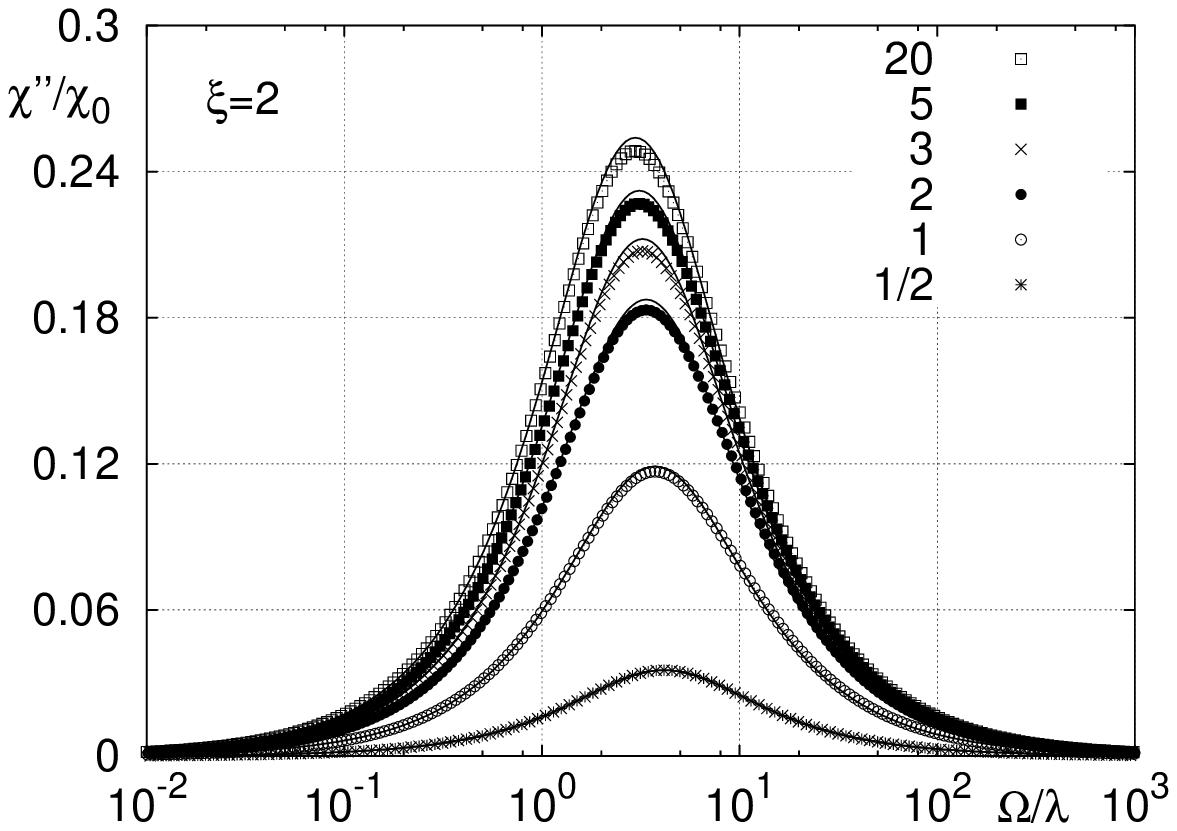}
}
\vspace*{-1.ex}
\centerline{
\includegraphics[width=7.2cm]{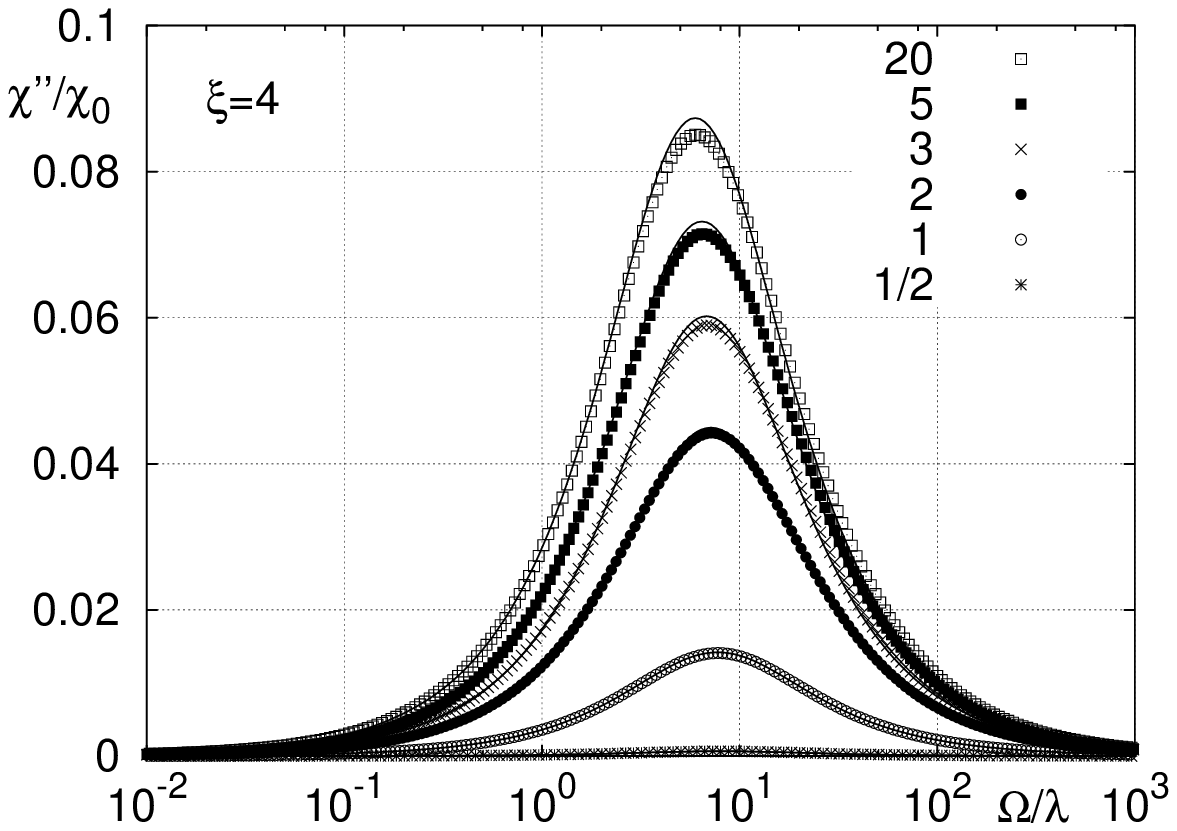}
\hspace*{-3.ex}
\includegraphics[width=7.2cm]{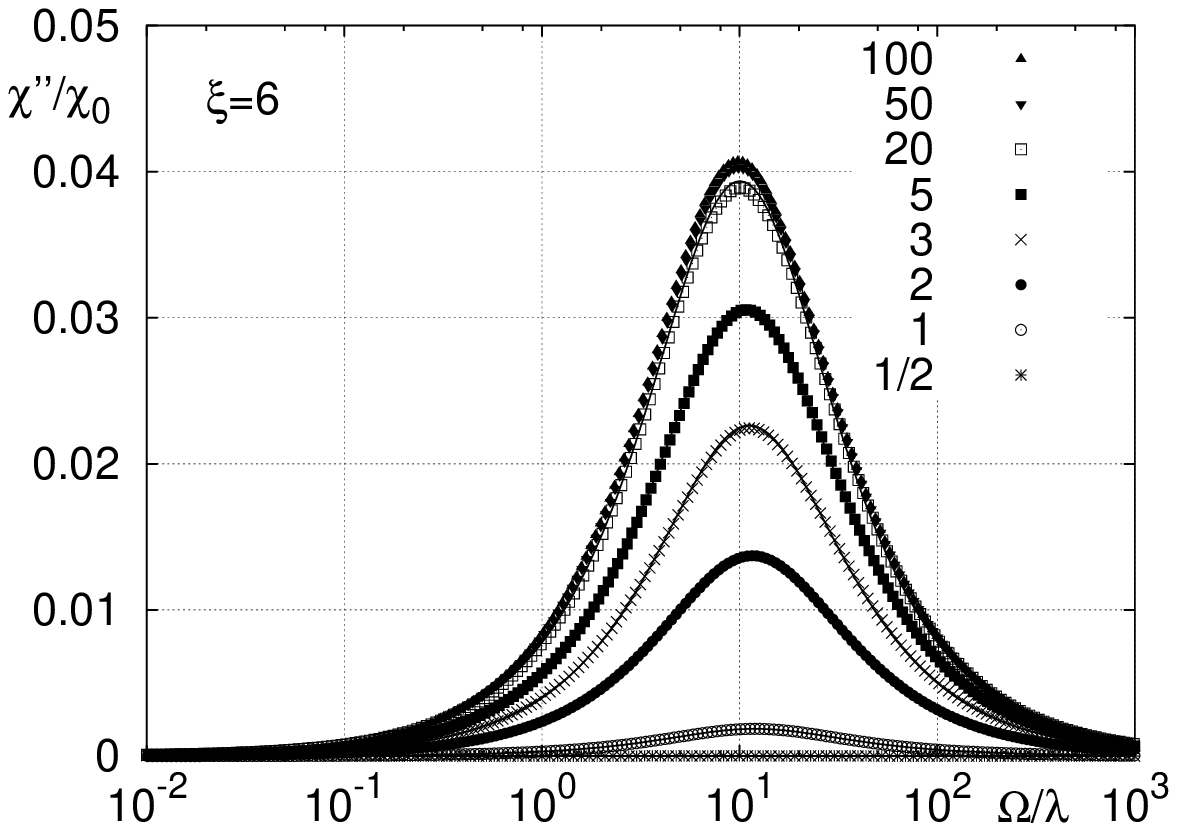}
}
\vspace*{-1.ex}
\caption{
Imaginary part of the susceptibility for $S=1/2$, $1$, $2$, $3$, $5$,
$20$.
Again $\kT=1$ whereas the fields are chosen to fix $\xi=S\Bz/\kT=0$,
$2$, $4$, $6$ for the different $S$.
The solid lines are the single-relaxation time
approximation~(\ref{chi:SRT}) and the symbols continued-fraction
calculations [normalised by the Curie $\chi_{0}=S(S+1)/3\kT$].
For $\xi=6$ we also display numerical results for $S=50$ and $100$
(rhombi), indicating convergence to a classical limit.
}
\label{fig:chi:iso:S:SRT}
\end{figure}


\subsubsection{
Longitudinal response.
}

Figure~\ref{fig:chi:iso:S0p5-S1:w} displays susceptibility spectra for
a spin $S=1/2$ in a number of fields, showing the agreement between
the analytical and continued-fraction results (again numerically
exact).
The curves go down as the field is increased.
This is mainly due to the reduction of the equilibrium part of the
response $\Mz'/\kT$ [Eq.~(\ref{chi:1/2})], which is the slope of the
magnetisation and decreases with increasing $\Bz$.
The peaks of the imaginary part $\chi''$, and the maximum slope of the
real part $\chi'$, occur at $\w\sim\rate$.
Then, as the relaxation rate $\rate=\Wu(-y)+\Wu(y)$ increases with
$\Bz$ (\ref{app:Ws}), the curves shift to higher frequencies.
Finally, as the response comprises a single Debye factor
$\chi\sim1/(1+\iu\,x)$, plotting imaginary vs.\ real parts (Cole-Cole
or Argand representation) gives perfect semicircles \cite{bottcherII}.

The longitudinal response for $S=1$ is also shown in
Fig.~\ref{fig:chi:iso:S0p5-S1:w}.
In comparison with spin one-half, the susceptibilities are higher and
more sensitive to the bias field.
This can be related with the magnetisation curves of
Fig.~\ref{fig:Mz:iso}, which for $S=1$ have larger slopes and change
faster with $\Bz$ (the parameter coupling to the field, i.e., $S$, is
larger now).
On the other hand, as the deviation of Eq.~(\ref{chi:1}) from a single
Debye is small, the Cole-Cole plots are nearly semicircular.

For arbitrary $S$, finally, we compare the continued-fraction results
with the heuristic formula $\chi_{\|}(\w)=\chi_{\|}/(1+\iu\w\tint)$.
Figure~\ref{fig:chi:iso:S:SRT} shows that the agreement is very good
in general, implying that the relaxation modes $\Lambda_{i}$ are quite
grouped (on a logarithmic scale).
There are only small deviations at the peaks in intermediate fields
($\xi\sim2$--$4$), in accordance with Garanin's findings in
Ref.~\cite{gar91llb}.
On the other hand, this figure (and Fig.~\ref{fig:Mz:iso}) illustrates
that in order to reach the classical asymptotes, the quantities need
to be scaled appropriately.%
\footnote{
We used the scalings $\Bz\to S\Bz/\kT$, $\Mz\to\Mz/S$, and
$\chi\to\chi/S(S+1)$.
In general, we increase $S$ keeping the amount of Zeeman energy and
the anisotropy barriers constant (and hence finite at $S\to\infty$).
For the Hamiltonian $\Hs=-\K\,\Sz^{2}-\B\cdot\vS$ this amounts to fix
$\K\,S^{2}$ and $S\,B$, while introducing more levels with $S$ (their
spacing decreases as $\tf\sim1/S$, \ref{app:energetics}).
Correspondingly \cite[App.~A]{gardat2005,zuegar2006}, frequency and
damping are scaled as $\w\propto1/S$ and $\Wo\propto1/S$ (recall that
$\Wo\sim\lambda_{\rm LL}/2S$).
Alternatively one can use the scaling combination $\w/\Wo$ (or some
$\w\tau$), as we did here.
} 
%


\subsubsection{
Paramagnetic resonance.
}
\label{SR:iso}

We conclude with the response to transverse probing fields.
These incorporate $\Scpm$ in the Hamiltonian, not commuting with the
dominant $\Sz$-dependent part, and provoking transitions
$\mket\to|\m\pm1\rangle$ between the unperturbed levels.
These transitions result in peaks at the frequencies
$\tf_{\m,\m\pm1}=\el_{\m}-\el_{\m\pm1}$
in the imaginary part of the susceptibility \cite{dattagupta}
(absorption line-shape).
Classically the phenomenon corresponds to the matching of the
oscillating field with the Larmor precession of the spin.
Finally, as the quantum transverse response
$\langle\Scpm\rangle
=
\sum_{\m}\lf_{\m}^{\pm}\langle X_{\m\pm1}^{\m}\rangle$
involves off-diagonal elements of $\dm$ (coherences), one refers to
the precession in transverse fields as {\em coherent\/} dynamics.

\begin{figure}[!tb]
\centerline{
\includegraphics[width=7.2cm]{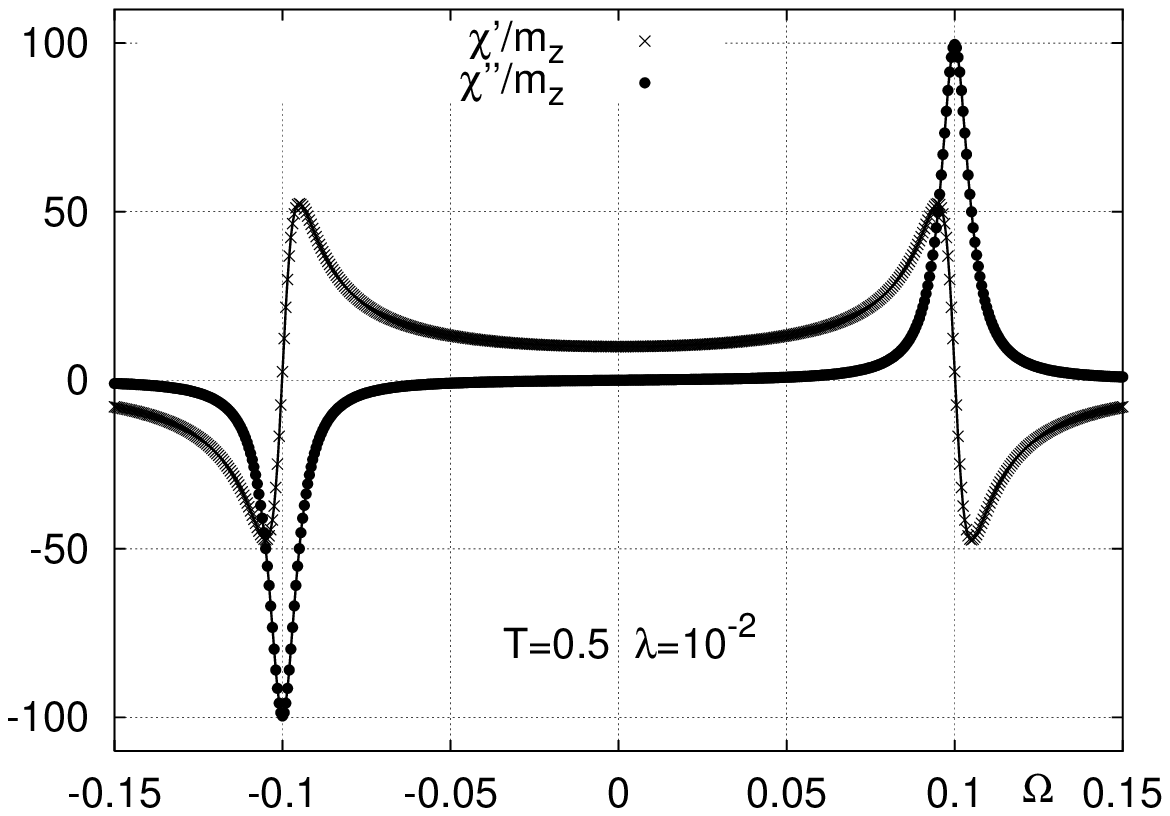}
\hspace*{-3.ex}
\includegraphics[width=7.2cm]{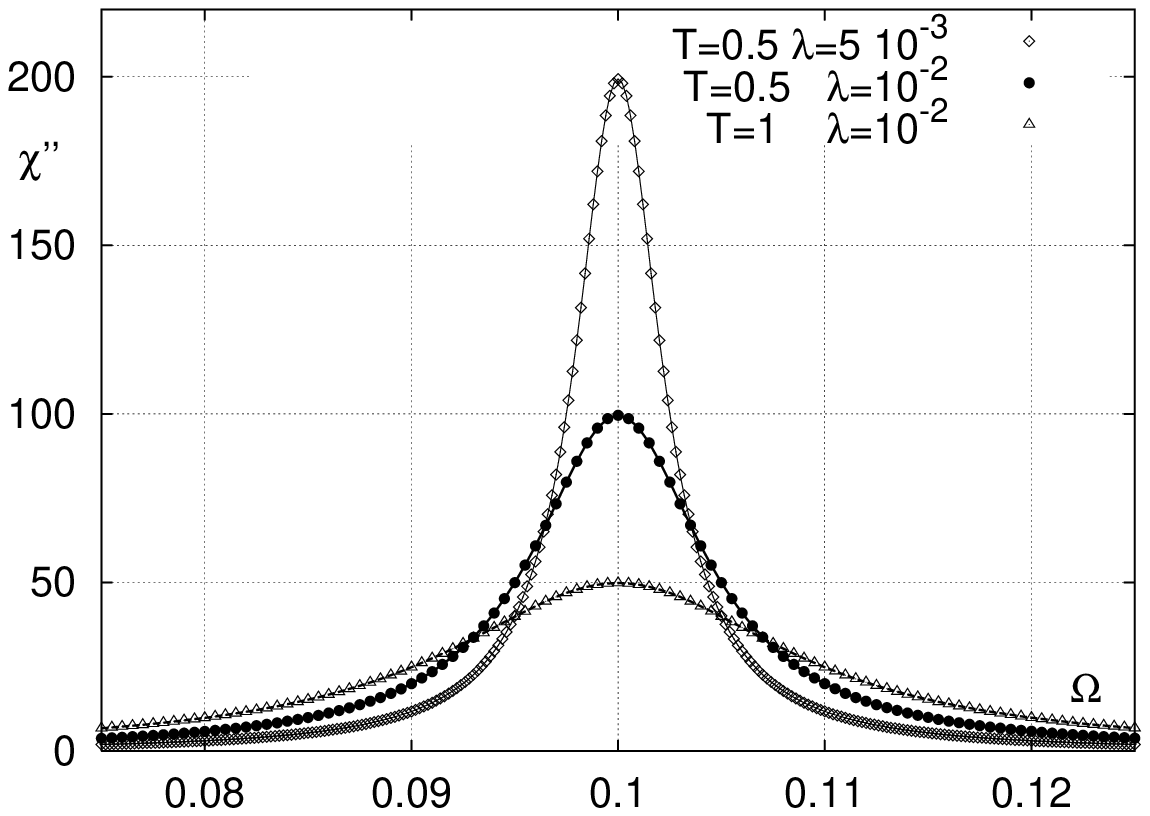}
}
\vspace*{-1.ex}
\caption{
Transverse susceptibility $\chi_{\perp}(\w)$ of an isotropic spin
$S=1/2$ in a magnetic field $\Bz=0.1$.
The lines are Eq.~(\ref{chi:1/2:perp}) and the symbols
continued-fraction results.
Left: real and imaginary parts for $\Wo=0.01$ at $\kT=0.5$.
Right: detail of the effects of the damping (halved to $\Wo=0.005$)
and of the temperature (doubled to $T=1$) on the absorption line-shape
$\chi_{\perp}''(\w)$.
}
\label{fig:peaks:SR:iso}
\end{figure}

As the level spacings of an isotropic spin are all equal
$\tf_{\m,\m\pm1}=\pm\Bz$, the phenomenology for the different $S$ is
qualitatively similar.
Figure~\ref{fig:peaks:SR:iso} shows the susceptibility
$\chi_{\perp}\equiv\chi_{xx}$ for $S=1/2$, for which we have exact
formulae \cite{shi80I}
%
\begin{equation}
\label{chi:1/2:perp}
\fl
\chi_{\perp}(\w)
=
\frac{\Mz}{2}
\,
\Big(
\frac{1}{\Bz-\w+\iu\ratetwo}
+
\frac{1}{\Bz+\w-\iu\ratetwo}
\Big)
\;,
\qquad
\ratetwo
=
\tfrac{1}{2}
\Wiso
(1+\e^{-y})
\end{equation}
[To get this equation one has to solve the full density-matrix
equation~(\ref{DME:Siso}); for $S=1/2$ the form coincides with that
obtained from phenomenological Bloch equations identifying
$T_{2}=1/\ratetwo$.]
The continued-fraction results and Eq.~(\ref{chi:1/2:perp}) agree to
all computed figures.
In particular, the numerical results duly fulfill the basic relation
$\chi(-\w)=[\chi(\w)]^{\ast}$, yielding even $\chi'(\w)$ and odd
$\chi''(\w)$.
The imaginary part shows peaks at $|\w|=\Bz$, the level separation,
accompanied by zig-zag with sign change of the real part.
Decreasing the spin-bath coupling $\Wo$ the absorption peaks become
narrower and higher, as in a forced and damped oscillator.
Here, this can also be attained by changing $T$ (right panel).
These behaviours, captured by the line-width $\ratetwo$ in
Eq.~(\ref{chi:1/2:perp}), reflect the ``smearing out'' of the energy
levels due to the bath coupling.
%

With this simple example we have introduced the basic phenomenology of
magnetic resonance and some factors influencing it.
On the other hand, the perfect agreement of the continued-fraction
results with the exact solution lends confidence in our handling of
the non-diagonal elements of $\dm$, required to compute
$\langle\Scpm\rangle
=
\sum_{\m}\lf_{\m}^{\pm}\dm_{\m,\m\pm1}$.
This is important for the subsequent application to spins in the
anisotropy potential, where there are less analytical expressions to
compare with.


\section{
Application to anisotropic spins (superparamagnets)
}
\label{sec:Sani}

Now we consider spins with a Hamiltonian $\Hs=-\K\,\Sz^{2}-\B\cdot\vS$
and a quadratic coupling $F\sim\Sz\Scpm$, motivated by the
spin-lattice interaction in (super)paramagnets
\cite{garchu97,dohful75}.
Correspondingly, we use the rate function
$\Wu(\tf)=\Wo\,\tf^{3}/(\e^{\tf/\kT}-1)$ of a $\io=3$ super-Ohmic
phonon bath.
At variance with the uniform Zeeman spectrum, the anisotropy results
in non-equispaced levels (Fig.~\ref{fig:levels}) and hence in several
rates $\Wu_{\m\tm\m\pm1}=\Wu(\tf_{\m,\m\pm1})$.
Thus this problem is a spin analogue of translational Brownian motion
in non-harmonic potentials.


\subsection{
Elements of the density-matrix recurrences
}
\label{sec:Q:Sani}

To solve the density-matrix equation~(\ref{DME:Sani}) by continued
fractions we convert it, as explained in Sec.~\ref{sec:forcing}, into
a vector 3-term recurrence of the form
$\mQ_{\n}^{-}\mc_{\n-1}
+
\mQ_{\n}\mc_{\n}
+
\mQ_{\n}^{+}\mc_{\n+1}
=
-\mF_{\n}$,
with  $(\mc_{\n})_{\m}=\dm_{\m\n}$.
The matrix coefficients $\mQ_{\n}$ comprise Hamiltonian and
relaxational contributions, which for this problem follow comparing
Eqs.~(\ref{DME:Sani}) and~(\ref{DME:expl}) [we include
$\tf_{\n\m}=-[\K(\n+\m)+\Bz](\n-\m)$]:
\begin{eqnarray*}
\fl
\mQ_{\n}^{-}
&
\left\{
\begin{array}{lclc}
\Q_{\n,\n-1}^{\m,\m-1}
&=&
\half
\,
\lfb_{\n-1}\lfb_{\m-1}
\;
(\Wu_{\n\tm\n-1}+\Wu_{\m\tm\m-1})
\\[0.5ex]
\Q_{\n,\n-1}^{\m,\m}
&=&
-(\iu/2)\Bcp\lf_{\n-1}
\\[0.5ex]
\Q_{\n,\n-1}^{\m,\m+1}
&\equiv&
0
\end{array}
\right.
\\
\fl
\mQ_{\n}
&
\left\{
\begin{array}{lclc}
\Q_{\n,\n}^{\m,\m-1}
&=&
(\iu/2)
\Bcm
\lf_{\m-1}
\\[0.5ex]
\Q_{\n,\n}^{\m,\m}
&=&
-
\iu\,
[\K(\n+\m)+\Bz](\n-\m)
\\[0.25ex]
& &
-
\half
\big(
\lfb_{\n}^{2}
\Wu_{\n+1\tm\n}
+
\lfb_{\m}^{2}
\Wu_{\m+1\tm\m}
\big)
-
\half
\big(
\lfb_{\n-1}^{2}
\Wu_{\n-1\tm\n}
+
\lfb_{\m-1}^{2}
\Wu_{\m-1\tm\m}
\big)
\\[0.5ex]
\Q_{\n,\n}^{\m,\m+1}
&=&
(\iu/2)\Bcp\lf_{\m}
\end{array}
\right.
\\
\fl
\mQ_{\n}^{+}
&
\left\{
\begin{array}{lclc}
\Q_{\n,\n+1}^{\m,\m-1}
&\equiv&
0
&
\\[0.5ex]
\Q_{\n,\n+1}^{\m,\m}
&=&
-(\iu/2)\Bcm\lf_{\n}
\\[0.5ex]
\Q_{\n,\n+1}^{\m,\m+1}
&=&
\half
\,
\lfb_{\n}\lfb_{\m}
\;
(\Wu_{\n\tm\n+1}+\Wu_{\m\tm\m+1})
\end{array}
\right.
\end{eqnarray*}
Recall that the replacement $\lf_{\m}\to\lfb_{\m}=(2\m+1)\lf_{\m}$ in
the relaxation parts comes from $\Sz$ in $F\sim\{\Sz,\,\Scpm\}$,
leading to ``position-dependent'' damping.
The field derivatives of these coefficients give the source terms
$\mF_{\n}$ for the treatment of probing fields (see \ref{app:Ws} for
$\drm\Wu/\drm\Bz$); then $\mQ_{\n}\to\mQ_{\n}-\iw\iu\w\,\mI$.
%
%
With the density-matrix recurrence~(\ref{RR:matrix}) so specified we
can now apply the continued-fraction algorithm of \ref{app:RR-CF} to
solve it.


\subsection{
Thermal-equilibrium response
}
\label{sec:Sani:static}

Again we begin discussing briefly the static properties in a
longitudinal field.
Compact expressions for the magnetisation
$\Mz=\Z^{-1}\sum_{\m}\m\,\e^{-\bEm}$ follow for small spins.
Introducing $d=\K/\kT$ and $y=\Bz/\kT$, we have
$-\bEm=d\,\m^{2}+y\,\m$.
Then for spin one-half
%
%
$\Mz=\half(\e^{y/2}-\e^{-y/2})/(\e^{y/2}+\e^{-y/2})$,
equal to the isotropic result (as $\K$ enters in the two levels
equivalently).
However, for $S=1$ and $2$ the magnetisations read
%
\begin{equation}
\label{mz:uni:S=1:2}
\langle\Sz\rangle
=
\frac
{2\,\e^{d}\,\sh y}
{1+2\,\e^{d}\,\ch y}
\qquad
\langle\Sz\rangle
=
\frac
{2\,\e^{d}\,\sh y+4\,\e^{4d}\,\sh(2y)}
{1+2\,\e^{d}\,\ch y+2\,\e^{4d}\,\ch(2y)}
\;,
\end{equation}
which are valid for both $\K>0$ (easy-axes anisotropy) and $\K<0$
(easy plane; then the energy levels of Fig.~\ref{fig:levels} are
turned upside-down).
Notice that the $\m=0$ level does not contribute to
$\sum_{\m}\m\,\e^{-\bEm}$ but contributes ``phase space'' in
$\Z=\sum_{\m}\e^{-\bEm}$.

\begin{figure}[!tb]
\centerline{
\includegraphics[width=7.2cm]{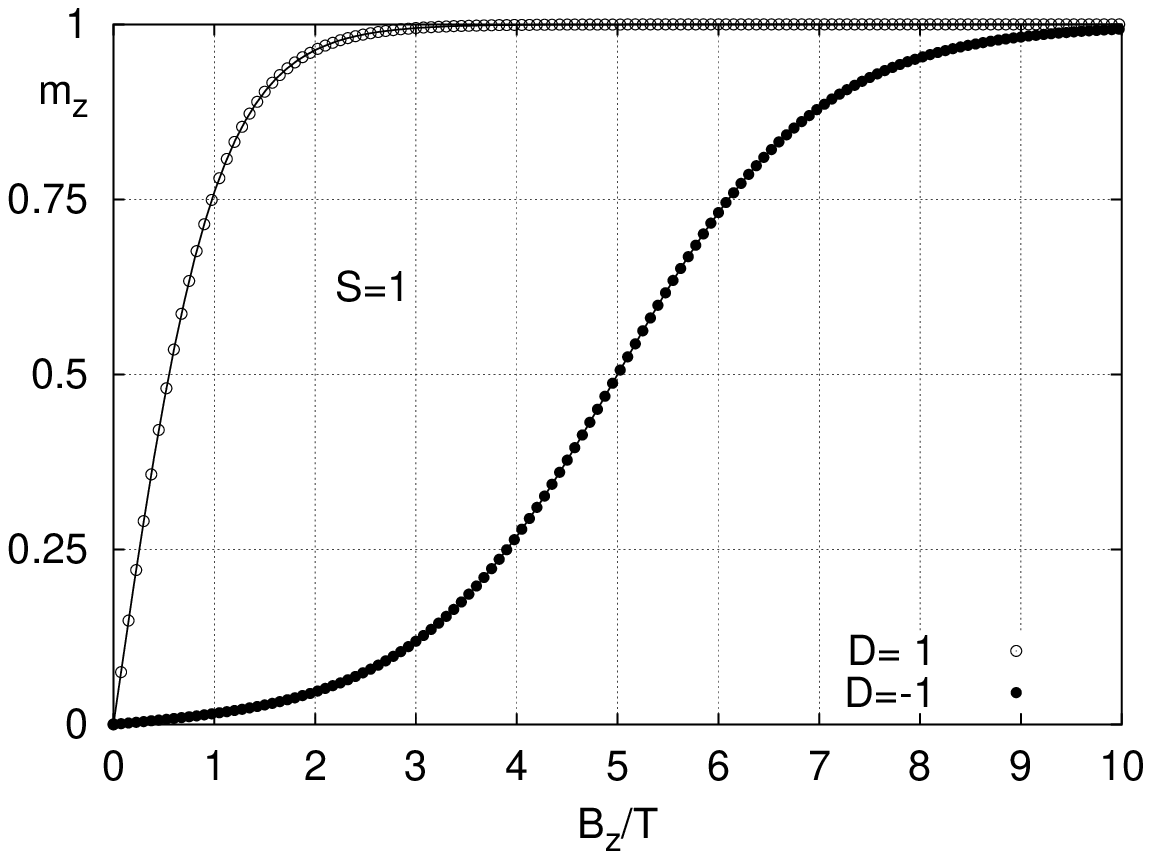}
\hspace*{-3.ex}
\includegraphics[width=7.2cm]{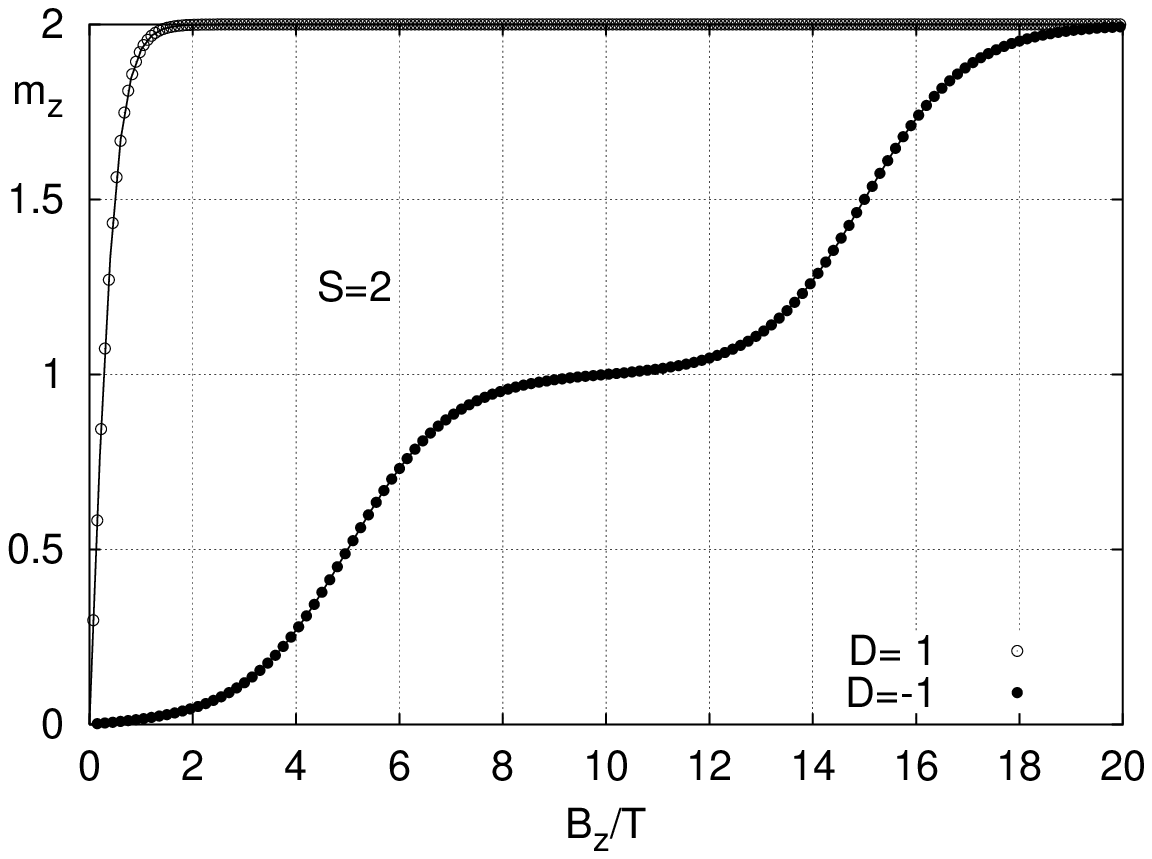}
}
\vspace*{-1.ex}
\caption{
Magnetisation curves of anisotropic spins with $S=1$ (left) and $S=2$
(right) both for positive and negative $\K$ at $\kT=0.2$.
Lines, Gibbsian formulae~(\ref{mz:uni:S=1:2}); symbols,
continued-fraction results (with $\Wo=10^{-9}$; vd.\ footnote in
Sec.~\ref{sec:Siso:static}).
}
\label{fig:Mz:ani}
%
\centerline{
\includegraphics[width=7.2cm]{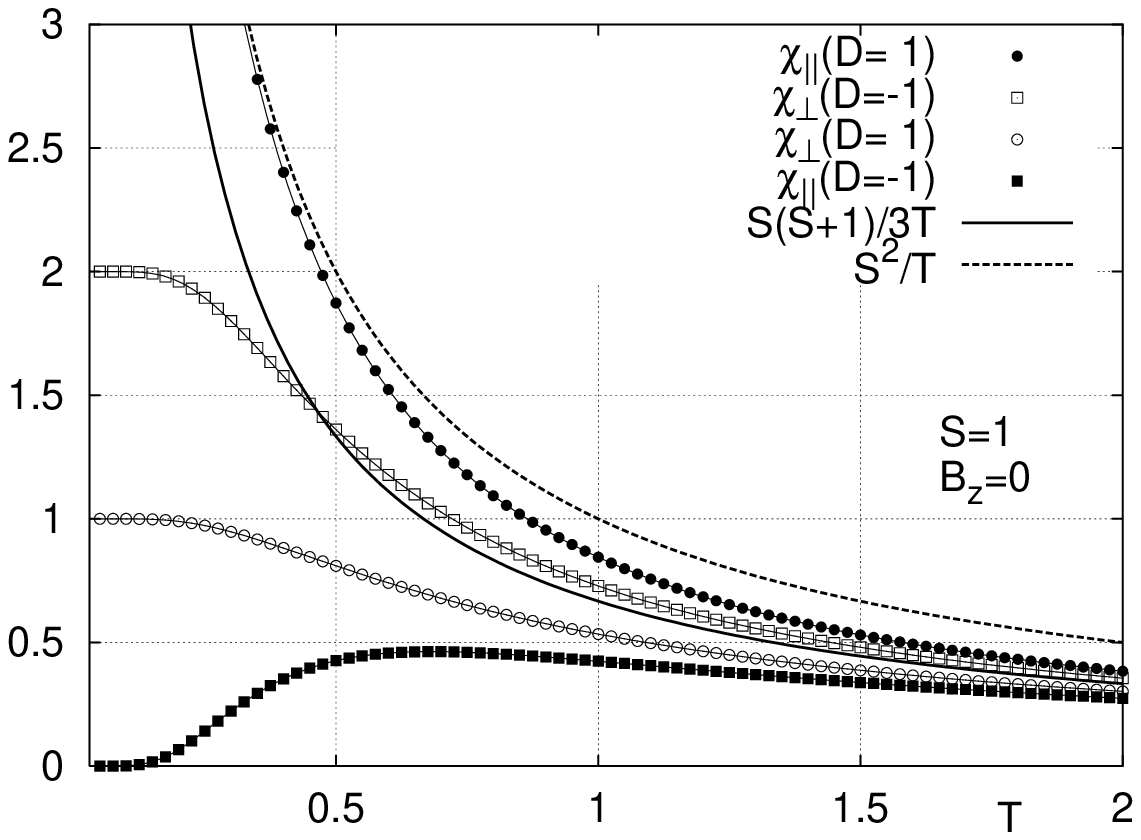}
}
\vspace*{-1.ex}
\caption{
Equilibrium susceptibilities vs.\ temperature of $S=1$ spins with
$\K=1$ (circles) and $\K=-1$ (squares).
We show longitudinal responses (filled symbols) and tranverse (open),
together with Eqs.~(\ref{chi:uni:S=1}) (thin lines).
The thick solid line is Curie law, approached by all curves at high
temperatures, and the dashed line the Ising asymptote [for
$\chi_{\|}(\K>0)$].
}
\label{fig:chi:ani}
\end{figure}

For $\K>0$ the states with $\m=\pm S$ have the lowest energies in
weak fields and are only separated by $2\Bz S$.
Then the magnetisation curves have the convex features of the
isotropic-spin case (Fig.~\ref{fig:Mz:ani}, open symbols).
In contrast, for $\K<0$ the curves depart from zero slowly
(exponentially); the low-field ground state is then $\m=0$, well
separated from the first excited level (by $|\K|-\Bz$).
Indeed, for $S=2$ and $\K<0$, when the field makes $\m=1$ the new
ground state, the magnetisation is again stabilised at $\Mz\simeq1$
until it ``jumps'' to $\Mz\simeq2$.
The jumps become steeper as $T$ is decreased.

The longitudinal susceptibility for $S=1$ follows differentiating
$\Mz$ in Eq.~(\ref{mz:uni:S=1:2}); the transverse response
$\chi_{\perp}$ can be obtained from Van Vleck's formula
(Sec.~\ref{sec:Siso:static}).
The results for the initial (zero-bias) susceptibilities are
%
\begin{equation}
\label{chi:uni:S=1}
\chi_{\|}
=
\frac{1}{\kT}
\,
\frac{2}{2+\e^{-\K/\kT}}
\;,
\qquad
\chi_{\perp}
=
\frac{2}{\K}
\,
\frac{1-\e^{-\K/\kT}}{2+\e^{-\K/\kT}}
\;,
\qquad
(S=1)
\;.
\end{equation}
At high $T$ both expressions recover the isotropic susceptibility:
$\chi_{\|}|_{d=0}=\chi_{\perp}|_{d=0}=2/3\kT$
[here $S(S+1)=2$].
At low temperature one has $\chi_{\|}|_{d\to\infty}=1/\kT$ for $\K>0$
(2-state like) in accordance with
$\Mz
\to
\thrm y
\sim
\Mz|_{S=1/2}$.
Thus in both limit ranges $\chi_{\|}$ obeys $1/\kT$ laws, with the
factor $1/(1+\frac{1}{2}\e^{-\K/\kT})$ governing the intermediate $T$
crossover between the isotropic-spin and Ising regimes.
In contrast, for $\K<0$, the longitudinal susceptibility goes to zero
exponentially at low $T$.
Again this is due to the $\m=0$ ground state for easy-plane
anisotropy.
Finally, as for the transverse response at low $T$, it tends to the
constant limits $\chi_{\perp}|_{d\to\infty}=1/\K$ and
$\chi_{\perp}|_{d\to-\infty}=2/|\K|$.
%

Textbook examples of the longitudinal and transverse susceptibilites
\cite{carlin} are displayed in Fig.~\ref{fig:chi:ani}, showing
the agreement between the analytical expressions and the
continued-fraction results (the longitudinal obtained as
$\chi_{\|}=(\langle\Sz^{2}\rangle-\langle\Sz\rangle^{2})/\kT$
and the transverse from $\chi_{\perp}(\w)$ using a small
$\w/\Wo=10^{-3}$; cf.\ Sec.~\ref{sec:Siso:static}).
This agreement, together with the magnetisation curves of
Fig.~\ref{fig:Mz:ani}, indicates that we are handling properly a
quantum system with non-equispaced levels, as well as its transverse
response, which requires off-diagonal density-matrix elements.


\newpage
\subsection{
Dynamical response
}
\label{sec:Sani:dynamic}


\subsubsection{
Analytical results.
}
\label{sec:Sani:S1}

Again exact expressions can be obtained for small $S$ from the balance
equations~(\ref{DME:mm}), with coefficients
$\R_{\m}^{-}=\Wu_{\m\tm\m-1}\lfb_{\m-1}^{2}$,
$\R_{\m}=-(\lfb_{\m}^{2}\Wu_{\m+1\tm\m}+\lfb_{\m-1}^{2}\Wu_{\m-1\tm\m})$,
and $\R_{\m}^{+}=\Wu_{\m\tm\m+1}\lfb_{\m}^{2}$.
For $S$ one-half the coupling model considered does not produce
relaxation (see \ref{app:ladder}).
The first non-trivial case is $S=1$, whose longitudinal susceptibility
comprises two Debye factors \cite{zuegar2006}
\begin{equation}
\label{chi:1:ani}
\chi_{\|}(\w)
=
\frac{\Mz'}{\kT}
\Big(
\frac{\coefa}{1+\iu\w/\Lambda_{1}}
+
\frac{1-\coefa}{1+\iu\w/\Lambda_{2}}
\Big)
\;,
\qquad
\coefa
=
\frac
{\Lambda_{2}-\Lambda_{\rm eff}}
{\Lambda_{2}-\Lambda_{1}}
\;.
\end{equation}
Here $\Lambda_{i}$ are eigenvalues of the balance-equations matrix
%
\begin{equation}
\label{lambdas}
\Lambda_{0}
=
0
\;,
\qquad
\Lambda_{1,2}
=
(\Gamma_{+}+\Gamma_{-})
\mp
\sqrt{
(\Gamma_{+}-\Gamma_{-})^{2}
+
4 w_{+}w_{-}
}
\;,
\end{equation}
with the rates
$\Gamma_{\pm}
=
2[\Wu(\tf_{\pm1,0})+\Wu(-\tf_{\pm1,0})]$
%
%
and transition probabilities
$w_{\pm}=\Wt_{\pm1\tm0}=2\,\Wu(\tf_{\pm1,0})$
[recall that $\Wu(\tf)=\Wo\,\tf^{\io}/(\e^{\tf/\kT}-1)$,
$\tf_{\pm1,0}=-(\K\pm\Bz)$, and $\lfb_{\pm1,0}=\pm\sqrt{2}$].
The amplitude $\coefa\in[0,\,1]$ controlling the weights of the two
summands involves $\Lambda_{1,2}$ and $\Lambda_{\rm
eff}=(w_{+}+w_{-})/\Z\Mz'$, the initial decay rate of the
magnetization \cite{zuegar2006}.
As $\Z\Mz'=2(\ch y+2\e^{d})/(2\ch y+\e^{-d})$ we have $\Lambda_{\rm
eff}\to\rateone$ as $\K/\kT\to0$, recovering the susceptibility of
isotropic $S=1$ spins [Eq.~(\ref{chi:1})].
We are not aware of exact results for $\chi_{\|}(\w)$ of larger spins
or for the transverse dynamical response with $\K\neq0$.
%
%


\subsubsection{
Longitudinal response.
}

As the range of parameters that can be explored is wide, we
concentrate on low temperatures (and eventually, weak fields), the
experimentally most interesting range in superparamagnets.
It is convenient to introduce {\em reduced\/} anisotropy and field
parameters
%
\begin{equation}
\label{sigma-xi}
\sigma
=
|\K|\,S^{2}/\kT
\;,
\qquad
\xi
=
S\,\Bz/\kT
\;,
\qquad
h
=
\xi/2\sigma
\;.
\end{equation}
The latter is $\Bz$ in units of $2|\K|S$ which is of the order of the
anisotropy field at the minima or the field for barrier disappearance
(\ref{sec:Banis}).
As mentioned before, when comparing different $S$ we scale parameters
keeping $\sigma$ and $\xi$ fixed.
\begin{figure}[!tb]
\centerline{
\includegraphics[width=7.2cm]{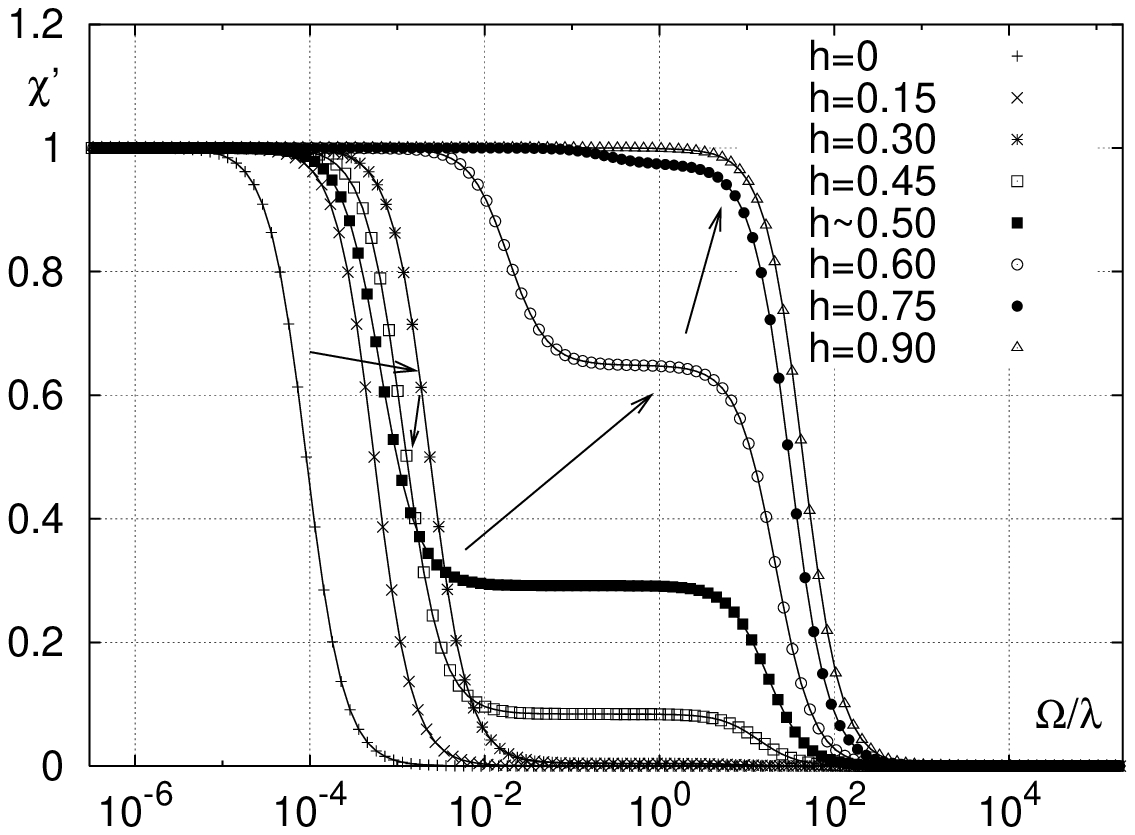}
\hspace*{-3.ex}
\includegraphics[width=7.2cm]{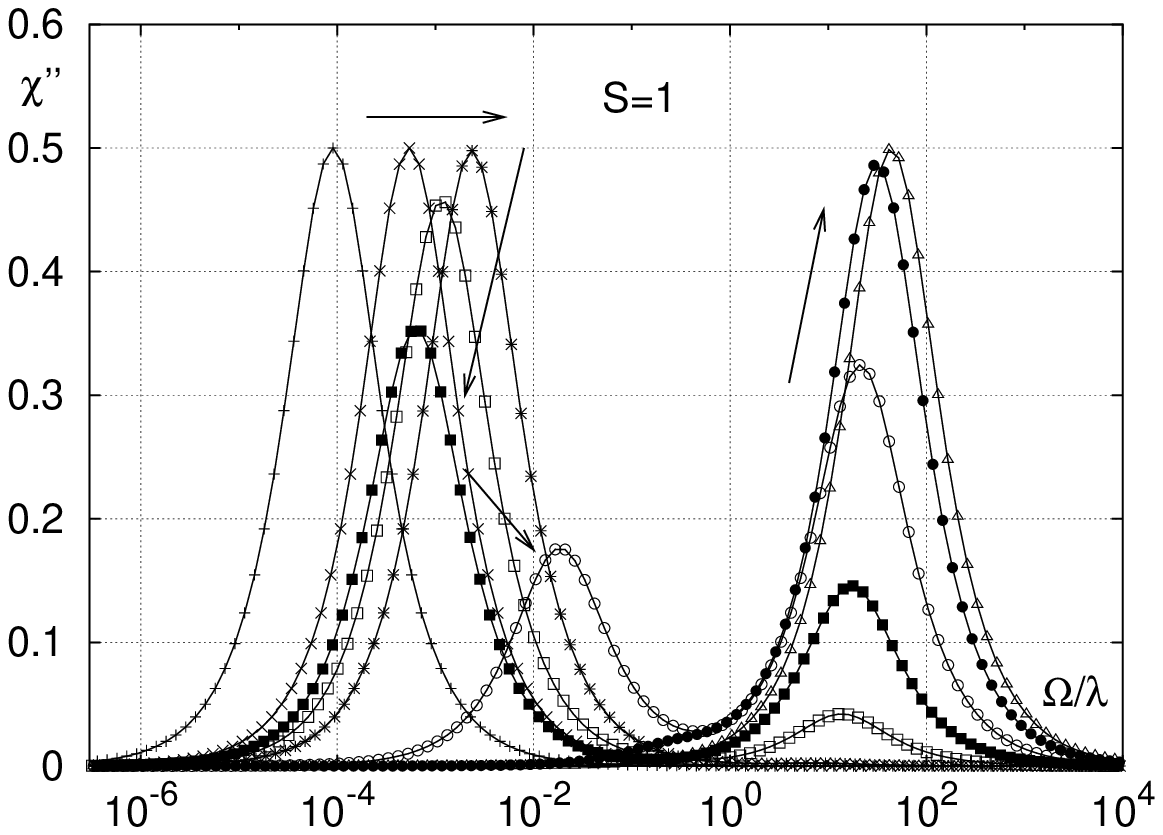}
}
\vspace*{-1.ex}
\caption{
Longitudinal susceptibility spectra of an anisotropic $S=1$ spin with
$\K=1$ and $\Wo=10^{-9}$ at $\kT=0.1$ and various $h=\Bz/2DS$, below
and above barrier disappearance $B_{\rm c}=\K$ ($\sigma=10$,
$\xi=h\cdot2\sigma=0\to18$).
Lines, exact two-mode Eq.~(\ref{chi:1:ani}); symbols,
continued-fraction results; the arrows trace the field evolution.
The susceptibility is normalised by the equilibrium $\w\to0$ value.
%
%
}
\label{fig:chi:ani:S1:w}
%
\centerline{
\includegraphics[width=7.2cm]{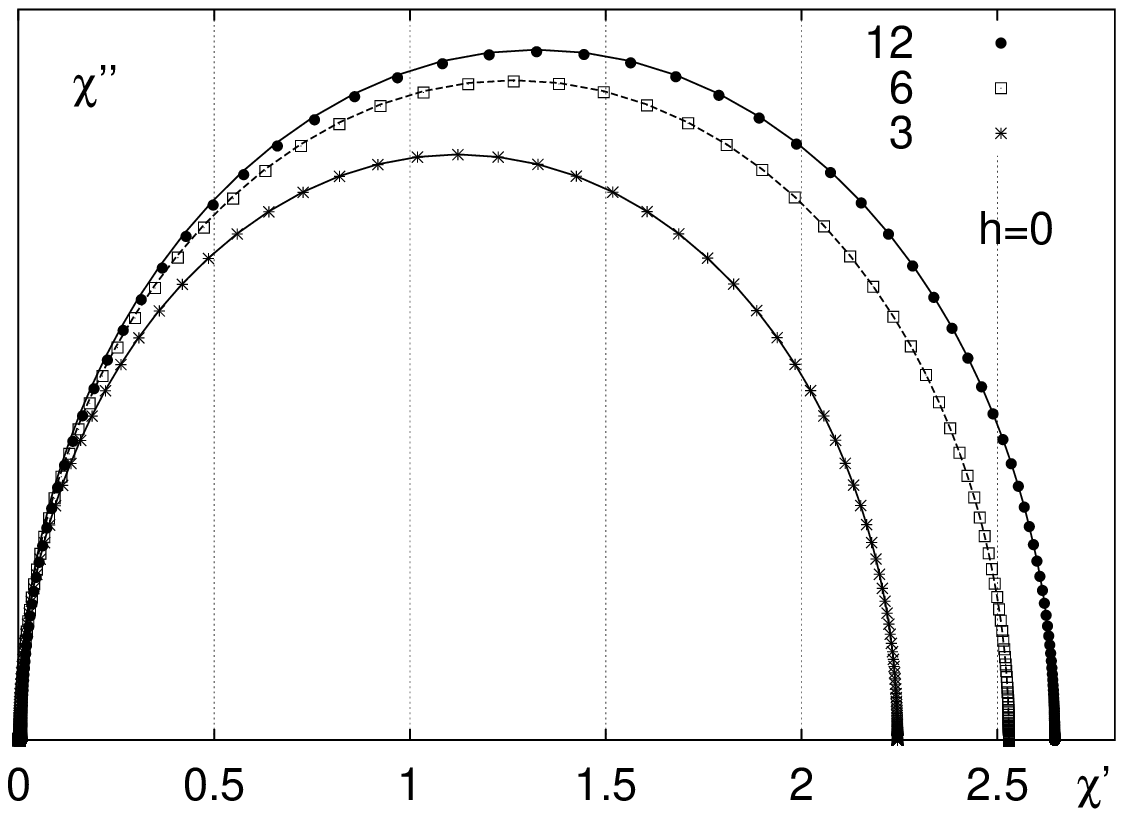}
\hspace*{-3.ex}
\includegraphics[width=7.2cm]{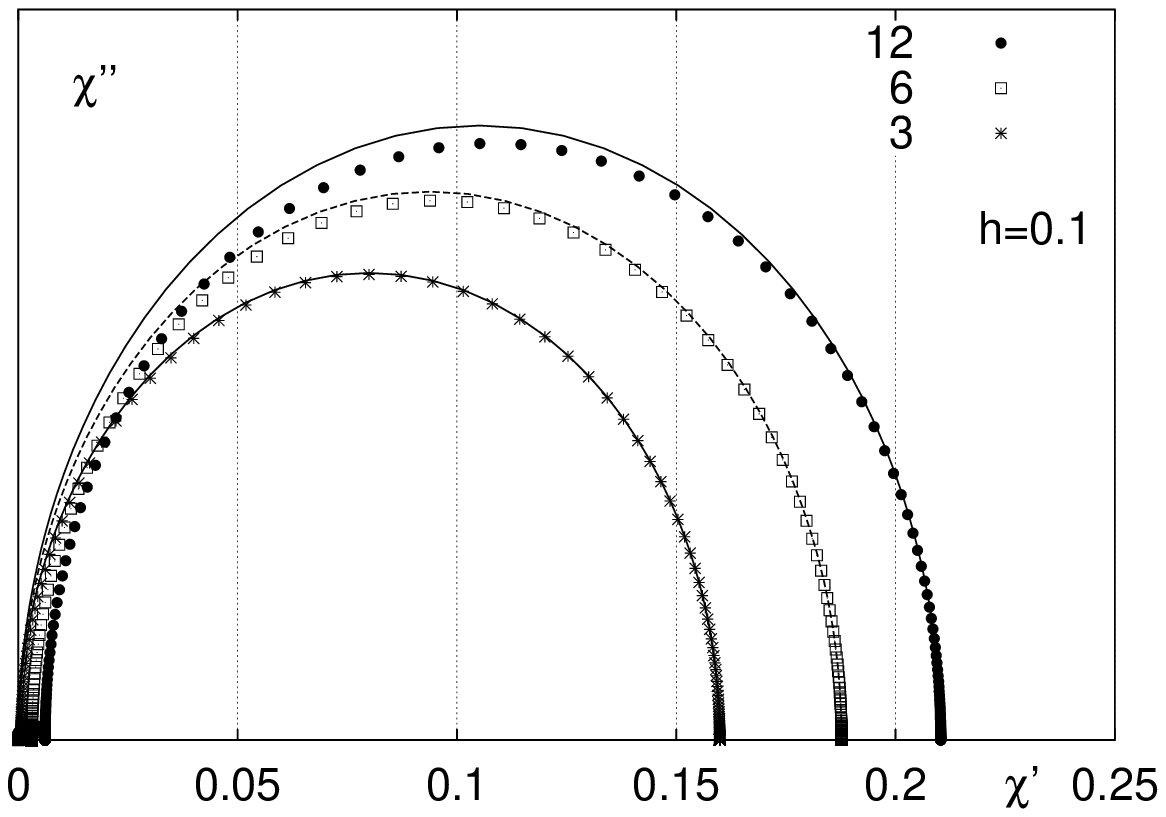}
}
\vspace*{-1.ex}
\centerline{
\includegraphics[width=7.2cm]{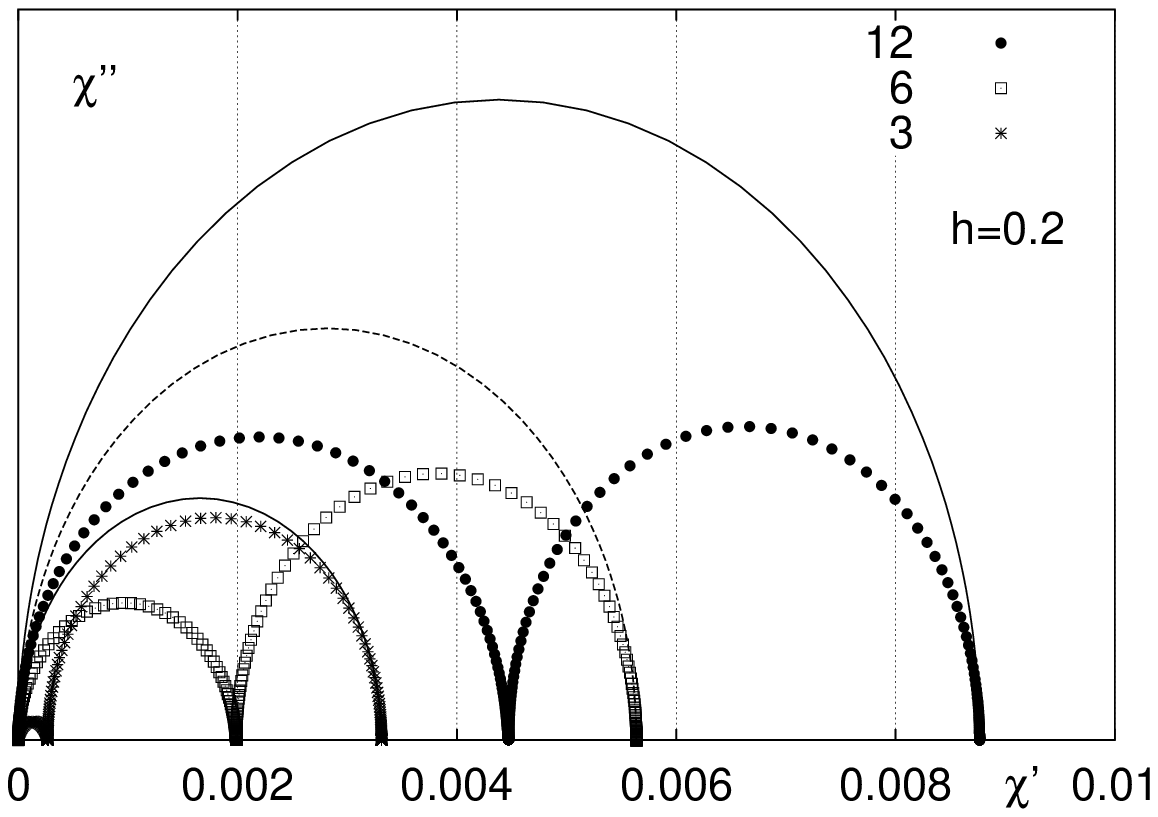}
\hspace*{-3.ex}
\includegraphics[width=7.2cm]{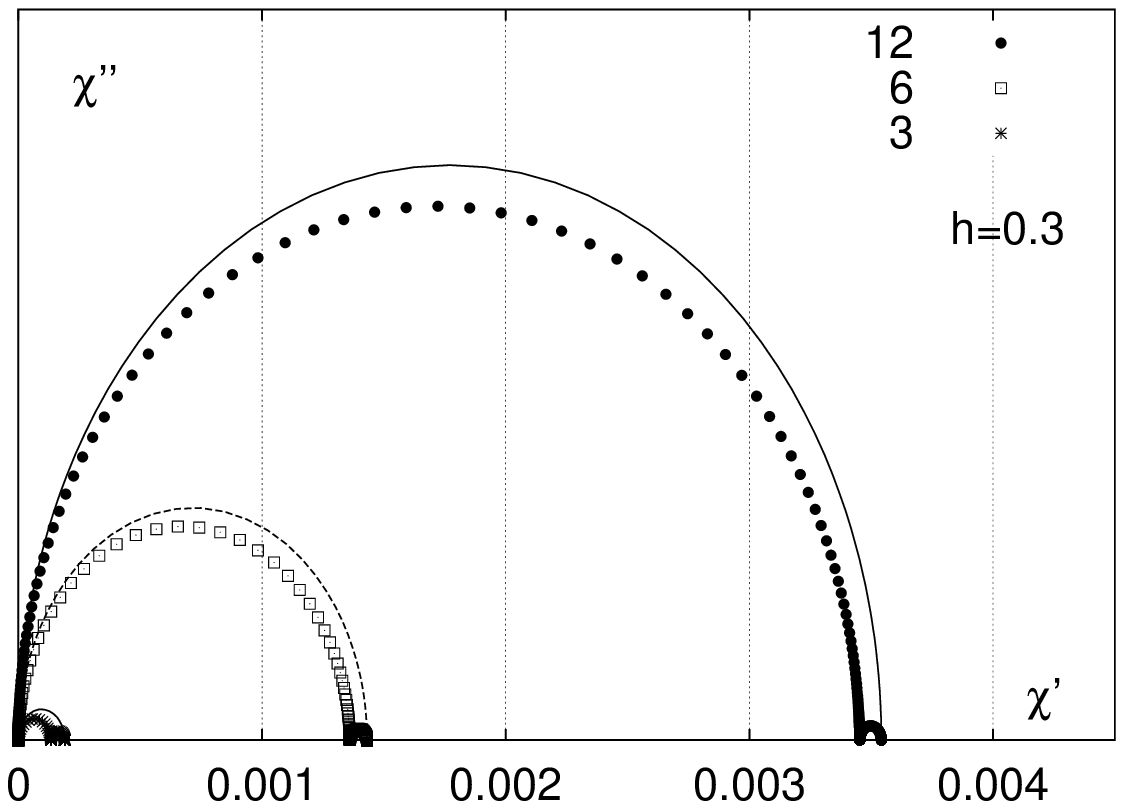}
}
\vspace*{-1.ex}
\caption{
Cole-Cole plot of the longitudinal susceptibility of spins $S=3$, $6$,
$12$. 
The anisotropy parameter $\K$ is adjusted to fix
$\sigma=\K\,S^{2}/\kT=10$ and the fields such that $\xi=S\Bz/\kT=0$,
$2$, $4$, $6$ ($h=0$, $0.1$, $0.2$, $0.3$).
The lines are single-Debye approximations and the symbols
continued-fraction results.
The susceptibilities are scaled by $\chi_{0}=S(S+1)/3\kT$ and the
$y$ axis range is half the $x$ axis one.
}
\label{fig:chi:ani:S:SRT}
\end{figure}

Let us begin with $S=1$.
Its dynamical susceptibility is shown in Fig.~\ref{fig:chi:ani:S1:w}
and the full agreement between the analytical and numerical results.
The curves evidence two relaxation modes [$(2S+1)-1$, the equilibrium
$\Lambda_{0}=0$].
Inspecting the structure of the corresponding eigenvectors
\cite{zuegar2006}, the low-frequency mode, $\Lambda_{1}$, can be
associated to over-barrier crossings and the faster mode,
$\Lambda_{2}$, to transitions between neighbouring levels
($\pm1\leftrightarrow0$, intra-well dynamics).
The over-barrier process dominates the response at weak fields; the
intra-well is active but by symmetry its contribution to
$\langle\Sz\rangle$ practically cancels out (but not to
$\langle\Sz^{2}\rangle$, the Kerr relaxation observable).
Increasing $\Bz$ the spectrum losses the potential barrier at $B_{\rm
c}=\K$ ($h_{\rm c}=1/2$) and the fast transitions between adjacent
states take over.
For $\Bz\gsim\K$ the two modes are still separated, because the levels
are not equispaced yet.
Finally at high enough fields a Zeeman spectrum is approached and the
isotropic susceptibility recovered [Eq.~(\ref{chi:1})]; then the two
modes are close (in frequency) and $\chi(\w)$ approaches again a
single Debye form.

Let us now address the response of larger spins.
Figure~\ref{fig:chi:ani:S:SRT} shows that, although we should be
finding $2S$ modes in $\chi(\w)$, they appear gathered in two main
groups: the over-barrier mode $\Lambda_{1}$ and a bunch of
high-frequency modes, related to intra-well transitions.
This, in turn, leads to a phenomenology akin to that of $S=1$.
In this figure we have chosen the Argand plot ($\chi''$ vs.\ $\chi'$)
where competing modes are resolved in two neat semicircles.
They evolve into one in the limits of low and high field.
In the low $\Bz$ regime the response is dominated by the over-barrier
dynamics and there is a good agreement with a single Debye form
$\chi(\w)=\chi/(1+\iu\w\tau)$, with $\tau\sim1/\Lambda_{1}$.
But in contrast to $S=1$, the intra-well modes are clearly manifested
at fields $h^{\ast}$ well below the field of barrier disapearance
$h_{\rm c}=1-1/2S$ [Eq.~(\ref{Bcrit:Q})].
Indeed, in the classical limit it was estimated that above only
$h^{\ast}\sim0.17$ the fast modes can significantly compite with the
over-barrier process \cite{gar96} due to the thermal depopulation of
the upper well (cf.\ Fig.~\ref{fig:levels}).
For large $S$ (say $S\gsim10$) we obtain $h^{\ast}$ approaching such
classical result.
Besides, we see that the onset of the intra-well modes depends on the
spin value.
That is, $h^{\ast}=h^{\ast}(S)$, increasing for decreasing
$S$.
This seems natural because the results should recover
$h^{\ast}\sim0.5$ as $S\to1$.
Equivalently, at a fixed $h$ the semicircle of the fast modes (the
left one) is less developed the smaller the $S$ is (panel $h=0.2$).
Eventually, at large $\Bz$ a single-Debye again describes $\chi(\w)$
for all $S$.
(For further discussion on $h^{\ast}(S)$, the modes interplay, and
analytical approximations, see Ref.~\cite{zuegar2006}.)


\subsubsection{
Superparamagnetic blocking.
}

Finally we present the results at low fields, the range most studied
in nanomagnets ($2\K S\sim10$\,T in Mn$_{12}$) in a way closer to
experiment \cite{baretal96,gometal98}.
There one varies $T$ at a fixed $\w$, because the exponential
dependence of the relaxation time $\tau\propto\exp(\Delta U/\kT)$
permits to span various decades in $\w\tau$ in an easier way.
The results show the phenomenology of superparamagnetic blocking---a
maximum in the magnitude of the response at some intermediate
$\w$-dependent temperature (Fig.~\ref{fig:chi-ani-SB}).
\begin{figure}[!t]
\centerline{
\includegraphics[width=7.2cm]{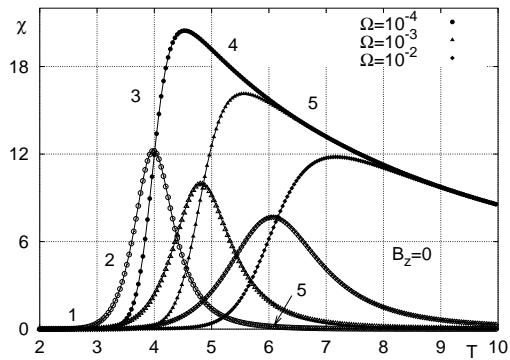}
}
\vspace*{-1.ex}
\caption{
Dynamical susceptibility $\chi_{\|}$ vs.\ temperature of an
anisotropic spin $S=10$ with $\K=0.5$ at several frequencies (in units
of $\Wo/\K^{2}$).
Real parts (solid symbols), imaginary parts (open); single-Debye
Eq.~(\ref{chi:SRT}) (lines).
The numbers by the $\w=10^{-4}$ curves correspond to the regimes
discussed in the text.
Considering $\K$ and $T$ given in Kelvin, the values used are close to
those of Mn$_{12}$.
}
\label{fig:chi-ani-SB}
\end{figure}
This is different in nature from the maxima exhibited by the {\em
equilibrium\/} $\chi_{\|}$ for $\K<0$, or for $\K>0$ in an external
field (Figs.~\ref{fig:chi-eq-iso} and~\ref{fig:chi:ani}).
The dynamical blocking is due to the competition of $\tau$ with $1/\w$
and the two-fold rolex played by $T$.
It unblocks the over-barrier transitions at low temperatures, enabling
the spins to follow the oscillating field $\dB\cos(\w t)$, but
sufficiently high $T$ also provokes the thermal misalignment of the
spins from the field direction, degrading the response.

Let us follow the process in some detail.
(1) At low temperatures, $\tau\gg1/\w$, the probability of
over-barrier crossings is negligible and the dynamics consists of
transitions at the bottom of the wells (with a small averaged
projection onto the field).
(2) Increasing $T$ the spins appreciably depart from the minima, and
the response starts to rise with $T$.
However, as the thermo-activation is not efficient enough, the
response
$\langle\Sz\rangle(t)\sim\dB\,(\chi'\cos\w t+\chi''\sin\w t)$
sizable lags behind the field, as manifested by the large $\chi''$.
(3) At higher temperatures the over-barrier mechanism becomes more
efficient; the reponse continues increasing, but becoming more
in-phase with the excitation ($\chi'$ dominates).
(4) If $T$ is further increased, however, the thermal agitation also
provokes misalignment of the spins from the field direction.
Then the magnitude of the response exhibits a maximum and starts to
decrease; this occurs at a temperature $T_{\rm b}$ such that
$\tau(T_{\rm b})\sim1/\w$.
(5) Eventually, at high $T$, the spins quickly adjust to the
equilibrium distribution corresponding to the instantaneous field.
Then $\chi'$ goes over the equilibrium susceptibility ($\propto1/T$)
while $\chi''$ drops to zero.
Note finally that, in agreement with the previous subsection, this
low-field phenomenology is well described by a single Debye form, as
Fig.~\ref{fig:chi-ani-SB} shows.%
\footnote{
%
For low $\w$ we have found numerical instabilities of the
continued-fraction results with very large $\Delta U/\kT$.
Some accuracy problems also arised in isotropic spins at very large
$\Bz/\kT$.
They can be attributed to the exponential dependences of the
relaxation rates, giving tiny numbers at certain critical places; this
is known to happen in the classical case under the same limit
conditions.
%
%
} 
%


\subsubsection{
Paramagnetic resonance of anisotropic spins.
}
\label{sec:Sani:S:FMR}

We conclude with the spin-resonance behaviour of quantum
superparamagnets \cite{gardat2005}.
Recall that a field oscillating perpendicular to the anisotropy axis
provokes transitions $\mket\to|\m\pm1\rangle$ between the unperturbed
levels.
Computing the response along such field 
$\langle\Scpm\rangle=\sum_{\m}\lf_{\m}^{\pm}\dm_{\m,\m\pm1}$
requires off-diagonal density-matrix elements and falls outside a
Pauline master equation for the populations $\dm_{\m\m}$.

The induced transitions result in peaks in the absorption line-shape
$\chi''(\w)$ at the frequencies $\tf_{\m,\m\pm1}$.
For an isotropic spin all level differences were equal
$\tf_{\m,\m+1}=\Bz$.
Here, however, the anisotropy yields non-equispaced levels
$\tf_{\m,\m+1}=\K(2\m+1)+\Bz$ and one would expect multiple peaks in
$\chi''(\w)$, with the corresponding zig-zags in the real part
$\chi'(\w)$.
At zero field the $2S+1$ levels are degenerated by pairs, $\m$ with
$-\m$ (Fig.~\ref{fig:levels}), and we should find only $S$ peaks at
the locations $\w=\K(2\m+1)$.
The largest frequencies ($\tf_{\mathrm{w}}\sim2\K S$) correspond to
transitions at the wells ($|\m|\sim S$), while those near the barrier
top appear at low $\w$ ($|\m|\sim0$, $\tf_{\mathrm{b}}\sim\K$).
Finally, as we saw in the isotropic $S=1/2$ spin
(Fig.~\ref{fig:peaks:SR:iso}), the absorption peaks have finite width
and height due to the damping and the temperature.
\begin{figure}[!tb]
\centerline{
\includegraphics[width=7.2cm]{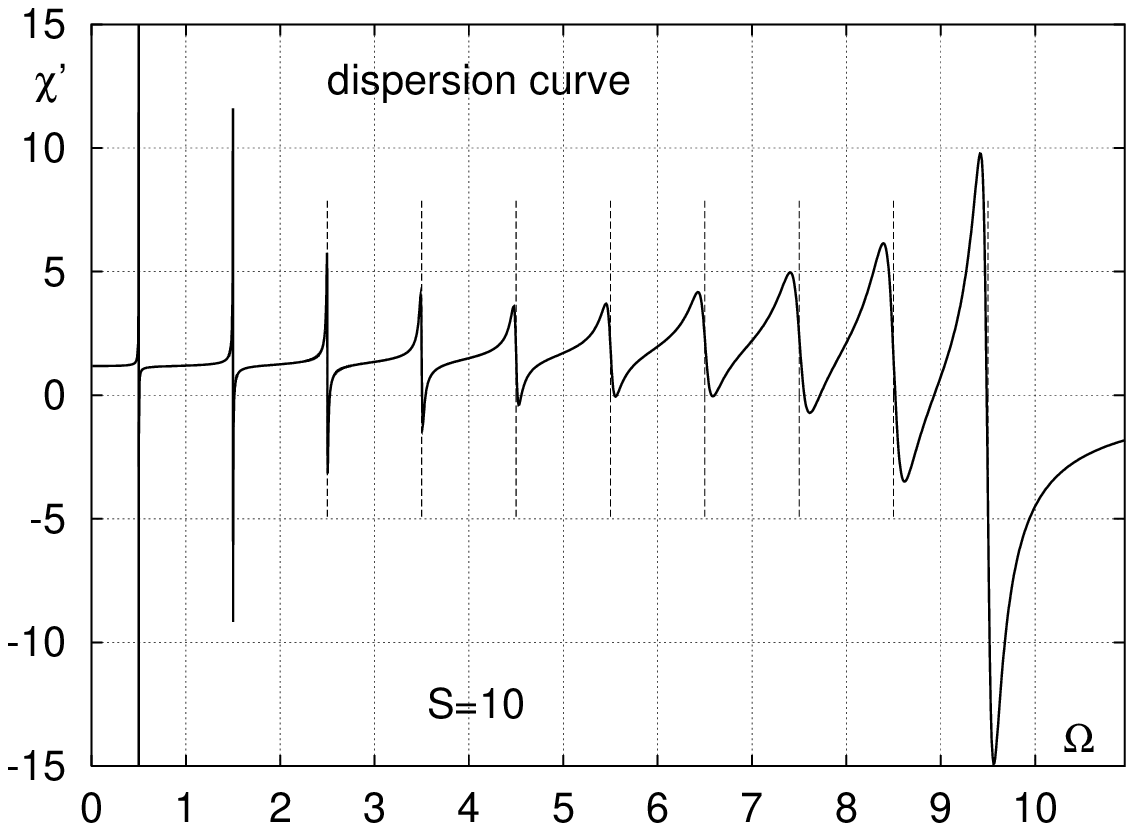}
\hspace*{-3.ex}
\includegraphics[width=7.2cm]{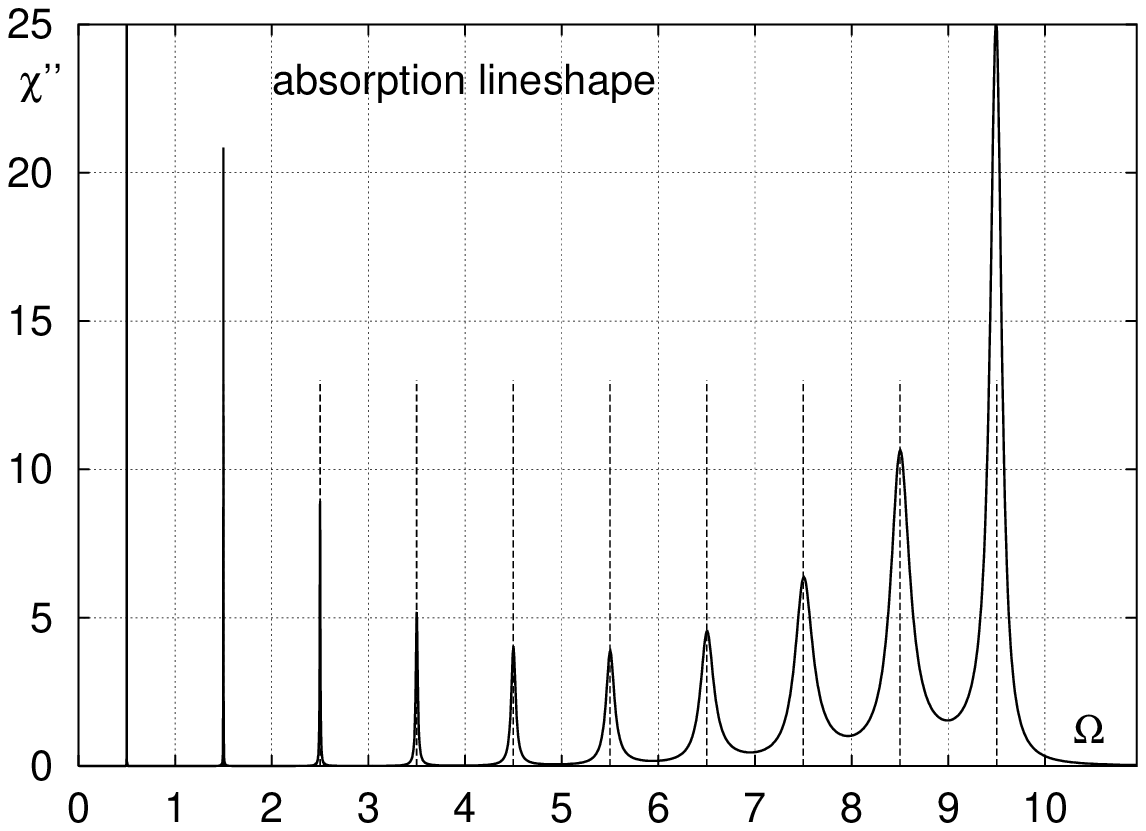}
}
\vspace*{-1.ex}
\caption{
Continued-fraction results (lines) for $\chi_{\perp}(\w)$ of an
anisotropic $S=10$ spin with $\K=0.5$ and $\Wo/\K^{2}=3\cdot10^{-8}$.
The calculations were done at zero field and $\kT=10$ ($\sigma=5$).
Real part, left panel; imaginary part, right.
Vertical lines: loci of the transition frequencies
$\tf_{\m,\m+1}=\K(2\m+1)$, $\m=0,1,\cdots,9$.
%
%
}
\label{fig:peaks:SR}
\end{figure}

Figure~\ref{fig:peaks:SR} shows these features for $S=10$.
Starting from the rightmost $\chi''$ peak, associated to the
ground-state transitions, the intensity of the peaks decreases with
$\w$, as they progresively involve transitions between higher levels,
thermally less populated.
On the other hand, the peak width is not uniform in $\w$.
This is due to the $\Sz$-dependent spin-phonon interaction,
$F\sim\Sz\Scpm$, which gives an extra $\m$ dependence of the
relaxation term (compared to the coupling $F\sim\Scpm$).
This enters via the modified ladder factors
$\lfb_{\m}^{2}=(2\m+1)^{2}\lf_{\m}^{2}$ and can be seen as an
effective level-dependent ``damping'',
$\Wo_{{\rm eff}}(\m)\sim\Wo\,(2\m+1)^{2}$.
Therefore the transitions between upper levels (lower $\m$ and $\w$)
correspond to a reduced effective damping, and those peaks become
narrower and higher.
Overall, the competition of the thermal depopulation and the reduced
damping yields peak heights initially decreasing as $\w$ is reduced
and rising again at low frequencies.%
\footnote{
Note that $\Wu(\tf_{\m,\m\pm1})$ can also add to the dependence on
$\m$ \cite{gardat2005}, except for an Ohmic bath at high $T$ (then
$\Wu\simeq\Wo\,\kT$).
However, the bare factor $\lf_{\m}^{2}=S(S+1)-\m(\m+1)$ does not.
It is geometrical, giving the factor $(1-z^{2})$ in the classical
Fokker--Planck equation \cite{zuegar2006} which accounts for the
increased phase space at large angles $z=\cos\vartheta$.
} 

\begin{figure}[!tb]
\centerline{
\includegraphics[width=7.2cm]{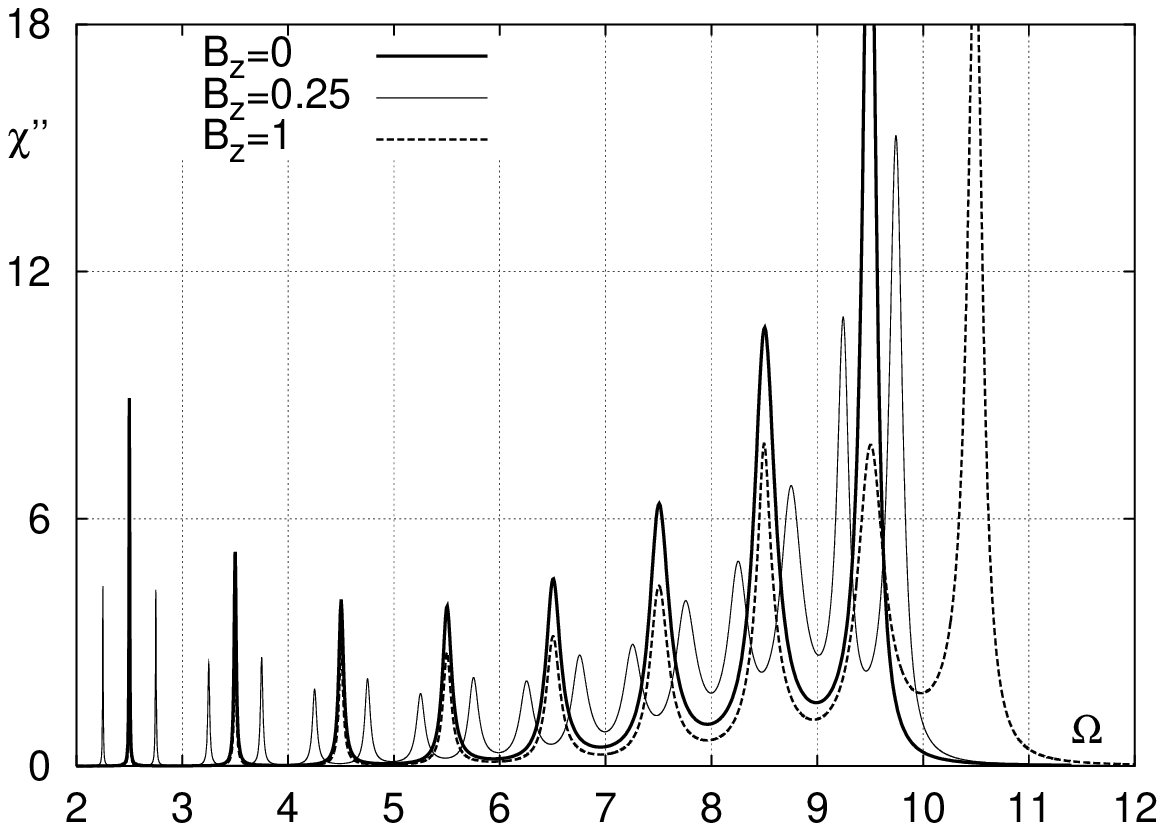}
\hspace*{-3.ex}
\includegraphics[width=7.2cm]{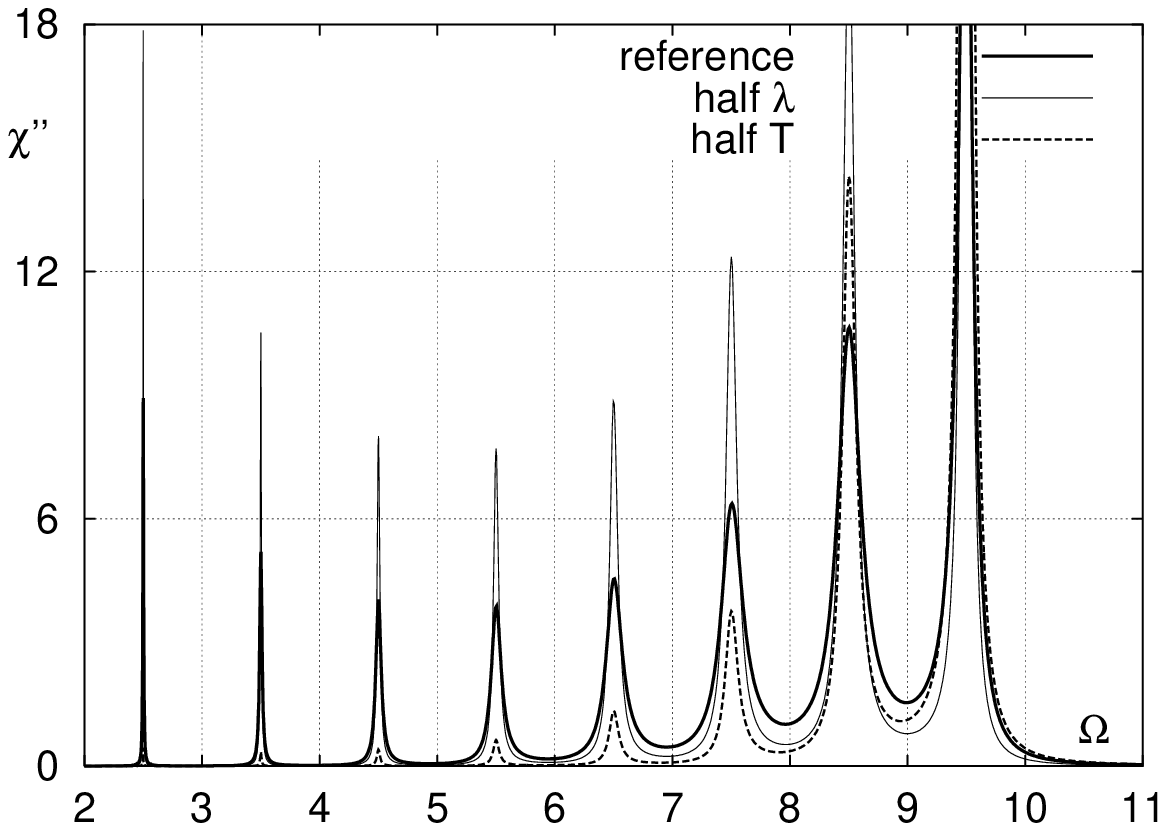}
}
\vspace*{-1.ex}
\caption{
Effects of the external field, damping, and temperature on the
line-shape $\chi_{\perp}''(\w)$ of anisotropic $S=10$ spins with
$\K=0.5$.
Left: results for $\Wo/\K^{2}=3\cdot10^{-8}$ and $\sigma=5$ at $\Bz=0$
(as in Fig.~\ref{fig:peaks:SR}), $\Bz=0.25$ ($=\K/2$) and $\Bz=1$
($=2\K$; cf.\ Fig.~\ref{fig:levels}).
Right: reference zero-field line of left together with results halving
the damping at the same $T$, and halving $T$ while keeping $\Wo$.
}
\label{fig:peaks:SR:effects}
\end{figure}

We conclude with various effects on the line shape $\chi''(\w)$.
The application of a field lifts the degeneracy by pairs and the peaks
split (Fig.~\ref{fig:peaks:SR:effects}, left).
At the first resonance $\Bz=\K$ the levels become degenerated again
($\m=0$ and $\m=-1$, $\m=1$ and $\m=-2$,\dots; Fig.~\ref{fig:levels})
and the energy differences, $\tf=0$, $2\K$, $4\K$,\dots, are just
half-way those of zero field, $\tf=1\K$, $3\K$,\dots, $\K(2S-1)$.
Then the ``side peak'' to the right merges with the one to the left of
the neighbouring peak (curve not shown).
For $\Bz>\K$ the peaks split again (they simply crossed) and at the
first even resonance $\Bz=2\K$ they merge again but on the original
locations (the spacings correspond to the $\Bz=0$ ones plus
$\tf=\K(2S-1)+2\K$ from the ground state; Fig.~\ref{fig:levels}).
Figure~\ref{fig:peaks:SR:effects} also shows the sharpening of the
peaks when decreasing the damping (as in the isotropic $S=1/2$ spin,
with the new feature of non-uniform widths) and the dramatic reduction
with temperature of the intensity of the lower $\w$ lines.
Those transitions involve higher levels whose thermal population gets
exponentially reduced as $T$ is lowered.


\section{
Summary and discussion
}
\label{sec:summary}

In the field of quantum dissipative systems one usually works with the
reduced density operator $\dm$ of the subsystem of interest (tracing
over the bath).
In several problems, due to the weak system-bath coupling, one can
derive perturbatively a closed equation of motion for $\dm$ --- a
quantum master equation.
This plays the role of the Fokker--Planck kinetic equation in
classical problems.
We have addressed one such system, a quantum spin with arbitrary $S$
in a dissipative environment, and solved exactly the corresponding
density-matrix equation by implementing continued-fraction methods.
As the density matrix is obtained in full, coherent dynamics is
included along with the relaxation and thermoactivation.

The continued-fraction method belongs to a family of exact methods in
condensed matter and statistical physics and has been fruitfully
exploited in problems of Brownian motion in external potentials.
Here we took advantage of the index-recurrence structure of the
density-matrix equation to bring it in the form of a few-term
recurrence relation, suitable to apply continued fractions.
For simple spin problems this had been done by Shibata and co-workers,
exploiting the decoupling into independent equations for the
diagonals,
$\dot{\dm}_{\m,\m+k}
\sim
F(\dm_{\m',\m'+k})$,
and solving them by {\em scalar\/} continued fractions.
In general such decoupling does not take place (e.g., in the presence
of transverse fields), and {\em matrix\/} continued fractions are
required.
This has been the contribution of this work, allowing the obtainment
of numerically exact solutions of the full density-matrix equation for
quite generic spin problems.
Besides, compared with previous exact techniques, the spin values
affordable have been increased significantly (up to $S\sim100$--$200$
on an old laptop).
This range of $S$ should be enough for studying the evolution to the
classical limit in many problems (one of the central issues in open
quantum systems).
Reaching large $S$ is also important when effective spin models are
used to describe interacting 2-level systems.

Technically, we have worked within a Hubbard formalism (Heisenberg
equations of motion for the operator basis $\Xnm=\nket\mbra$), whose
main advantage is being compact.
This is not essential, however, and was used as intermediate step; the
equations for $\Xnm$ are linear and their averaging gives directly the
standard equations for the density-matrix elements
$\dm_{\n\m}=\nbra\dm\mket$.
%
%
We have focused on stationary responses, for the obtainment of which
this numerical method is specially suited (in contrast with
path-integral propagation schemes affected by sign problems at long
times).
Our starting point was to convert the time-dependent master equation
into a perturbative chain of stationary density-matrix equations with
each step solvable by continued fractions.
We have worked it in full for the linear dynamical susceptibility, but
the extension to get non-linear responses is systematic.
Besides, upon Laplace transformation, a number of time propagation
problems could also be tackled
(of the type ``evolution between stationary states'', not
``system-meets-bath'' problems requiring time-dependent coefficients).

From the outset the implementation was done in its general form (with
matrix continued fractions), not taking advantage of the splitting
into diagonals of simple spin problems.
This made the initial tests tougher while it allowed proceeding
smoothly to more general problems, not enjoying such decoupling.
The implementation has been simpler than in a Fokker--Planck-type
approach with pseudo-distributions, as it avoids the transformation of
the density operator into some phase-space representation, the
expansion in complete sets of functions, and eventually the
manipulation of the coefficient recurrences (as in quantum Brownian
motion in phase space).
Similarly, the implementation has been easier than in the
continued-fraction solution of rotational Fokker--Planck equations for
{\em classical\/} spins and dipoles, as some aspects are simplified in
the quantum case.
For instance, the finite number of discrete levels results in {\em
finite\/} continued fractions, and convergence or termination problems
are fortunately by-passed.
Thus, some numerical instabilities found are to be attributed to
accuracy problems when handling tiny numbers; actually they appeared
in parameter ranges already problematic in the classical case (very
low $T$ and $\w$).
On the other hand, the finite number of steps in the algorithm can be
carried out by hand for small spins.
The approach can actually be called semi-analytic, which is the reason
behind the numerically exact agreements found with explicit solutions.

In this frame and with these tools we addressed the statics and
dynamics of spins with arbitrary $S$ in contact with a thermal bath.
We have considered the familiar isotropic spin, $\Hs=-\Bz\Sz$, and
spins in a bistable anisotropy potential $\Hs=-\K\,\Sz^{2}-\B\cdot\vS$
(superparamagnets).
The first one, with its equispaced spectrum, is a rotational
counterpart of the quantum harmonic oscillator, while the anisotropic
spin corresponds to problems of translational Brownian motion in
non-harmonic potentials (double-well or periodic).
The coupling to the bosonic bath considered have the structures
$\Ham_{\mathrm{sb}}\sim\boldsymbol{\eta}\cdot\vS$
(bilinear) and
$\Ham_{\mathrm{sb}}\sim\{\Sz,\,\boldsymbol{\eta}\cdot\vS\}$
(non-linear).
The former may describe Kondo coupling to electron-hole excitations
and the latter interaction with phonons, two important mechanisms in
solids.
Classically they correspond to field-type and anisotropy-type
fluctuations in the spin Langevin equations.

Both for isotropic and anisotropic spins we have given examples of
static response, the dynamical susceptibility (to analyse the
contribution of the different felaxation modes), and of spin resonance
in transverse fields, which is very sensitive to the level spectrum
and to the structure of the spin-bath coupling.
Recall that effects like the spin resonance, or tunnel in transverse
fields, demand the solution of the full density-matrix; such coherent
dynamics involves off-diagonal elements and is not captured by a Pauli
balance equation for the level populations.
Finally, in some examples we used parameters close to actual quantum
superparamagnets and typical experiments.

We touched in passing the issue of the validity range of a
master-equation description.
Several limitations are inherited from the approximations required to
derive quantum master equations
(factorizing initial conditions, weak system-bath coupling, high-$T$
or semiclassical bath, etc.)
along with manipulations specific of the problem addressed (secular or
rotating-wave approximations, decoupling or adiabatic elimination of
off-diagonal elements, etc.).
This issue, however, is independent of the question of resolvability
of quantum master equations by continued fractions, method which could
in principle be applied to improved equations.


\ack
Work supported by DGES, project BFM2002-00113, and DGA, project {\sc
  pronanomag} and grant B059/2003.
Part of the writing was done in the stimulating and friendly
atmosphere of the S.~N.~Bose National Centre for Basic Sciences in
Calcutta.


\appendix


\section{
Energy-level spacings and effective fields
}
\label{app:energetics}

In this appendix we calculate the level separations and related
quantities (effective fields) for a uniaxial spin in a longitudinal
field
$\Hs_{\rm d}
=
-\K\,\Sz^{2}-\Bz\Sz$ (our unperturbed Hamiltonian).
Then the eigenstates of $\Sz$ are eigenstates of $\Hs_{\rm d}$ too,
i.e., $\Hs_{\rm d}\mket=\el_{\m}\mket$, with
$\el_{\m}=-\K\,\m^{2}-\Bz\m$.
The energy differences between these levels
%
\begin{equation}
\label{energy:differences}
\tf_{\n\m}
=
\el_{\n}
-
\el_{\m}
\quad
\leadsto
\quad
\tf_{\n\m}
=
-
\left[
\K(\n+\m)
+
\Bz
\right]
(\n-\m)
\;,
\end{equation}
appear in the unitary part of the Heisenberg equation for the Hubbard
operators [Eq.~(\ref{DME:closed})] and control the $\m\to\n$
transition rate $\Wu_{\n\tm\m}=\Wu(\tf_{\n\m})$ [Eq.~(\ref{W:def})].


\paragraph{
Level spacings.
}
\label{sec:level:spacings}

In the master equations considered only transitions between adjacent
levels enter.
For $\n=\m\pm1$ the energy differences~(\ref{energy:differences})
give
%
\begin{equation}
\label{level:spacings}
\tf_{\m,\m\pm1}
=
\pm
\left[
\K(2\m\pm1)+\Bz
\right]
\;.
\end{equation}
In contrast with the equispaced Zeeman spectrum, the level spacings
depend on $\m$ due to the anisotropy.
For succesive pairs they are related by
$\tf_{\m-1,\m}
=
\tf_{\m,\m+1}-2\K$,
decreasing as the barrier top is approached (Fig.~\ref{fig:levels}).
To illustrate, for integer $S$ at zero field the evolution from wells
to barrier is
%
\begin{equation}
\label{level:evolution}
\fl
\tf_{\mathrm{w}}
\equiv
\tf_{S-1,S}
=
\K(2S-1)
\to
\K(2S-3)
\to
\cdots
\to
3\K
\to
\K
=
\tf_{0,\pm1}
\equiv
\tf_{\mathrm{b}}
\;.
\end{equation}
The boundaries coincide for $S=1$, while
$\tf_{\mathrm{w}}\sim2\K S$
for large $S$.
For $\K\sim0.5$\,K and $S=10$ (as in Mn$_{12}$) we have the limit
energy scales $\tf_{\mathrm{w}}\sim10$\,K and
$\tf_{\mathrm{b}}\sim0.5$\,K.
Finally, when proceeding towards the classical limit fixing the
anisotropy barrier $\K\,S^{2}$ and Zeeman energy $S\,B$, the levels
approach a continuum as $\tf\sim1/S$. 
%
%


\paragraph{
Effective, anisotropy, and critical fields.
}
\label{sec:Banis}

Classically one defines the effective field
$B_{\eff}\equiv-(\partial\Hs/\partial \mz)$, with $\mz=S\cos\vartheta$
(the spin polar angle).
This quantity enters in the Landau-Lifshitz precession equation.
For a Hamiltonian including only the anisotropy term $\Hs=-\K\Sz^{2}$,
this definition gives the anisotropy field $B_{\anis}=2\K
S\cos\vartheta$.
In the quantum case the $\mz$-derivative is naturally replaced by a
finite difference
$B_{\eff}\equiv-(\el_{\m+1}-\el_{\m})/\Del\m$.
Then, $\Del\m=1$ plus Eq.~(\ref{level:spacings}) gives the effective
and anisotropy fields of the quantum problem
%
\begin{equation}
\label{Beff:Q}
B_{\eff}(\m)
=
\K(2\m+1)+\Bz
\quad
\stackrel{\Bz=0}{\Rightarrow}
\quad
B_{\anis}(\m)
=
2\K
\big(\m+\half\big)
\;.
\end{equation}
For $S\gg1$, we have
$B_{\anis}\simeq2\K S(\m/S)\to 2\K S\cos\vartheta$, 
recovering the classical result.

To conclude, the critical field $B_{\rm c}$ is that at which the
barrier disappears (or equivalently the $\Bz$ that zeroes the last
effective field $B_{\eff}(-S)$; see Fig.~\ref{fig:levels}).
Equating to zero the spacing between the last two levels,
$\tf_{-S+1,-S}(B_{\rm c})=0$,
one gets
%
\begin{equation}
\label{Bcrit:Q}
B_{\rm c}
=
\K(2S-1)
\;,
\qquad
\big(\simeq B_{\anis}|_{\rm wells}\big)
\;.
\end{equation}
This gives $B_{\rm c}=0$ for $S=1/2$, where there is no barrier, while
the classical value matches the anisotropy field at the wells,
$2\K S=B_{\anis}|_{\vartheta=0}$ ($\sim10$\,T in Mn$_{12}$).


\section{
Angular-momentum ladder factors
}
\label{app:ladder}

Here we discuss some properties of the ladder factors
$\lf_{\m}^{\pm}=[S(S+1)-\m(\m\pm1)]^{1/2}$.
In the standard basis of eigenstates of $\vS^{2}$ and $\Sz$, where
$\vS^{2}\mket=S(S+1)\mket$ and $\Sz\mket=\m\mket$, the
$\lf_{\m}^{\pm}$ characterize the action of the raising/lowering
operators $\Scpm\mket=\lf_{\m}^{\pm}|\m\pm1\rangle$.
In addition they are the expansion coefficients of
$\Scpm=\Sx\pm\iu\Sy$ on the Hubbard operator basis
$\Scpm=\sum_{\m}\lf_{\m}^{\pm}X_{\m\pm1}^{\m}$
%
%
[Eq.~(\ref{S:hubbard})].


\paragraph{
Ordinary factors $\lf_{\m}$.
}

\begin{figure}[!t]
\centerline{
\includegraphics[width=5.cm]{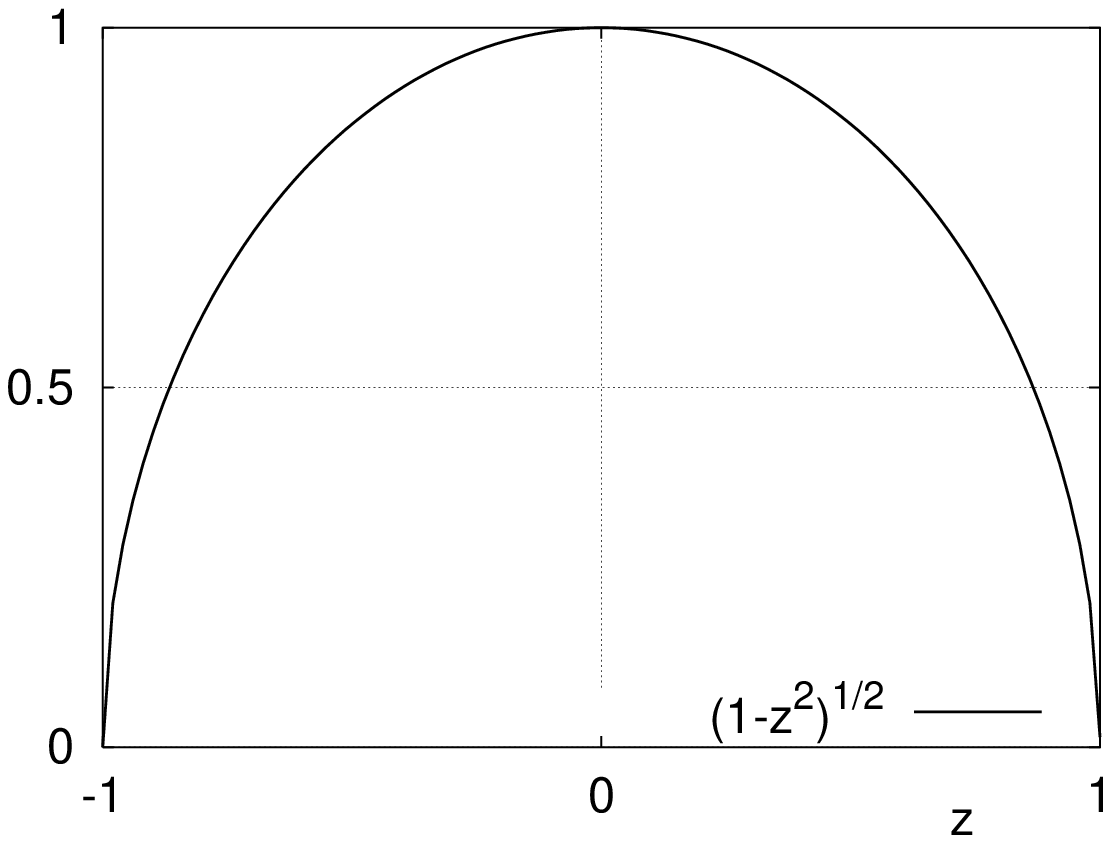}
\hspace*{-4.ex}
\includegraphics[width=5.cm]{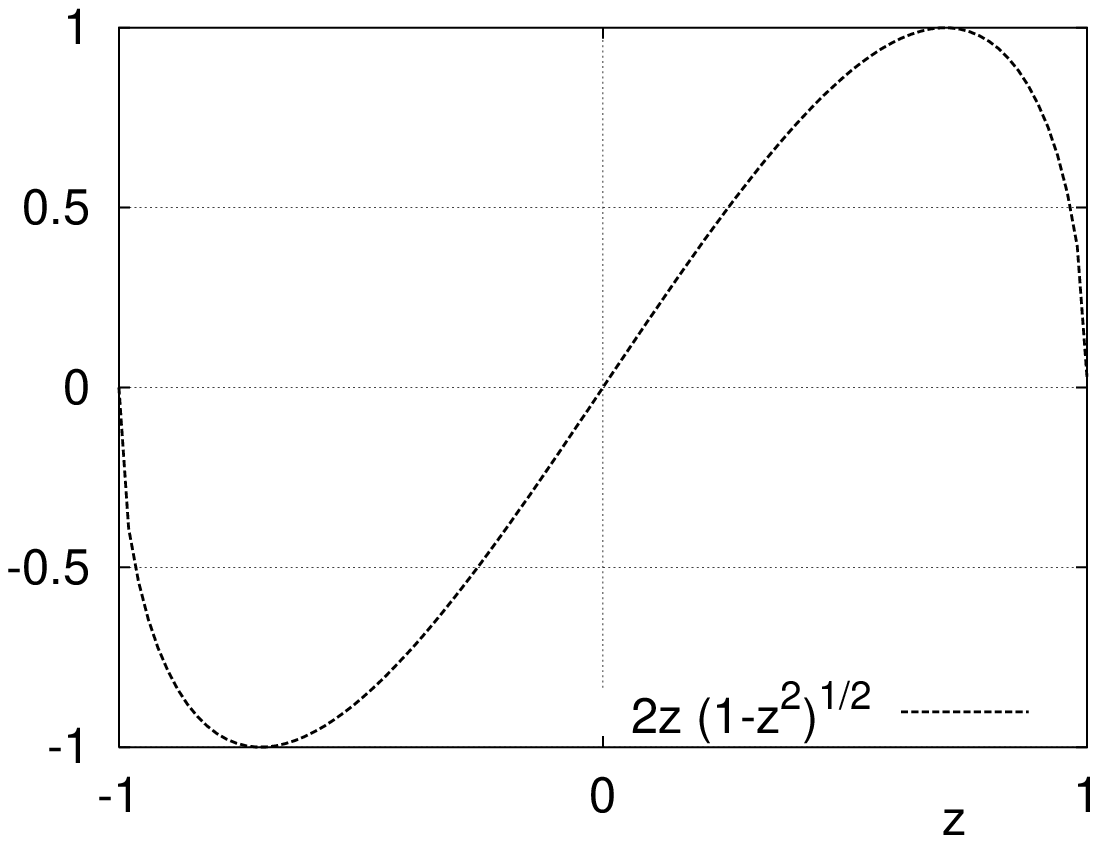}
\hspace*{-4.ex}
\includegraphics[width=5.cm]{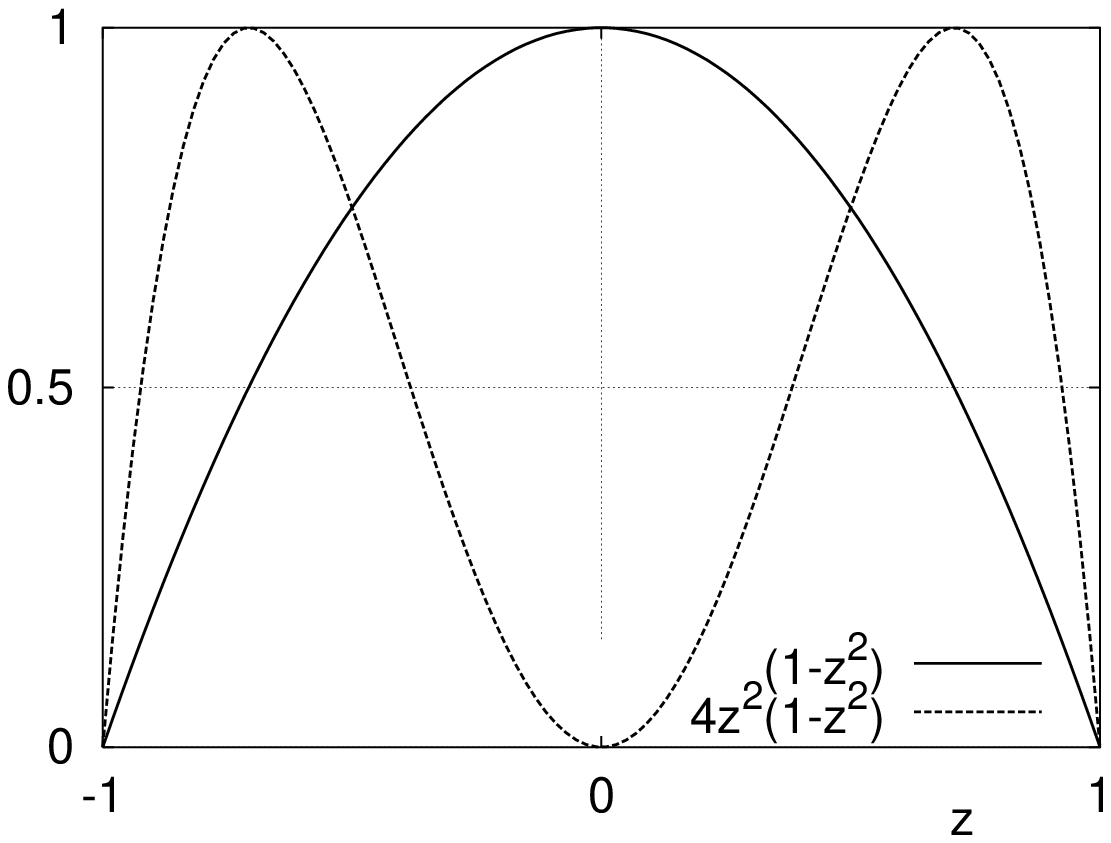}
}
\vspace*{-1.ex}
\caption{
Functions $\lf(z)=\sqrt{1-z^{2}}$ and $\lfb(z)=2z\sqrt{1-z^{2}}$,
sketching the dependence of the ladder factors $\lf_{\m}$ and
$\lfb_{\m}$ on $z\sim\m/S$.
Right: squared factors $\lf^{2}(z)=(1-z^{2})$ and
$\lfb^{2}(z)=(2z)^{2}(1-z^{2})$, as they enter in the relaxation term.
}
\label{fig:lf}
\end{figure}

We introduce several alternative notations convenient in different
contexts.
The 2-index form is
$\lf_{\m,\m'}=[S(S+1)-\m\,\m']^{1/2}$,
whence
%
\begin{equation}
\label{ladder:factor:alt}
\lf_{\m}^{\pm}
=
\lf_{\m,\m\pm1}
=
\sqrt{S(S+1)-\m(\m\pm1)}
=
\lf_{\m\pm1,\m}
\;.
\end{equation}
This gives an explicit index connection in the ladder action
$\Scpm\mket=\lf_{\m,\m\pm1}|\m\pm1\rangle$
as well as in
$\Scpm=\sum_{\m}\lf_{\m,\m\pm1}X_{\m\pm1}^{\m}$.
%
%
Both $\lf_{\m,\m\pm1}$ can be expressed in terms of a single ladder
factor $\lf_{\m}$
%
\begin{equation}
\label{ladder:factor}
\fl
\lf_{\m,\m+1}
=
\lf_{\m}
\;,
\quad
\lf_{\m,\m-1}
=
\lf_{\m-1}
\qquad
\lf_{\m}^{2}
=
S(S+1)-\m(\m+1)
\;.
\end{equation}
At $\m$ and $-(\m+1)$ the $\lf_{\m}$ take equal values:
$\lf_{\m}=\lf_{-\m-1}$.
Thus
$\lf_{S}=\lf_{-S-1}=0$ (end points),
$\lf_{S-1}=\lf_{-S}=\sqrt{2S}$,
$\lf_{S-2}=\lf_{-S+1}=\sqrt{2(2S-1)}$,
\dots,
$\lf_{S-k}=\lf_{-(S+1)+k}=\sqrt{k(2S+1-k)}$.
%
%
Identifying $\m/S\sim z=\cos\vartheta$, we have the behaviour
$\lf_{\m}^{2}\sim(1-z^{2})$, which is the factor accounting for the
reduction of the configuration space as the poles are approached
(Fig.~\ref{fig:lf}).


\paragraph{
The ``bar'' factors $\lfb_{\m}$.
}

For the spin-bath coupling $F\sim\{\Sz,\,\Scpm\}$ we come across some
modulated ladder factors in the master equation:
$\lfb_{\m,\m\pm1}=(2\m\pm1)\lf_{\m,\m\pm1}$.
Using $2\m\pm1=\m+(\m\pm1)$ we can write in a symmetric way
%
\begin{equation}
\label{bar:ladder:factor:uni}
\lfb_{\m,\m'}
=
\lfb_{\m',\m}
=
(\m+\m')
\lf_{\m,\m'}
\;.
\end{equation}
Again we can introduce a compact single factor $\lfb_{\m}$:
%
\begin{equation}
\label{bar:ladder:factor}
\lfb_{\m,\m+1}
=
\lfb_{\m}
\;,
\quad
\lfb_{\m,\m-1}
=
\lfb_{\m-1}
\;,
\qquad
\lfb_{\m}
\equiv
(2\m+1)
\lf_{\m}
\;.
\end{equation}
The symmetry $\lf_{-\m}=\lf_{\m-1}$, together with
$(-2\m+1)=-(2\m-1)$, yields the corresponding (anti)symmetry of the
bar factors, namely $\lfb_{-\m}=-\lfb_{\m-1}$.
Then
$\lfb_{S}=-\lfb_{-S-1}=0$ (boundaries),
$\lfb_{S-1}=-\lfb_{-S}=(2S-1)\sqrt{2S}$,
$\lfb_{S-2}=-\lfb_{-S+1}=(2S-3)\sqrt{2(2S-1)}$,
\dots,
$\lfb_{S-k}=\lfb_{-(S+1)+k}=(2S+1-2k)\sqrt{k(2S+1-k)}$.
%

For half-integer spin, $\lfb_{\m}$ vanishes at $\m=-1/2$.
In general $\lfb_{\m}$ goes close to zero for small $\m$ (barrier top)
as it has opposite signs for positive and negative $\m$
(Fig.~\ref{fig:lf}, middle).
For $S=1/2$ all relevant $\lfb_{\m}$ vanish, reflecting that the
coupling $F\sim\{\Sz\,,\Scpm\}$ does not produce relaxation on a
$S=1/2$ spin; physically this $F$ arises from the modulation of the
anisotropy $-\K\Sz^{2}$ by the lattice vibrations \cite{garchu97}, but
$\Sz^{2}$ does not change the energy of $S=1/2$ [recall the
equilibrium result $\Mz=\half\thrm(\half\,\Bz/\kT)$].
%


\section{
The Hubbard (level-shift) operators
}
\label{app:hubbard}

In this appendix we discuss some properties of the operators
$\Xnm=\nket\mbra$ and derive their Heisenberg equations of
motion in the conservative case.
They form a complete set; if we think of a spin operator as a
$(2S+1)\times(2S+1)$ matrix, then $\Xnm$ is the matrix with
zeroes everywhere, except one $1$ at the position $(\n,\m)$
\cite[Ch.~1]{gamlev95}.
In this space of linear operators, they form an orthonormal system with
respect to the scalar product $(X,\,Y)\equiv\Tr(X^{\adj}Y)$.


\paragraph{
Properties of $\Xnm$.
}

We begin demonstrating several useful results.

\begin{itemize}

\item
{\em Expressing an operator $\opA$ in the Hubbard basis}
%
\begin{equation}
\label{A:hubbard}
\opA
=
\tsum_{\n\m}
A_{\n\m}
\Xnm
\;,
\qquad
A_{\n\m}
=
\nbra\opA\mket
\;.
\end{equation}
Proof: using twice the closure relation
$\textrm{I}=\sum_{\m}\mket\mbra$:
%
\begin{equation*}
\fl
\opA
=
\tsum_{\n}\nket\nbra
\,
\opA
\,
\tsum_{\m}\mket\mbra
=
\tsum_{\n\m}
\nbra
\opA
\mket
\,
\nket\mbra
=
\tsum_{\n\m}
A_{\n\m}
\,
\Xnm
\,.
\end{equation*}

\item
{\em Equal-time relation}
%
\begin{equation}
\label{hubbard:ETR}
X_{\n}^{k}X_{l}^{\m}
=
\delta_{kl}\Xnm
\;.
\end{equation}
Proof: using the $\mket$ basis orthonormality,
$X_{\n}^{k}X_{l}^{\m}
=
\nket
\langle k|l\rangle
\mbra
=
\delta_{kl}\Xnm$. 

\item
{\em Commutator}
%
\begin{equation}
\label{hubbard:comm}
\big[
X_{\n}^{k}
\,,
X_{l}^{\m}
\big]
=
\delta_{kl}\Xnm
-
\delta_{\n\m}X_{l}^{k}
\;.
\end{equation}
Proof: 
$[X_{\n}^{k}
\,,
X_{l}^{\m}]
=
X_{\n}^{k}X_{l}^{\m}
-
X_{l}^{\m}X_{\n}^{k}$
plus the equal-time relation~(\ref{hubbard:ETR}).

\item
{\em Adjoint}
(recall the zeroes plus $1$ at $(\n,\m)$ representation of
$\Xnm$)
%
\begin{equation}
\label{hubbard:adj}
\big(\Xnm\big)^{\adj}
=
X_{\m}^{\n}
\;.
\end{equation}
Proof: we use the auxiliary notation $\langle\psi,\phi\rangle$ for the
scalar product:
%
\begin{equation*}
\fl
\big\langle(\Xnm)^{\adj}\psi,\phi\big\rangle
=
\big\langle\psi,\Xnm\phi\big\rangle
=
\langle\psi\nket
\mbra\phi\rangle
=
\big(
\langle\phi\mket
\nbra\psi\rangle
\big)^{\ast}
=
\big\langle\phi,X_{\m}^{\n}\psi\big\rangle^{\ast}
=
\big\langle X_{\m}^{\n}\psi,\phi\big\rangle
\;.
\end{equation*}
The validity of this result $\forall~\psi~\&~\phi$ gives
Eq.~(\ref{hubbard:adj}).
%

\item
{\em Relation with the density matrix}
(note the index ordering)
%
\begin{equation}
\label{dm:hubbard:avg}
\llangle\Xnm\rrangle
=
\dm_{\m\n}
\;.
\end{equation}
Proof: use the level-shift property
$\Xnm|k\rangle
=
\delta_{\m k}\nket$
and
$\llangle A\rrangle\equiv\Tr(\dm\,\opA)$:
%
%
\begin{equation*}
\Tr\big(\dm\,\Xnm\big)
=
\tsum_{k}
\langle k| 
\dm\,
\Xnm
|k\rangle
=
\tsum_{k}
\langle k|\dm\nket
\delta_{\m k}
=
\dm_{\m\n}
\;.
\end{equation*}
As corollaries, replacing $\dm\to\textrm{I}$ one gets the {\em trace
formula\/} $\Tr\big(\Xnm\big)=\delta_{\n\m}$.
Then one can prove the orthonormality of the Hubbard basis:
$(\Xnm,\,\Xkl)
\equiv
\Tr[(\Xnm)^{\adj}\Xkl]
=
\Tr(X_{\m}^{\n}\Xkl)
=
\Tr(\delta_{\n k}X_{\m}^{l})
=
\delta_{\n k}
\delta_{\m l}$.

\end{itemize}


\paragraph{
Heisenberg equation of motion.
}
\label{hubbard:EOM}

As the Hubbard operators $\Xnm$ do not depend explicitly on $t$, their
Heisenberg dynamical equation is simply
$\iu
\dot{X}_{\n}^{\m}
=
[\Xnm\,,\Hs]$.
Here we derive it explicitly for a Hamiltonian comprising a diagonal
part $\Hs_{\rm d}(\Sz)$ (in the standard basis)
%
%
plus a transverse field:
$\Hs
=
\Hs_{\rm d}(\Sz)
-
\half
(\Bcp\Scm+\Bcm\Scp)$.

Let us commute the different parts of $\Hs$ with $\Xnm$.
For $\Hs_{\rm d}$ we can write
$\Hs_{\rm d}=\sum_{k}\el_{k}X_{k}^{k}$,
since its matrix elements are
$\nbra\Hs_{\rm d}\mket=\el_{\m}\delta_{\n\m}$
[see Eq.~(\ref{A:hubbard})].
Then, we work $[\Xnm\,,\Hs_{\rm d}]$ out with help from
Eq.~(\ref{hubbard:comm}):
%
\begin{equation*}
\fl
\big[
\Xnm
\,,
\Hs_{\rm d}
\big]
=
\tsum_{k}
\el_{k}
\big[
\Xnm
\,,
X_{k}^{k}
\big]
=
\tsum_{k}
\el_{k}
\big(
\delta_{\m k}X_{\n}^{k}
-
\delta_{k\n}X_{k}^{\m}
\big)
=
-
\tf_{\n\m}
\Xnm
\;,
\end{equation*}
where $\tf_{\n\m}=\el_{\n}-\el_{\m}$.
For the transverse components we employ the representation
$\Scpm
=
\sum_{k}
\lf_{k,k\pm1}
X_{k\pm1}^{k}$
[Eq.~(\ref{S:hubbard})] and again commutator~(\ref{hubbard:comm}):
%
\begin{eqnarray*}
\fl
\big[
\Xnm
\,,
\Scpm
\big]
&=
\tsum_{k}
\lf_{k,k\pm1}
\big[
\Xnm
\,,
X_{k\pm1}^{k}
\big]
=
\tsum_{k}
\lf_{k,k\pm1}
\big(
\delta_{\m,k\pm1}X_{\n}^{k}
-
\delta_{\n,k}X_{k\pm1}^{\m}
\big)
\\
\fl
&=
\lf_{\m\mp1,\m}
X_{\n}^{\m\mp1}
-
\lf_{\n,\n\pm1}
X_{\n\pm1}^{\m}
=
\lf_{\m}^{\mp}
X_{\n}^{\m\mp1}
-
\lf_{\n}^{\pm}
X_{\n\pm1}^{\m}
\;,
\end{eqnarray*}
where we have used
$\lf_{\m,\m'}=\lf_{\m',\m}$
and
$\lf_{\m,\m\pm1}=\lf_{\m}^{\pm}$ (\ref{app:ladder}).
%
%
Finally gathering these partial results, including $-\Bcpm/2$, and
multiplying across by $-\iu$
%
\begin{equation*}
\dot{X}_{\n}^{\m}
%
=
-
\iu
\big[
\Xnm
\,,
\Hs_{\rm d}
\big]
+
(\iu/2)
\Bcp
\big[
\Xnm
\,,
\Scm
\big]
+
(\iu/2)
\Bcm
\big[
\Xnm
\,,
\Scp
\big]
\;,
\end{equation*}
we arrive at the unitary evolution equation~(\ref{DME:closed}) for the
Hubbard operators.

If we had used the exact eigenstates of the total Hamiltonian, the
term $\iu\tf_{\n\m}\Xnm$ would have been the only term in the equation
(with $\Bcpm$ entering via the $\el_{\m}$).
Using the angular-momentum basis instead (for example, when the exact
eigenstructure is not known or, for convenience, to treat
time-dependent $\Bcpm$), one needs to add explicitly the contribution
of the transverse terms, as we have done here.


\section{
The relaxation term
}
\label{app:Rmarkov}

In contact with the bath the closed equation of
motion~(\ref{DME:closed}) is augmented by the relaxation
term~(\ref{Rnonmarkov}).
Here starting from such $R_{\n}^{\m}$ we derive the time-local
relaxation term~(\ref{Rmarkov}) and then we particularise it to
spin-bath couplings via $\Scpm$.
Finally we discuss the adoption of the secular or rotating-wave
approximation and the issue of the positivity of the reduced
description.


\subsection{
Derivation of the time-local relaxation term~(\ref{Rmarkov})
}
\label{app:Rmarkov:gen}

We start from
$-R_{\n}^{\m}
=
\int_{-\infty}^{t}
\!\drm{\tp}
\{
\Ker(\tp-t)
\,
F(\tp)
[F\,,\Xnm]
-
\Ker(t-\tp)
\,
[F\,,\Xnm]
F(\tp)
\}$,
%
%
where operators without argument are evaluated at $t$.
The formal dependence on the previous evolution enters through
$F(\tp)$.
We expand it on the Hubbard basis, $F(\tp)=\sum_{kl}F_{kl}\Xkl(\tp)$,
and replace the retarded dependences by the (dominant) $\Sz$-part of
the conservative evolution
$\Xkl(\tp)=\e^{-\iu(t-\tp)\tf_{kl}}\Xkl(t)$,
whence
$F(\tp)=\sum_{kl}F_{kl}\e^{-\iu(t-\tp)\tf_{kl}}\Xkl(t)$.
Next, the change of variable $s=t-\tp$ brings the integral relaxation
term into the form
%
\begin{equation*}
\fl
-R_{\n}^{\m}
=
\tsum_{kl}
F_{kl}
\int_{0}^{\infty}\!\drm{s}
\Big\{
\Ker(-s)
\,
\e^{-\iu s\tf_{kl}}
\,
\Xkl
\big[
F
\,,
\Xnm
\big]
-
\Ker(s)
\,
\e^{-\iu s\tf_{kl}}
\,
\big[
F
\,,
\Xnm
\big]
\Xkl
\Big\}
\;.
\end{equation*}
The $s$-dependences only occur in the kernel $\Ker$ and the
oscillating factors.
On integrating them brings into scene the relaxation rates $\Wu_{k\tm
l}=\Wu(\tf_{kl})$, with
$\Wu(\tf)
=
\int_{0}^{\infty}\!\drm{s}\,
\e^{-\iu s\tf}
\Ker(s)$
[Eq.~(\ref{W:def})].
Then, using $\Ker(-\tp)=\big[\Ker(\tp)\big]^{\ast}$ and
$\tf_{lk}=-\tf_{kl}$, gives
%
\begin{equation}
\label{Rnonmarkov:Fkl}
-R_{\n}^{\m}
=
\tsum_{kl}
F_{kl}
\Big\{
\Wu_{l\tm k}^{\ast}
\;
\Xkl
\big[
F
\,,
\Xnm
\big]
-
\Wu_{k\tm l}
\;
\big[
F
\,,
\Xnm
\big]
\Xkl
\Big\}
\;.
\end{equation}
This completes the task of getting a time-local relaxation term.
Note however that $R_{\n}^{\m}$ still depends on the $X$ in a
non-explicit way (since $F$ also contains them).%
\setcounter{footnote}{1} 
\footnote{
%
%
It is assumed that at the initial time, $\tini=-\infty$, system and
bath were decoupled.
If this is assumed to occur at some finite $\tini$, the rates adquire
a time dependence
$\Wu(\tf;t)
=
\int_{0}^{t-\tini}\!\drm{s}\,
\e^{-\iu s\tf}
\Ker(s)$.
However, as in most problems system and bath had coexisted for a very
long time, one shifts $\tini\to-\infty$ and the transients due to such
decoupled initial conditions are washed away.
} 

The rest of the calculation consists of simplifying the $X$ structure
of the above term.
First, we expand the inner $F$ as
$F=\sum_{\ik\il}F_{\ik\il}X_{\ik}^{\il}$
and use
$[X_{\ik}^{\il}\,,\Xnm]
=
\delta_{\n\il}X_{\ik}^{\m}-\delta_{\m\ik}X_{\n}^{\il}$
to work the commutators.
Then we multiply by $\Xkl$ (from the left and right) and use the
equal-time relation $\Xkl\Xnm=\delta_{l\n}X_{k}^{\m}$ to get
expressions linear in the Hubbards.
Eventually, the result of the left and right multiplications is in
turn multiplied by $\Wu_{l\tm k}^{\ast}F_{kl}$ and $\Wu_{k\tm
l}F_{kl}$ respectively [Eq.~(\ref{Rnonmarkov:Fkl})] and summed over
$k$ \& $l$:
\begin{eqnarray*}
\fl
\tsum_{kl}
\Wu_{l\tm k}^{\ast}
F_{kl}
\,
\Xkl
\,
\big[
F\,,
\Xnm
\big]
&=
\tsum_{kl}
\Wu_{l\tm k}^{\ast}
F_{kl}
F_{l\n}
\,
X_{k}^{\m}
-
\tsum_{k\il}
\Wu_{\n\tm k}^{\ast}
F_{k\n}
F_{\m\il}
\,
X_{k}^{\il}
\\
\fl
\tsum_{kl}
\Wu_{k\tm l}
F_{kl}
\big[
F\,,
\Xnm
\big]
\,
\Xkl
&=
\tsum_{l\ik}
\Wu_{\m\tm l}
F_{\m l}
F_{\ik\n}
\,
X_{\ik}^{l}
-
\tsum_{kl}
\Wu_{k\tm l}
F_{kl}
F_{\m k}
\,
X_{\n}^{l}
\;.
\end{eqnarray*}
To clarify the structure and to arrive at the sought form
$\sum_{\n'\m'}
\R_{\n,\n'}^{\m,\m'}
X_{\n'}^{\m'}$
we make several index changes.
First line, first term:
$l\to\is$ (summed index),
$k\to\n'$,
and we introduce
$\sum_{\m'}\delta_{\m\m'}$.
First line, second term:
$\il\to\m'$
and
$k\to\n'$.
Second line, first term:
$\ik\to\n'$
and
$l\to\m'$.
Second line, second term:
$k\to\is$ (summed index),
$l\to\m'$,
and introduce
$\sum_{\n'}\delta_{\n\n'}$.
Then we obtain for the above right-hand sides
\begin{eqnarray*}
& &
\tsum_{\n'\m'}
\big[
\delta_{\m\m'}
\big(
\tsum_{\is}
\Wu_{\is\tm\n'}^{\ast}
F_{\n'\is}
F_{\is\n}
\big)
-
\Wu_{\n\tm\n'}^{\ast}
F_{\n'\n}
F_{\m\m'}
\,
\big]
X_{\n'}^{\m'}
\\
& &
\tsum_{\n'\m'}
\big[
\Wu_{\m\tm\m'}
F_{\m\m'}
F_{\n'\n}
-
\delta_{\n\n'}
\big(
\tsum_{\is}
\Wu_{\is\tm\m'}
F_{\is\m'}
F_{\m\is}
\big)
\big]
\,
X_{\n'}^{\m'}
\;.
\end{eqnarray*}
Substraction of these lines according to Eq.~(\ref{Rnonmarkov:Fkl})
gives finally the relaxation term~(\ref{Rmarkov}).


\subsection{
Relaxation term for couplings via $\Scpm$
}
\label{app:Rmarkov:lin}

Here we particularise the relaxation term~(\ref{Rmarkov}) to the
coupling~(\ref{F:lin}), with $F(\vS)$ ``linear'' in $\Scpm$.
We start expanding the anticommutators and introducing
$F_{\pm}
=
\eta_{\mp}
[\Vz(\Sz)\,\Scpm
+
\Scpm\,\Vz(\Sz)]$,
so that $F=F_{-}+F_{+}$.
To obtain the matrix elements $F_{\n\m}=\nbra F\mket$ we compute first
those of $F_{-}$, using $\Scm\mket=\lf_{\m,\m-1}|\m-1\rangle$:
%
\begin{equation}
\label{lfb:gen}
\fl
(F_{-})_{\n\m}
=
\LF_{\m,\m-1}
\,
\delta_{\n,\m-1}
\;,
\qquad
\LF_{\m,\m'}
=
\eta_{+}
[\Vz(\m)+\Vz(\m')]
\lf_{\m,\m'}
\;.
\end{equation}
Note that the extended ladder factor $\LF_{\m,\m'}$ is in general
complex.
Then by means of $F_{+}=(F_{-})^{\adj}$ we arrive at
$F_{\n\m}
=
\LF_{\m,\m-1}\delta_{\n,\m-1}
+
\LF_{\m+1,\m}^{\ast}\delta_{\n,\m+1}$
[i.e., Eq.~(\ref{Fnm})].

Let us proceed to do the sums in the relaxation term~(\ref{Rmarkov})
with these matrix elements and some care.
For the {\em third line\/} we need
$\sum_{\is\m'}
\Wu_{\is\tm\m'}
F_{\m\is}
F_{\is\m'}
\,
X_{\n}^{\m'}$.
The sum over $\m'$ gives
\begin{eqnarray*}
\tsum_{\m'}
\Wu_{\is\tm\m'}
F_{\is\m'}
\,
X_{\n}^{\m'}
&=&
\LF_{\is+1,\is}
\Wu_{\is\tm\is+1}
\,
X_{\n}^{\is+1}
+
\LF_{\is,\is-1}^{\ast}
\Wu_{\is\tm\is-1}
\,
X_{\n}^{\is-1}
\;.
\end{eqnarray*}
Multiplying by
$F_{\m\is}
=
\LF_{\m+1,\m}\delta_{\is,\m+1}+\LF_{\m,\m-1}^{\ast}\delta_{\is,\m-1}$
and summing over $\is$ we obtain
%
\begin{eqnarray*}
\mathbf{3rd~line~}
&
=
\LF_{\m,\m-1}^{\ast}
\LF_{\m-1,\m-2}^{\ast}
\,
\Wu_{\m-1\tm\m-2}
&
X_{\n}^{\m-2}
\\
&
+
\big(
\;\;
|\LF_{\m,\m-1}|^{2}
\;
\Wu_{\m-1\tm\m}
+
|\LF_{\m+1,\m}|^{2}
\;
\Wu_{\m+1\tm\m}
\;\;
\big)
&
\Xnm
\\
&
+
\LF_{\m+2,\m+1}
\LF_{\m+1,\m}
\,
\Wu_{\m+1\tm\m+2}
&
X_{\n}^{\m+2}
\;.
\end{eqnarray*}
The perturbative paths are clear, e.g., $\m+2\to\m+1\to\m$ to connect
$X_{\n}^{\m+2}$ with $\Xnm$.
Inspection of Eq.~(\ref{Rmarkov}) reveals that the {\em first line\/}
follows from the third by exchanging $\m\leftrightarrow\n$,
conjugating, and permuting upper and lower indices in $X$
(adjointing):
%
\begin{eqnarray*}
\mathbf{1st~line~}
&
=
\LF_{\n,\n-1}
\LF_{\n-1,\n-2}
\,
\Wu_{\n-1\tm\n-2}^{\ast}
&
X_{\n-2}^{\m}
\\
&
+
\big(
\;\;
|\LF_{\n,\n-1}|^{2}
\;
\Wu_{\n-1\tm\n}^{\ast}
+
|\LF_{\n+1,\n}|^{2}
\;
\Wu_{\n+1\tm\n}^{\ast}
\;\;
\big)
&
\Xnm
\\
&
+
\LF_{\n+2,\n+1}^{\ast}
\LF_{\n+1,\n}^{\ast}
\,
\Wu_{\n+1\tm\n+2}^{\ast}
&
X_{\n+2}^{\m}
\;.
\end{eqnarray*}
Now we are left with
$\sum_{\n'\m'} (\Wu_{\n\tm\n'}^{\ast}+\Wu_{\m\tm\m'})
F_{\n'\n}F_{\m\m'} \,X_{\n'}^{\m'}$,
the central line of Eq.~(\ref{Rmarkov}).
Exclude $F_{\n'\n}$ and do first the sum over $\m'$, then multiply by
$F_{\n'\n}$ written in the form
$F_{\n'\n}
=
\LF_{\n,\n-1}\delta_{\n',\n-1}+\LF_{\n+1,\n}^{\ast}\delta_{\n',\n+1}$,
sum over $\n'$, and reverse signs; you should get
%
\begin{eqnarray*}
\mathbf{2nd~line~}
=
&
-
\LF_{\n,\n-1}
\LF_{\m,\m-1}^{\ast}
\,
(\Wu_{\n\tm\n-1}^{\ast}+\Wu_{\m\tm\m-1})
&
X_{\n-1}^{\m-1}
\\
&
-
\LF_{\n+1,\n}^{\ast}
\LF_{\m,\m-1}^{\ast}
\,
(\Wu_{\n\tm\n+1}^{\ast}+\Wu_{\m\tm\m-1})
&
X_{\n+1}^{\m-1}
\\
&
-
\LF_{\n,\n-1}
\LF_{\m+1,\m}
\,
(\Wu_{\n\tm\n-1}^{\ast}+\Wu_{\m\tm\m+1})
&
X_{\n-1}^{\m+1}
\\
&
-
\LF_{\n+1,\n}^{\ast}
\LF_{\m+1,\m}
\,
(\Wu_{\n\tm\n+1}^{\ast}+\Wu_{\m\tm\m+1})
&
X_{\n+1}^{\m+1}
\;.
\end{eqnarray*}
Collecting the three contributions we finally obtain the
specialisation of the Markovian relaxation term~(\ref{Rmarkov}) to the
coupling~(\ref{F:lin}):
%
\begin{eqnarray}
\label{Rmarkov:lin}
\fl
R_{\n}^{\m}
&=&
\LF_{\n,\n-1}
\LF_{\m,\m-1}^{\ast}
\;
(\Wu_{\n\tm\n-1}^{\ast}+\Wu_{\m\tm\m-1})
\quad
X_{\n-1}^{\m-1}
\nonumber\\*
\fl
&-&
\big(\;\;\;
|\LF_{\n+1,\n}|^{2}
\;
\Wu_{\n+1\tm\n}^{\ast}
+
|\LF_{\m+1,\m}|^{2}
\;
\Wu_{\m+1\tm\m}
\nonumber\\*
\fl
& &
\;\;
+
|\LF_{\n,\n-1}|^{2}
\;
\Wu_{\n-1\tm\n}^{\ast}
+
|\LF_{\m,\m-1}|^{2}
\;
\Wu_{\m-1\tm\m}
\;\;
\big)
\quad
\Xnm
\\*
\fl
&+&
\LF_{\n+1,\n}^{\ast}
\LF_{\m+1,\m}
\;
(\Wu_{\n\tm\n+1}^{\ast}+\Wu_{\m\tm\m+1})
\quad
X_{\n+1}^{\m+1}
\nonumber\\*[1.ex]
\fl
&-&
\LF_{\n,\n-1}
\LF_{\n-1,\n-2}
\,
\Wu_{\n-1\tm\n-2}^{\ast}
\;
X_{\n-2}^{\m}
-
\LF_{\n+2,\n+1}^{\ast}
\LF_{\n+1,\n}^{\ast}
\,
\Wu_{\n+1\tm\n+2}^{\ast}
\;
X_{\n+2}^{\m}
\qquad \leftarrow
\nonumber\\*
\fl
&+&
\LF_{\n,\n-1}
\LF_{\m+1,\m}
\,
(\Wu_{\n\tm\n-1}^{\ast}+\Wu_{\m\tm\m+1})
\quad
X_{\n-1}^{\m+1}
\nonumber\\*
\fl
&+&
\LF_{\n+1,\n}^{\ast}
\LF_{\m,\m-1}^{\ast}
\,
(\Wu_{\n\tm\n+1}^{\ast}+\Wu_{\m\tm\m-1})
\quad
X_{\n+1}^{\m-1}
\nonumber\\*
\fl
&-&
\LF_{\m,\m-1}^{\ast}
\LF_{\m-1,\m-2}^{\ast}
\,
\Wu_{\m-1\tm\m-2}
\;
X_{\n}^{\m-2}
-
\LF_{\m+2,\m+1}
\LF_{\m+1,\m}
\,
\Wu_{\m+1\tm\m+2}
\;
X_{\n}^{\m+2}
\nonumber
\;.
\end{eqnarray}
On invoking the secular approximation (see below) terms involving
$\LF\times\LF$ or $\LF^{\ast}\times\LF^{\ast}$ are dropped (last four
lines) and only the terms of the type $\LF\times\LF^{\ast}$ are
retained (first four).
This gives the secularized relaxation term~(\ref{Rmarkov:lin:invk})
in the main text.


\subsection{
The non-secular terms and the positivity issue
}
\label{app:Rmarkov:sec}

Neglecting the ``non-secular'' terms $\LF\times\LF$ and
$\LF^{\ast}\times\LF^{\ast}$ corresponds to the rotating-wave
approximation of quantum optics, where counter-rotating terms are
dropped.
Specifically, writting schematically the coupling as
$\Ham_{\mathrm{sb}}
\sim
\Scm(\ap+\epsilon\,\am)+\Scp(\am+\epsilon\,\ap)$,
one can see that setting $\epsilon=0$ the last four lines of
Eq.~(\ref{Rmarkov:lin}) dissappear \cite{desher95}.
To justify this manipulation, it is argued that
$\ap\Scp\sim\e^{\iu(\omega+\tf)t}$ oscillates faster than
$\ap\Scm\sim\e^{\iu(\omega-\tf)t}$ and can be averaged out.
A word of caution though.
While this reasoning may apply to systems with monotonous spectrum
(harmonic oscillator, isotropic spin), the justification is not
satisfactory if the sign of $\tf_{\m,\m+1}$ depends on $\m$ (upon
$\tf\to-\tf$ it would be $\ap\Scm$ the term oscillating fast and
$\ap\Scp$, which is dropped, the slow one).

Keeping all terms in Eq.~(\ref{Rmarkov:lin}) does not pose big
problems for a continued-fraction solution.
Only the line marked with the arrow (fifth one) breaks the 3-term
recurrence in $\n$, giving
$\dot{\mc}_{\n}
\sim
\mQ_{\n,\n-2}\mc_{\n-2}
+
\mQ_{\n,\n-1}\mc_{\n-1}
+
\mQ_{\n,\n}\mc_{\n}
+
\mQ_{\n,\n+1}\mc_{\n+1}
+
\mQ_{\n,\n+2}\mc_{\n+2}$,
as equation of motion for $(\mc_{\n})_{\m}=\langle\Xnm\rangle$.
But this can be treated by $5\to3$ recurrence folding with block
vectors and matrices \cite{risken,garzue2004}
\begin{equation*}
%
\mathbf{C}_{\n}
=
\left(
\begin{array}{l}
\mc_{2\n}
\\
\mc_{2\n+1}
\end{array}
\right)
\qquad
\mathbf{Q}_{\n,\n'}
=
\left(
\begin{array}{ll}
\mQ_{2\n,2\n'}
&
\mQ_{2\n+1,2\n'}
\\
\mQ_{2\n,2\n'+1}
&
\mQ_{2\n+1,2\n'+1}
\end{array}
\right)
\;,
\end{equation*}
recovering the canonical form
$\dot{\mathbf{C}}_{\n}
\sim
\mathbf{Q}_{\n,\n-1}\mathbf{C}_{\n-1}
+
\mathbf{Q}_{\n,\n}\mathbf{C}_{\n}
+
\mathbf{Q}_{\n,\n+1}\mathbf{C}_{\n+1}$.
Alternatively, one can handle $\mc_{\n\pm2}$ in an iterative way
(keeping the ordinary vectors and matrices) which converges quickly
for weak damping.
Notwithstanding this, we preferred to illustrate the
continued-fraction treatment of the master equation with a simple yet
generic case.

Incidentally, the density-matrix equation without non-secular terms is
of the so-called Lindblad type \cite{spo80} for isotropic spins.
In general, however, Eq.~(\ref{Rmarkov:lin}) does not ensure that for
arbitrary initial conditions $\dm$ is a positive operator at all
times (which is required for its probabilistic interpretation).
Although this sounds alarming, no real problem exists when one recalls
the assumptions under which the master equation is derived and the
meaning of $\dm$ as partial trace of $\dm_{\rm tot}$.

First, to derive the master equation one assumes that at some initial
time $\tini$ system and bath were decoupled
$\dm_{\rm tot}(\tini)=\dm(\tini)\otimes\dm_{\rm b}(\tini)$.
The resulting evolution equation for $\dm=\Tr_{\rm b}(\dm_{\rm tot})$
is precisely Eq.~(\ref{Rmarkov:lin}) but with some time-dependent
coefficients (recall the last footnote).
But as this master equation comes from the solution of
$\iu
(\drm\dm_{\rm tot}/\drm t)
=
[\Hs\,,\dm_{\rm tot}]$,
which preserves positivity of $\dm_{\rm tot}$ (and hence of any
partial trace), then $\dm(t)>0$ should hold to the accuracy considered
in the system-bath coupling.
Actually, no positivity violation has been reported when using the
proper time-dependent coefficients.
The cases of $\dm<0$ explicitly shown \cite{blades96,mungar96}
correspond to use the master equation with the asymptotic
$t$-independent coefficients (valid at long times), to address
arbitrary initial condition problems.
However, in the asymptotic regime, system and bath have been
interacting for a long time and the possible
$\dm(t=0)=\Tr_{\rm b}(\dm_{\rm tot})$
cannot be chosen at our will.
%

Except in problems like the atom inserted in an electromagnetic cavity
or the like, decoupled initial conditions are just an mathematical
trick to facilitate the obtaining of the form of the master equation.
For problems like a spin in a solid, a Brownian particle in a fluid,
etc., it is not natural to assume that system and bath just met, but
that they were in contact for a long time.
Thus, one shifts the initial time $\tini\to-\infty$, so that at finite
times (say $t\geq0$) those factorizing initial conditions have been
forgotten (except maybe in marginal cases) and system-plus-bath have
evolved into some joint stationary state by their internal dynamics.
Then the coefficients to be used are the asymptotic ones.
But then it is natural that the initial conditions for $\dm(t=0)$
cannot be set arbitrarily, but only those compatible with the
evolution of the coupled system-plus-bath.
It is only overlooking this that the positivity becomes a issue.

Finally, in the asymptotic $t\geq0$ regime, one can manipulate the
system with fields or forces to prepare ``initial conditions'' for the
problems of interest.
This will only result in some time dependences of the master equation
coefficients.
In our case they arise when evolving $X(s)$ back in time in the memory
kernel, which now has to be done as
$X(s)=X(t)\exp[-\iu\int_{t}^{s}\!\drm\,u\,\tf(u)]$, with the proper
time-dependent Hamiltonian evolution \cite{kohdithan97}.
If the manipulation on the system is slow compared to the kernel width
(bath correlation time), one can use the leading term
$X(s)\simeq X(t)\,\e^{-\iu(s-t)\tf(t)}$.
Then one obtains precisely the ordinary master equation, with the
asymptotic form of the coefficients, but parametrically modulated by
the time-dependent fields or forces.

Preparing the system in this way, the initial conditions follow from
the master equation dynamics (and there is a single continuous process
from $\tini=-\infty$, instead of ``initial conditions'' being set
twice).
Then, to the accuracy in the coupling considered, the possitivity,
hermiticity, normalization, etc., are preserved inasmuch as
$\iu
(\drm\dm_{\rm tot}/\drm t)
=
[\Hs\,,\dm_{\rm tot}]$
underlies the reduced dynamics for $\dm=\Tr_{\rm b}(\dm_{\rm tot})$.


\section{
Transition and relaxation rates
}
\label{app:Ws}

The rate function $\Wu(\tf)$, associated to the kernel $\Ker(\tp)$,
can be expressed in terms of the bath spectral density
[Eq.~(\ref{W:J})].
This follows integrating the relation~(\ref{kernel:J}) between
$\Ker(\tp)$ and $J(\omega)$ with
$\mathrm{Re}[\int_{0}^{\infty}\!\drm\tp\,\exp(-\iu\tp\tf)]
=\pi\delta(\tf)$.
In this appendix we consider $\Wu(\tf)$ for spectral densities of the
form $J(\omega)=\Wo\,\omega^{\io}$.
We address two cases: (i) Ohmic bath $\io=1$ and (ii) the custom
$\io=3$ super-Ohmic bath (photons or phonons in three dimensions).
Note that in both cases $\io$ is odd.


\paragraph{
Unified functional form.
}

\begin{figure}[!tb]
\centerline{
\includegraphics[width=7.2cm]{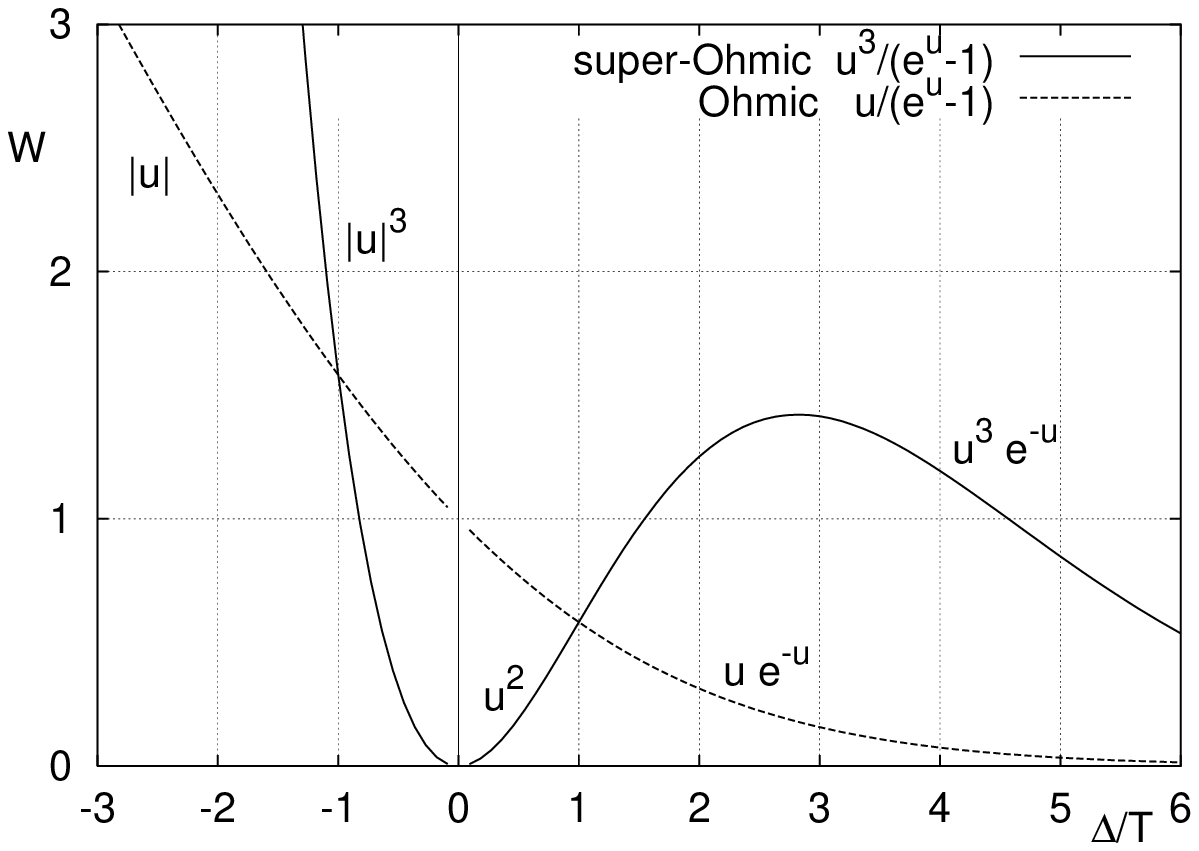}
\hspace*{-3.ex}
\includegraphics[width=7.2cm]{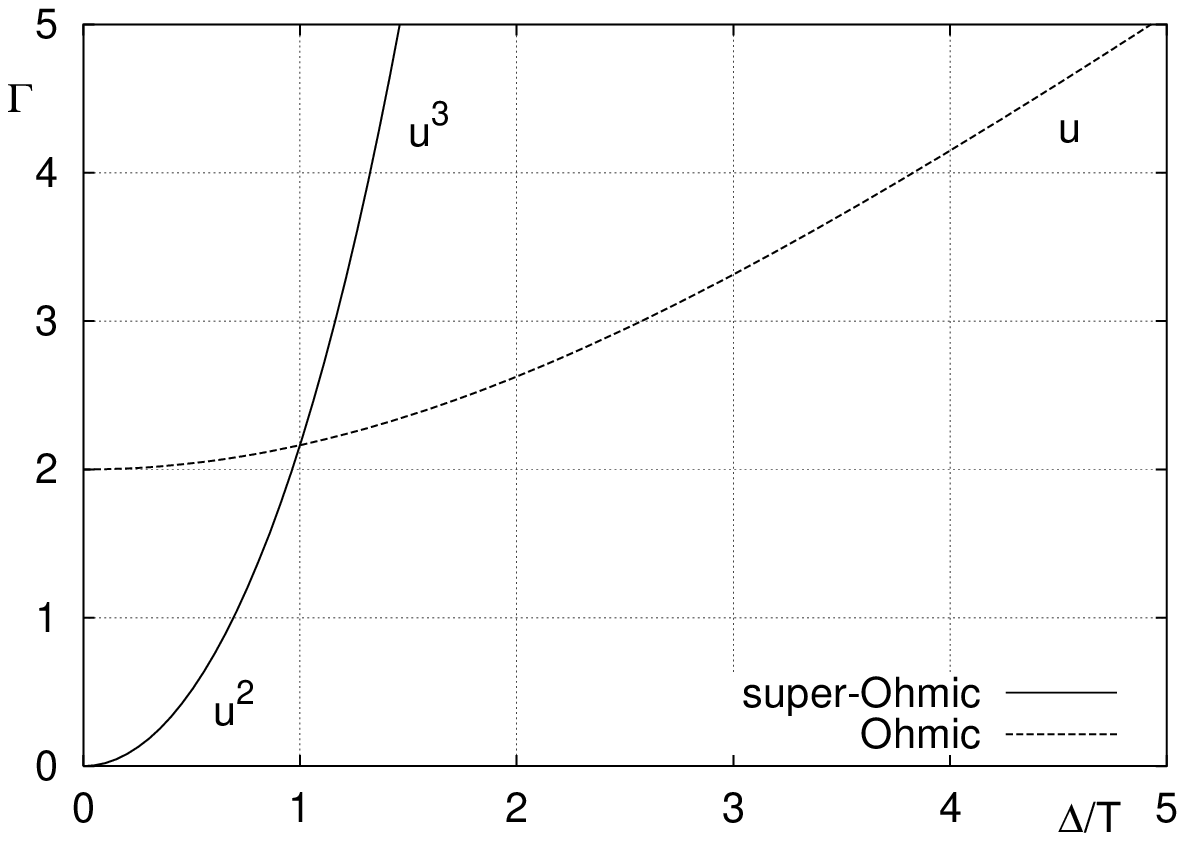}
}
\vspace*{-1.ex}
\caption{
Transition rate $\Wu(\y)$ (left panel) and relaxation rate
$\rate=\Wu(\y)+\Wu(-\y)$ (right).
Curves for both the Ohmic bath, $J(\omega)\propto\omega$ (dashed
lines), and the super-Ohmic bath $J(\omega)\propto\omega^{3}$ (solid)
are shown.
The limit functional dependences on $\y=\tf/\kT$ are marked.
Introducing a high-frequency cuf-off $\wD$ in $J(\omega)$, both
$\Wu(\y)$ and $\rate(\y)$ would eventually drop to zero for
$|\y|\gg\wD/\kT$, reflecting that no quanta of arbitrary large
$\omega$ are available to interchange energy with.
}
\label{fig:W}
\end{figure}

Introducing the reduced variable $\y=\tf/\kT$ and omitting
proportionality constants we write Eq.~(\ref{W:J}) as
$\Wu(\y>0)
=
\y^{\io}
\,
n_{\y}$
and
$\Wu(\y<0)
=
|\y|^{\io}
(n_{|\y|}+1)$
with the Bose factor $n_{\y}=1/(\e^{\y}-1)$.
Let us first demonstrate that a single functional form gives the rates
for both $\y<0$ and $\y>0$:
\begin{equation*}
\fl
\Wu(-\y)
\stackrel{\y>0}{=}
|-\y|^{\io}
\big(n_{|-\y|}+1\big)
=
\y^{\io}
\left(
\frac{1}{\e^{\y}-1}
+
1
\right)
=
\frac{\y^{\io}}{1-\e^{-\y}}
=
\frac{(-\y)^{\io}}{\e^{-\y}-1}
\;.
\end{equation*}
Note that we have used the oddness of the spectral index $\io$ to
absorb the sign change of the denominator (otherwise a
$\mathrm{sgn}(\y)$ appears).
Thus $\Wu(-\y)$ has the same functional dependence as the positive
$\y$ case, i.e., $(-\y)^{\io}\,n_{-\y}$, and we can write
%
\begin{equation}
\label{W:unified}
J(\omega)\propto\omega^{\io}
\;\;
({\rm odd}~\io)
\quad
\leadsto
\quad
\Wu(\y)
=
\frac{\y^{\io}}{\e^{\y}-1}
\;,
\quad
\forall
\y
\;.
\end{equation}
Figure~\ref{fig:W} displays the dependences of $\Wu$ on $\y=\tf/\kT$
for $\io=1$ and $3$.
For large and negative energy differences (emission of quanta) the
behaviour is $|\y|^{\io}$.
In the Ohmic case, the rate decreases monotonically from $\Wu(0)=1$
for positive energy differences (absorption).
The super-Ohmic $\Wu(\y)$, in contrast, after an initial parabolic
take-off from zero, has a maximum at $\y\sim2\sqrt{2}$ and decreases
exponentially dominated for large energy absorption.


\paragraph{
Derivatives of $\Wu(\y)$.
}

These are needed in the perturbative calculations
(Sec.~\ref{sec:forcing}).
Having a unified functional form [Eq.~(\ref{W:unified})] we can
differentiate without worrying about distinguishing between absorption
and emission processes:
\begin{equation}
\label{W:der}
\Wu'(\y)
=
\frac{\io\,\y^{\io-1}}{\e^{\y}-1}
+
\frac{\y^{\io}}{(\e^{\y}-1)(\e^{-\y}-1)}
%
%
\;.
\end{equation}
%
In the perturbative treatment of the probing field we actually
need the $\Bz$-derivatives of $\Wu_{\m\tm\m\pm1}$.
Applying the chain rule $\drm\Wu/\drm\Bz=\Wu'(\y)\,\drm\y/\drm\Bz$ we
obtain a $\pm$ sign depending on the transition ``lowering'' or
``raising'' the second index [see Eq.~(\ref{level:spacings})]:
\begin{equation*}
\fl
\Wu_{\m\tm\m\pm1}
\quad
\sim
\quad
\tf_{\m,\m\pm1}
=
\pm
\left[
\K(2\m\pm1)+\Bz
\right]
\quad
\leadsto
\quad
\partial_{\Bz}\tf_{\m,\m\pm1}
=
\pm
1
\;.
\end{equation*}
Finally, one must recall to multiply Eq.~(\ref{W:der}) also by
$1/\kT$ since $\y=\tf/\kT$.


\paragraph{
Relaxation rate.
}

The combination $\rate(\y)=\Wu(\y)+\Wu(-\y)$ appears is several
problems (isotropic $S=1/2$ and $S=1$ spins, Sec.~\ref{Siso:formulae},
anisotropic $S=1$, Sec.~\ref{sec:Sani:S1}).
Using Eq.~(\ref{W:unified}) this symmetrized rate can be written as
%
\begin{equation}
\label{rate}
\rate
\equiv
\Wu(\y)
+
\Wu(-\y)
=
\y^{\io}\,\cth(\half\y)
\;.
\end{equation}
This function behaves as $2\y^{\io-1}$ for small $\y$, then grows
monotonically and for large energy differences goes as $\y^{\io}$
(Fig.~\ref{fig:W}, right).
The monotony is proved differentiating Eq.~(\ref{rate}), whence
$\rate'=\half\y^{\io-1}(\io\sh\y-\y)/\sh^{2}(\half\y)>0$, for $\io\geq1$.


\section{Integral relaxation time}
\label{app:tauint}

The advantage of this quantity is that it can be obtained from the low
frequency response
$\chi(\w)\simeq\chi(1-\iu\w\tint+\ldots)$.
This result follows from the definition
$\tint\equiv\int_{0}^{\infty}\!\drm t\,\delta M(t)/\delta M(0)$
plus a low $\w$ expansion in the linear-response-theory relation
$\chi(\w)
=
\chi
\big(
1
-
\iu\,\w
\int_{0}^{\infty}
\!
\drm t\,
\e^{-\iu\,\w t}
[\delta M (t)/\delta M(0)]
\big)$.
Here we discuss a generalisation of the results for $\tint$ of
Refs.~\cite{gar91llb,garchu97,gar97} by considering a generic discrete
system obeying balance equations of the form \cite{zuegar2006}:
%
\begin{eqnarray}
\label{DME:pauli:Wt}
\fl
\dot{N}_{\m}
&=&
\left(
\Wt_{\m\tm\m-1}
N_{\m-1}
-
\Wt_{\m-1\tm\m}
N_{\m}
\right)
+
\left(
\Wt_{\m\tm\m+1}
N_{\m+1}
-
\Wt_{\m+1\tm\m}
N_{\m}
\right)
\;.
\end{eqnarray}
$\Wt_{\m\tm\m'}$ is the $\m'\to\m$ transition probability,
assumed to depend on the level spacings
$\Wt_{\m\tm\m'}=\Wt(\tf_{\m\m'})$ and to obey detailed balance
$\Wt_{\m\tm\m'}=\e^{-\tf_{\m\m'}/\kT}\Wt_{\m'\tm\m}$.
Besides the level separation is controlled by some bias $F$ as
$\tf=\tf^{(0)}+F$.

Next one augments $F$ by an oscillating probe
$F\to F+\half f\,\e^{\iu\w t}+\cc$
and seeks for a solution of Eq.~(\ref{DME:pauli:Wt}) of the form
$N_{\m}=N_{\m}^{(0)}+\half f\,(N_{\m}^{(1)}\,\e^{\iu\w t}+\cc)$.
The corresponding susceptibility is $\chi(\w)\sim\sum_{\m}\m
N_{\m}^{(1)}(\w)$.
At low frequencies the equations for $N_{\m}^{(1)}$ can be solved
analytically and one gets $\chi(\w\to0)$, whence $\tint$ follows as
\cite[App.~B]{zuegar2006}
\begin{equation}
\label{tauint:generic}
\tint
=
\frac{1}{M'}
\sum_{\m}\,
\frac{\Phi_{\m}^{2}}{N_{\m}^{(0)}\Wt_{\m+1\tm\m}^{(0)}}
\;,
\qquad
\Phi_{\m}
=
\!\!\!
\sum_{k=\m_{\mathrm{i}}}^{\m}
\!\!\!
N_{k}^{(0)}
\big(
M-k
\big)
\;.
\end{equation}
Here $M'=\drm M/\drm y$, with $y=F/\kT$ and
$M=\sum_{\m}\m\,N_{\m}^{(0)}$ is the static response.
The range of $\m$ is $[\m_{\mathrm{i}},\,\m_{\mathrm{f}}]$ and the
superscript $(0)$ indicates absence of probing field.
Note that the auxiliary function $\Phi_{\m}$, in contrast to
$\Wt_{\m\tm\m'}$, depends only on equilibrium properties.
%

To get $\tint$ for isotropic spins, we simply cast the corresponding
balance equations (Sec.~\ref{sec:DME:Siso}) into the
form~(\ref{DME:pauli:Wt}), by identifying
$\Wt_{\m\tm\m'}
=
\lf_{\m,\m'}^{2}
\Wu_{\m\tm\m'}$.
To do so we have taken advantage of the 2-index
form~(\ref{ladder:factor:alt}) of the ladder coefficients
($\lf_{\m-1}=\lf_{\m,\m-1}=\lf_{\m-1,\m}$, and
$\lf_{\m}=\lf_{\m,\m+1}=\lf_{\m+1,\m}$), and recalled the simplyfied
relations $\Wu_{\m\tm\m-1}=\Wiso$ and $\Wu_{\m-1\tm\m}=\Wiso\,\e^{-y}$
(Sec.~\ref{sec:DME:Siso}).
Thus $\tint$ in this case is given by
$\tint
=
(1/\Mz')
\sum_{\m=-S}^{S}\,
\big(\Phi_{\m}^{2}\big/N_{\m}^{(0)}\lf_{\m}^{2}\Wiso\big)$,
with
$\Phi_{\m}=\sum_{k=-S}^{\m}N_{k}^{(0)}(\Mz-k)$
\cite{gar91llb}.

To obtain $\tint$ for anisotropic spins we similarly bring their
balance equations into the above form (Sec.~\ref{sec:Sani:S1}).
We use the 2-index notation of the modified factors
$\lfb_{\m}=\lfb_{\m,\m+1}$ and $\lfb_{\m-1}=\lfb_{\m,\m-1}$ plus
$\lfb_{\m,\m'}=\lfb_{\m',\m}$ [Eq.~(\ref{bar:ladder:factor:uni})] and
identify
$\Wt_{\m\tm\m'}
=
\lfb_{\m,\m'}^{2}
\Wu_{\m\tm\m'}$.
%
%
Then the generic expression~(\ref{tauint:generic}) gives the
relaxation time of the anisotropic-spin problem as
$\tint
=
(1/\Mz')
\sum_{\m}
\big(
\Phi_{\m}^{2}\big/N_{\m}^{(0)}\lfb_{\m}^{2}\Wu_{\m+1\tm\m}^{(0)}
\big)$,
with the same expression for $\Phi_{\m}$
(cf.~Eq.~(5.13) in Ref.~\cite{garchu97} and Eq.~(16) in
Ref.~\cite{gar97}).
Recall finally that, in contrast to the isotropic case, no
simplification occurs in the rates $\Wu_{\m+1\tm\m}$ because of the
non-equispaced spectrum.


\section{
Continued-fraction methods to solve recurrence relations 
}
\label{app:RR-CF}

We conclude with a hands-on summary of the method of resolution of
$3$-term recurrences of the form
%
\begin{equation}
\label{RR}
\eQ_{\ir}^{-}\ec_{\ir-1}
+
\eQ_{\ir}\ec_{\ir}
+
\eQ_{\ir}^{+}\ec_{\ir+1}
=
-\eF_{\ir}
\;,
\qquad
\ir
=
1,2,3,\ldots
\end{equation}
Here the $\eQ_{\ir}$ are given coefficients ($\eQ_{1}^{-}\equiv0$)
and the $\eF_{\ir}$ forcing or source terms.
To obtain the unknown $\ec_{\ir}$ one inserts in Eq.~(\ref{RR}) the
ansatz \cite{risken,junris85,cofkalwal94}
%
\begin{equation}
\label{C:risken:com}
\ec_{\ir}
=
\eS_{\ir}\ec_{\ir-1}+\ea_{\ir}
\;,
\end{equation}
obtaining the following relations for the ladder coefficients
$\eS_{\ir}$ and shifts $\ea_{\ir}$:
%
\begin{equation}
\label{S:a}
\eS_{\ir}
=
{}-\frac
{\eQ_{\ir}^{-}}
{\eQ_{\ir}+\eQ_{\ir}^{+}\eS_{\ir+1}}
\;,
\qquad
\ea_{\ir}
=
{}-\frac
{\eF_{\ir}+\eQ_{\ir}^{+}\ea_{\ir+1}}
{\eQ_{\ir}+\eQ_{\ir}^{+}\eS_{\ir+1}}
\;.
\end{equation}
For finite recurrences $\ec_{\ir>\Itr}=0$ for some $\Itr$.
To enforce this we set $\eS_{\Itr+1}=0$, $\ea_{\Itr+1}=0$ and iterate
{\em downwards\/} in~(\ref{S:a}) getting all $\eS_{\ir}$ and
$\ea_{\ir}$ down to $\ir=2$.
Now, to reconstruct all $\ec_{\ir}$ from Eq.~(\ref{C:risken:com}), we
only need the anchor $\ec_{1}$, which obeys:
\begin{equation}
\label{condini}
\big(
\eQ_{1}
+
\eQ_{1}^{+}
\eS_{2}
\big)
\,
\ec_{1}
=
-
\big(
\eF_{1}
+
\eQ_{1}^{+}\ea_{2}
\big)
\;,
\end{equation}
[Eq.~(\ref{RR}) at $\ir=1$ plus $\ec_{2}=\eS_{2}\ec_{1}+\ea_{2}$].
Then, starting from the so-obtained $\ec_{1}$, we iterate {\em
$\ec_{\ir}=\eS_{\ir}\ec_{\ir-1}+\ea_{\ir}$ upwards}, getting the
solution of the recurrence~(\ref{RR}).

The above is the algorithmic form (easy to implement in a computer).
The solution can also be written as
$\ec_{\ir}
=
\Big(
\prod_{k=2}^{\ir}
\eS_{k}
\Big)
\ec_{1}
+
\sum_{k=2}^{\ir}
\Big(
\prod_{j=k+1}^{\ir}
\eS_{k}
\Big)
\ea_{k}$.
To illustrate the structure:
$\ec_{5}
=
\eS_{5}\eS_{4}\eS_{3}\eS_{2}\,\ec_{1}
+
\eS_{5}\eS_{4}\eS_{3}\,\ea_{2}
+
\eS_{5}\eS_{4}\,\ea_{3}
+
\eS_{5}\,\ea_{4}
+
\ea_{5}$.
For homogeneous recurrences ($\eF_{\ir}\equiv0$ $\Rightarrow$
$\ea_{\ir}=0$) the solution simply reads
$\ec_{\ir}
=
\big(\eS_{\ir}\eS_{\ir-1}\cdots\eS_{2}\big)\ec_{1}$.

As for the name of the method, note that $\eS_{\ir}$ is given in terms
of a fraction with $\eS_{\ir+1}$ in the denominator, which can in turn
be written as a fraction with $\eS_{\ir+2}$ in the denominator, and so
on.
This furnishes the structure of a {\em continued fraction}
%
\begin{equation}
\label{continued-fraction}
C
=
\frac{p_{1}}
{q_{1}
+
{\displaystyle
\frac{p_{2}}{q_{2}+\cdots}
}
}
\;.
\end{equation}
In simple problems one may identify the continued-fraction
representation of some known function \cite{wall}, getting explicit
analytical solutions.

{\em Differential\/} recurrences like
$\dot{\ec}_{\ir}
=
\eQ_{\ir}^{-}\ec_{\ir-1}
+
\eQ_{\ir}\ec_{\ir}
+
\eQ_{\ir}^{+}\ec_{\ir+1}
+
\eF_{\ir}$
can be handled analogously for $t$-independent $\eQ_{\ir}$.
Laplace transformation plus
$\tilde{\dot{g}}(s)=s\,\tilde{g}(s)-g(t=0)$ brings the differential
equation into the form
$\eQ_{\ir}^{-}
\widetilde{\ec}_{\ir-1}
+
(\eQ_{\ir}-s)
\widetilde{\ec}_{\ir}
+
\eQ_{\ir}^{+}
\widetilde{\ec}_{\ir+1}
=
-
[\tilde{\eF}_{\ir}
+
\ec_{\ir}(0)]$.
Then, introducing some modified forcings
$\hat{\eF}_{\ir}=\tilde{\eF}_{\ir}+\ec_{\ir}(0)$
and central coefficients
$\hat{\eQ}_{\ir}=\eQ_{\ir}-s$
[cf.\ Eq.~(\ref{RR:matrix})] the equation acquires the structure of
the ordinary recurrence relation~(\ref{RR}), and as such can be
solved.%
\footnote{
%
%
In general Laplace inversion to get the actual time evolution can be
numerically problematic.
Here we would have the advantage of (i) the $\widetilde{\ec}_{\ir}(s)$
being numerically exact and (ii) owing to the high efficiency of the
continued-fraction method, the possibility of computing them at many
$s$ points.
} 

Further, the quantities in the recursions can be $\Jtr$-vectors
($\mc_{\ir}$ and $\mF_{\ir}$) with $\Jtr\times\Jtr$-matrix
coefficients $\mQ_{\ir}$.
Then one proceeds analogously: inserting the ansatz
$\mc_{\ir}=\mS_{\ir}\mc_{\ir-1}+\ma_{\ir}$
in the recurrence
$\mQ_{\ir}^{-}\mc_{\ir-1}
+
\mQ_{\ir}\mc_{\ir}
+
\mQ_{\ir}^{+}\mc_{\ir+1}
=
-\mF_{\ir}$
and obtaining the coefficients $\mS_{\ir}$ and shifts $\ma_{\ir}$, as
in Eq.~(\ref{S:a}).
The only change is that the fraction bars stand now for matrix
inversion (``from the left'' $\mathbb{A}/\mathbb{B}\;\rightarrow\;
\mathbb{B}^{-1}\,\mathbb{A}$), and one speaks of {\em matrix\/}
continued fractions.
The algorithm involves $\Itr$ inversions of $\Jtr\times\Jtr$
matrices, so reducing the storage requeriments and number of
operations from the direct inversion (or diagonalisation) of the
underlying $(\Itr\times\Jtr)\times(\Itr\times\Jtr)$ matrix problem.

In the vector case, to iterate upwards
$\mc_{\ir}=\mS_{\ir}\mc_{\ir-1}+\ma_{\ir}$, the initial condition
$\mc_{1}$ obeys the matrix version of Eq.~(\ref{condini}).
In the absence of forcing ($\mF_{\ir}=0$ $\Rightarrow$ $\ma_{\ir}=0$;
e.g., the $0$th order Eq.~(\ref{DME:chain:0:ex})) the solution of such
$\Jtr\times\Jtr$ system involves an overall multiplicative constant.
One can add an extra equation to fix this, getting an augmented
$(\Jtr+1)\times\Jtr$ system \cite[App.~A]{garzue2004}, which can be
solved by methods appropriate for more equations than unknowns (e.g.,
$QR$-decomposition), yielding the required $\mc_{1}$.
In our spin problems the normalisation condition
$1=\sum_{\n}\dm_{\n\n}=\sum_{\n}(\mc_{\n})_{\n}$
involves all $\mc_{\n}$ and cannot be used as the extra equation for
$\mc_{\n=-S}$ (which plays the role of $\mc_{1}$).
However, we can fix arbitrarily one component, e.g.,
$(\mc_{-S})_{-S}=\mathrm{const}$ (the extra equation) and normalise
the solution at the end.
A practical choice is a Gibbsian weight
$(\mc_{-S})_{-S}\sim\Z^{-1}\exp(-\el_{-S}/\kT)$.

To conclude, as the indices $(\n,\m)$ can be half-integers (when $S$
is so), we employ in the numerical implementation some integer indices
$\ir=\n+(S+1)$ and $\jr=\m+(S+1)$, running from $1$ to $2S+1$
($=\Itr=\Jtr$).
For integer $S$ the equations can also be handled as two-sided
recurrences $-\Itr\leq\ir\leq\Itr$ (then the ansatz and initial
conditions are slightly modified; see
Refs.~\cite[App.~A]{risken,garzue2004}).
Although this may enhance stability is some problems, we have used the
general protocol allowing for non-integer spins.


\section*{References}



\end{document}